\documentclass[fleqn,usenatbib]{mnras}
\usepackage{newtxtext,newtxmath}
\usepackage[T1]{fontenc}
\usepackage{graphicx}	
\usepackage{amsmath}	
\usepackage{CJKutf8}
\usepackage{array}

\title{Large amplitude periodic outbursts and long period variables in the VVV VIRAC2-$\beta$ database}

\author[Z. Guo et al.]{Zhen Guo$^{1}$\thanks{E-mail: z.guo4@herts.ac.uk},
P. W. Lucas$^{1}$,
L. C. Smith$^{2}$,
 C. Clarke$^{2}$,
 C. Contreras Pe{\~n}a$^{1,3}$, 
 \newauthor
 A. Bayo$^{4, 5}$, C. Brice{\~n}o$^{6}$, J. Elias$^{7}$, 
 R. G. Kurtev$^{4,8}$, J. Borissova$^{4,8}$, J. Alonso-Garc\'{i}a$^{9,8}$, 
 \newauthor
 D. Minniti$^{10,11}$,
 M. Catelan$^{8,12,13}$, F. Nikzat$^{8,12}$, C. Morris$^{1}$ and N. Miller$^{1}$
\\
$^{1}$Centre for Astrophysics Research, University of Hertfordshire, Hatfield AL10 9AB, UK\\
$^{2}$Institute of Astronomy, University of Cambridge, Madingley Road, Cambridge, CB3 0HA, UK\\
$^{3}$Department of Physics and Astronomy, Seoul National University, 1 Gwanak-ro, Gwanak-gu, Seoul, 08826, Republic of Korea\\
$^{4}$Instituto de F{\'i}sica y Astronom{\'i}a, Universidad de Valpara{\'i}so, ave. Gran Breta{\~n}a, 1111, Casilla 5030, Valpara{\'i}so, Chile\\
$^{5}$N\'ucleo Milenio de Formaci\'on Planetaria (NPF), Casilla 5030, Chile\\
$^{6}$Cerro Tololo Inter-American Observatory/NSF’s NOIRLab, La Serena, Casilla 603, Chile\\
$^{7}$SOAR Telescope/NSF’s NOIRLab, La Serena, Casilla 603, Chile\\
$^{8}$Millennium Institute of Astrophysics, Nuncio Monse{\~n}or Sotero Sanz 100, Of. 104, Providencia, 7820436 Macul, Santiago, Chile\\
$^{9}$Centro de Astronom{\'i}a (CITEVA), Universidad de Antofagasta, Av. Angamos 601, 02800 Antofagasta, Chile\\
$^{10}$Departamento de Ciencias Fisicas, Universidad Andres Bello, Republica 220, 8320000 Santiago,
 Chile\\
$^{11}$Vatican Observatory, V00120 Vatican City State, Italy\\
$^{12}$Instituto de Astrof{\'i}sica, Facultad de F{\'i}sica, Pontificia Universidad Cat{\'o}lica de Chile, Av. Vicu{\~n}a Mackenna 4860, 7820436 Macul, Santiago, Chile\\
$^{13}$Centro de Astroingenier{\'i}a, Pontificia Universidad Cat{\'o}lica de Chile, Av. Vicuña Mackenna 4860, 7820436 Macul, Santiago, Chile\\
}

\date{Accepted XXX. Received YYY; in original form ZZZ}
\pubyear{2022}

\begin{document}
\label{firstpage}
\maketitle

\begin{abstract}

The VISTA Variables in the Via Lactea (VVV) survey obtained near-infrared photometry toward the Galactic bulge and the southern disc plane for a decade (2010 -- 2019). We designed a modified Lomb-Scargle method to search for large-amplitude ($\Delta K_{s, 2\%-98\%}$~>~1.5~mag) mid to long-term periodic variables ($P > 10$~d) in the 2nd version of VVV Infrared Astrometric Catalogue (VIRAC2-$\beta$). In total, 1520 periodic sources were discovered, including 59 candidate periodic outbursting young stellar objects (YSOs), based on the unique morphology of the phase-folded light curves, proximity to Galactic H{\sc ii} regions and mid-infrared colours. Five sources are spectroscopically confirmed as accreting YSOs. Both fast-rise/slow-decay and slow-rise/fast-decay periodic outbursts were found, but fast-rise/slow-decay outbursts predominate at the highest amplitudes. The multi-wavelength colour variations are consistent with a variable mass accretion process, as opposed to variable extinction. The cycles are likely to be caused by dynamical perturbations from stellar or planetary companions within the circumstellar disc.  An additional search for periodic variability amongst YSO candidates in published {\it Spitzer}-based catalogues yielded a further 71 candidate periodic accretors,  mostly with lower amplitudes. These resemble cases of pulsed accretion but with unusually long periods and greater regularity.  The majority of other long-period variables are pulsating dusty Miras with smooth and symmetric light curves. We find that some Miras have redder $W3 - W4$ colours than previously thought, most likely due to their surface chemical compositions.
\end{abstract}


\section{Introduction}
\label{sec:intro}
 The periodic nature of variable stars results from intrinsic or external mechanisms. In either case, the periodicity provides valuable information about the stellar properties, circumstellar matter, or a companion object. For instance, periodic variations of young stellar objects (YSOs) are often associated with stellar or disc rotation and dynamical perturbations by stellar or planetary companions. Time series photometry are applied to study the magnetic field \citep{Gregory2012}, angular momentum evolution \citep{Rebull2018}, and inner disc structures \citep{Bouvier2007} of pre-main-sequence (PMS) stars.

The PMS evolution is shaped by the mass accretion process, where the mass accretion rate is higher and more unstable during the protostellar stage \citep[e.g.][]{Hartmann1998}.  Periodic accretion bursts are rarely observed phenomena among YSOs: their periodic nature is in contrast to unpredictable EXor and FUor outbursts \citep[see review by][and references therein]{Audard2014}. Four cases of high-amplitude ``periodic outbursts'' (greater than 1.5 magnitudes) with period much longer than the stellar spin period (c. a day to a week) have been found \citep[][]{Hodapp1996, Hodapp2012, Muzerolle2013, Hodapp2015, YHLee2020, Dahm2020}. The physical mechanisms behind them are still unclear, but at least one of these is not a binary star system, so the perturbations may be due to a planet or low-mass brown dwarf \citep{Dahm2020}. In the case of L1634~IRS~7 \citep[$P=37$~d]{Hodapp2015}, authors argued that the circumstellar disc would be disrupted in a short timescale by a close-in companion, therefore a model in which the cyclical accretion is triggered by trapped disc material via the magnetic gating procedure \citep{DAngelo2012} was favoured.  In addition, a very few small-amplitude periodic bursting candidates have been found amongst Class I YSOs, such as source 533 in Cha I \citep{Flaherty2016} with P = 36~d and $\Delta [3.6] \sim 0.3$~mag. 

Cyclically modulated accretion is also detected in a few disc-bearing Class II young binary systems, with low to intermediate photometric amplitudes, such as the `pulsed accretion'' flares on TWA~3A \citep[P= 34.9 d]{Tofflemire2017b} and DQ~Tau \citep[P = 15.8 d]{Mathieu1997, Tofflemire2017, Kospal2018}. \citet{Muzerolle2019} proposed a model based on the optical to near-infrared colour variation to explain the variability of DQ~Tau, where the accretion flare is triggered by the merging of two circumstellar accretion discs during the closest approach of the binary orbits. On the theoretical side,  the current generation of simulations predicts that binaries in eccentric orbits are able to periodically modulate the mass accretion process by one order of magnitude:  pure hydrodynamic simulations of pulsed accretion were performed by \citet{Munoz2016} and magneto-hydrodynamic simulations of of wide binaries (P $> 20$ yr) by \citet{Kuruwita2020}. 

However, the total known sample of periodic accretors is extremely small given that most stars were born in rich clusters and binarity is common \citep{Lada2003, Tobin2016}. Moreover, recent observations detected newborn planets around pre-main sequence stars by H$\alpha$ emission from the circumplanetary disc \citep[e.g.][]{Haffert2019}, evidence of dynamical perturbation in the circumstellar disc \citep{Boccaletti2020}, and multiple ring-gap structures in the circumstellar disc of a 0.5 Myr Class I YSO \citep{SeguraCox2020}. Numerical simulations suggest that close-in hot-Jupiters on eccentric orbits are able to significantly modulate the mass accretion rate of the parent star \citep{Teyssandier2020}. Therefore, a larger sample of periodically accreting YSOs is desired to understand the effect on the star formation process of (sub)stellar companions or planets. Such a sample would also provide invaluable information on the behaviour of accretion and the evolution of circumstellar discs under dynamical perturbations, which might inform studies of aperiodic FUor and EXor events.

The VISTA Variables in the Via Lactea survey obtained time series near-infrared photometry of more than 700 million sources toward the Galactic bulge and the southern disc plane for a decade (2010 -- 2019) \citep{Minniti2010, Saito2012, Minniti2016}. Most of the data were taken in the $K_s$ band (2.15~$\mu$m), supplemented by a few multi-colour epochs of contemporaneous $J$ (1.25~$\mu$m), $H$ (1.65~$\mu$m) and $K_s$ (2.15~$\mu$m) data. With ability to monitor embedded long-term variables, the decade-long VVV near-infrared survey is ideal to discover long period variables \citep[LPVs,][]{Catelan2013}. Numerous periodic variables have been identified from the VVV survey, e.g. the VIVACE catalogu \citep{Molnar2021}, and previous works on individual Galactic regions \citep[e.g.][]{Molina2019, Medina2021, Alonso-Garcia2021}, searches for short-timescale pulsating sources \citep{Bhardwaj2017, Minniti2017} and eclipsing binaries \citep{Botan2021}. \citet[][hereafter Paper I]{Contreras2017} identified 816 variables ($\Delta K_s$~$>$~1~mag) from the first 4.5 years of the VVV data for the Galactic mid-plane region, including 247 long-term periodic variables ($P > 100$~d), among them 65 YSO candidates. In a more recent search of the VVV 4th Data Release\footnote{Available at \url{http://vsa.roe.ac.uk}}, 105 large-amplitude ($\Delta K_s > 3$~mag)  VVV variables were identified, including some periodic stars (Lucas et al., in prep). Coupled with the decade-long VVV/VVVX light curves and spectroscopic follow-up observations, \citep{Guo2021} found that 13 large-amplitude VVV variables are periodic / quasi-periodic YSOs, mostly having H{\sc I} recombination line signatures of magnetically controlled accretion \citep[see also][]{Contreras2017b, Guo2020}. Although lacking direct spectroscopic evidence at multiple phases of the accretion cycle, at least two of these YSOs (VVVv32 and DR4\_v55) were identified as probable cases of cyclical accretion, as opposed to variable extinction, via the changes in their spectral energy distributions \citep[SEDs, see][]{Guo2021}.

In this work, we will study the periodic outbursting behaviour on YSO candidates, as observational consequences of dynamical perturbation from low-mass companions in the young circumstellar disc. Two searches were conducted by our Lomb-Scargle based frequency analysis pipeline, targeting relatively large amplitude variables and previously identified YSO candidates in the up-to-date VVV photometric catalogue (VIRAC2-$\beta$, see \S\ref{sec: pre-selection}). Initial classification was performed using light curve morphology in order to distinguish candidate outbursting YSOs from post-main-sequence LPVs. Additional supporting evidence of PMS status was also required for sources in our final lists. The physical mechanisms behind these periodic variables were then further investigated via multi-wavelength photometry and the shape of phase-folded light curves.  

This paper is organised as follows. Pre-selection of targets and light curve analysis techniques are described in \S\ref{sec: lc}. Results of light curve analysis and classification of periodic sources are shown in \S\ref{sec:res}. In \S\ref{sec:dis}, we discuss the colour behaviour and phase-folded light curves of periodic sources, and we present some detailed discussions of several individual sources. The paper is concluded in \S\ref{sec: con}.

\section{Light curve analysis}
\label{sec: lc}
In this section, we will introduce the selection criteria of large-amplitude variables from VIRAC2-$\beta$ and the light curve analysis pipeline designed to identify periodic sources.

\subsection{Pre-selection of variable candidates}
\label{sec: pre-selection}

The VVV Infrared Astrometric Catalogue \citep[][and Smith et al., in prep]{Smith2018}, version 2, VIRAC2, provides time series photometry and five parameter astrometric fits for sources using VVV/VVVX observations from 2010 to 2019. Individual sources were extracted by profile fitting photometry using the DoPHOT program \citep[][]{Schechter1993, Garcia2018}, with improved photometric calibration.  In this work, we searched a development version of VIRAC2, hereafter VIRAC2-$\beta$, to identify mid-to-long term ($P > 10$~day) periodic variables. 

Several criteria were set before the periodicity search. In our first search, a minimum variation amplitude, $\Delta K_s$ (defined using the difference between the 2\% and 98\% percentiles of the ordered $K_s$ magnitudes) was set at 1.5 mag. This threshold selects stars having amplitudes comparable to episodic accretion outbursts \citep[e.g.][]{Dahm2020, YHLee2020} whilst keeping the number of candidates small enough for visual inspection. Some low-amplitude post-main-sequence variables (e.g. Cepheids and RR~Lyrae variables) were therefore eliminated.  We only selected targets detected in more than 100 $K_s$ epochs of observation in VIRAC2-$\beta$, with median $K_s$ ranging between 11 to 16 mag in order to avoid heavily-saturated and faint sources. Each epoch corresponds to a stacked pawprint image: there are typically at least two contemporaneous pawprint image stacks for each source, depending on location within the six point VISTA tiling pattern. Cuts based on the VIRAC2-$\beta$ astrometry were also used: sources with proper motion $\mu>0.5"/$yr, parallax $\varpi > 1"$, or unit weight error $uwe>1.4$\footnote{The unit weight error is the square-root of a reduced $\chi$-squared parameter. $uwe = (\sum({r_x}^2/{e_x}^2 + {r_y}^2/{e_y}^2) / (2n - 5) )^{0.5}$. For sources with 5 astrometric parameter solutions, $r_x$, $r_y$ are the residuals in RA and Dec, $e_x$, $e_y$ are the centroid errors in RA and Dec and n is the number of epochs. Stars with good astrometric solutions and centroid uncertainties have $uwe$ close to unity.} were excluded since inspection of the database has shown that most such entries in the database do not correspond to sources with reliable light curves or astrometry. 

In total, 208410 candidate variable sources were selected in the VIRAC2-$\beta$ catalogue, following a power-law distribution against $\Delta K_{s}$ ($N \propto \Delta K_s^{-1.24}$, see numbers in Table~\ref{tab:n_source}). Since the VVV/VVVX epochs are not evenly sampled, the VVV/VVVX $K_s$ light curves are initially averaged within 1~day bins to avoid over-sampling at short timescales. Data points lying more than 5$\sigma$ from the mean of individual bins are excluded (where $\sigma$ is the standard deviation in each bin when more than three detections per bin). Therefore, some sources have $\Delta K_s$ slightly smaller than 1.5 mag, as outliers were cleared. The light curve reduction and period searching pipeline was run on the University of Hertfordshire's high-performance computing facility\footnote{\url{http://uhhpc.herts.ac.uk}}. 

\begin{table}
\caption{Number of variable sources in VIRAC2-$\beta$}
\centering
\begin{tabular}{c c c c c}
\hline
\hline
$\Delta K_s$ (mag) & All & $\delta_{\rm phase} \le 0.3$ & ratio$^\star$ (\%)\\
\hline
$1.0 \le$ $\Delta K_s$ $<$ 1.5 &      2295081  &  - &  - \\
$1.5 \le$ $\Delta K_s$ $<$ 2.0 &      181703  &  1716 &  0.94\\
$2.0 \le$ $\Delta K_s$ $<$ 2.5 &       18795  &  831 &  4.4\\
$2.5 \le$ $\Delta K_s$ $<$ 3.0 &        5633  & 407 &  7.2\\
$3.0 \le$ $\Delta K_s$ $<$ 3.5 &        1700  & 146 &  8.6\\
$\Delta K_s$ $\ge$ 3.5 & 579 & 51 &  8.8\\
\hline
\hline
\end{tabular}
\flushleft{$\star$: the ratio between the number of all variables in particular amplitude range and the number of sources have $\delta_{\rm phase} \le 0.3$ in the same amplitude range. {  This table has been modified after the initial submission.}}

\label{tab:n_source}
\end{table}

\begin{figure}
\centering
\includegraphics[width=3.in,angle=0]{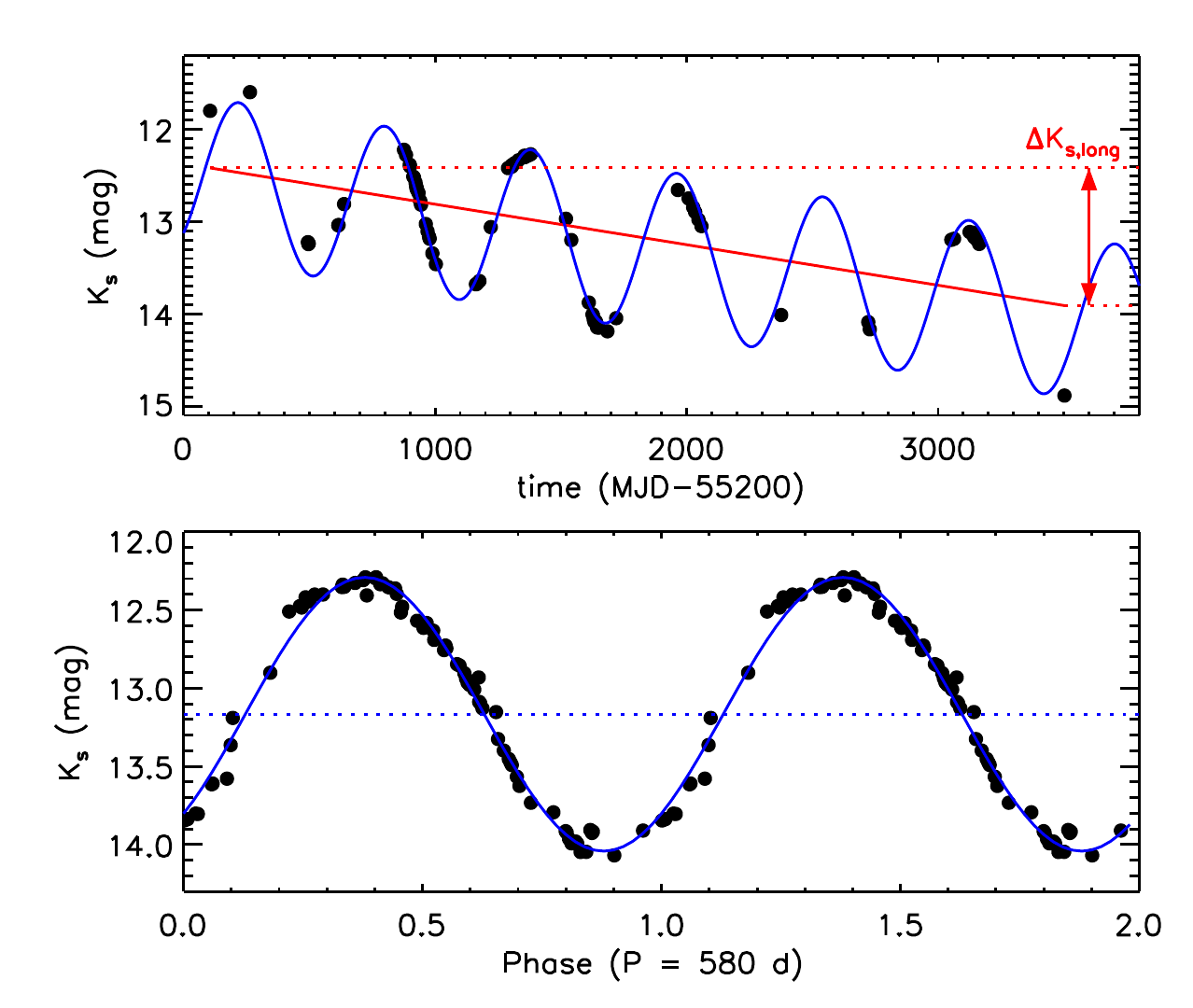}
\caption{An example of the period-extracting method developed in this paper.  {\it Top}: binned $K_s$ light curve with outliers removed. The linear fitting result of the general trend throughout the light curve is presented by the red solid line, along with the definition of $\Delta K_{s,\rm long}$. The modified sinusoidal fitting result is presented by the blue solid curve. {\it Bottom}: The linear trend removed phase-folded light curve, with the best-fitting sinusoidal template.}
\label{fig:example}
\end{figure}
\begin{figure*}
\includegraphics[height=2.5in,angle=0]{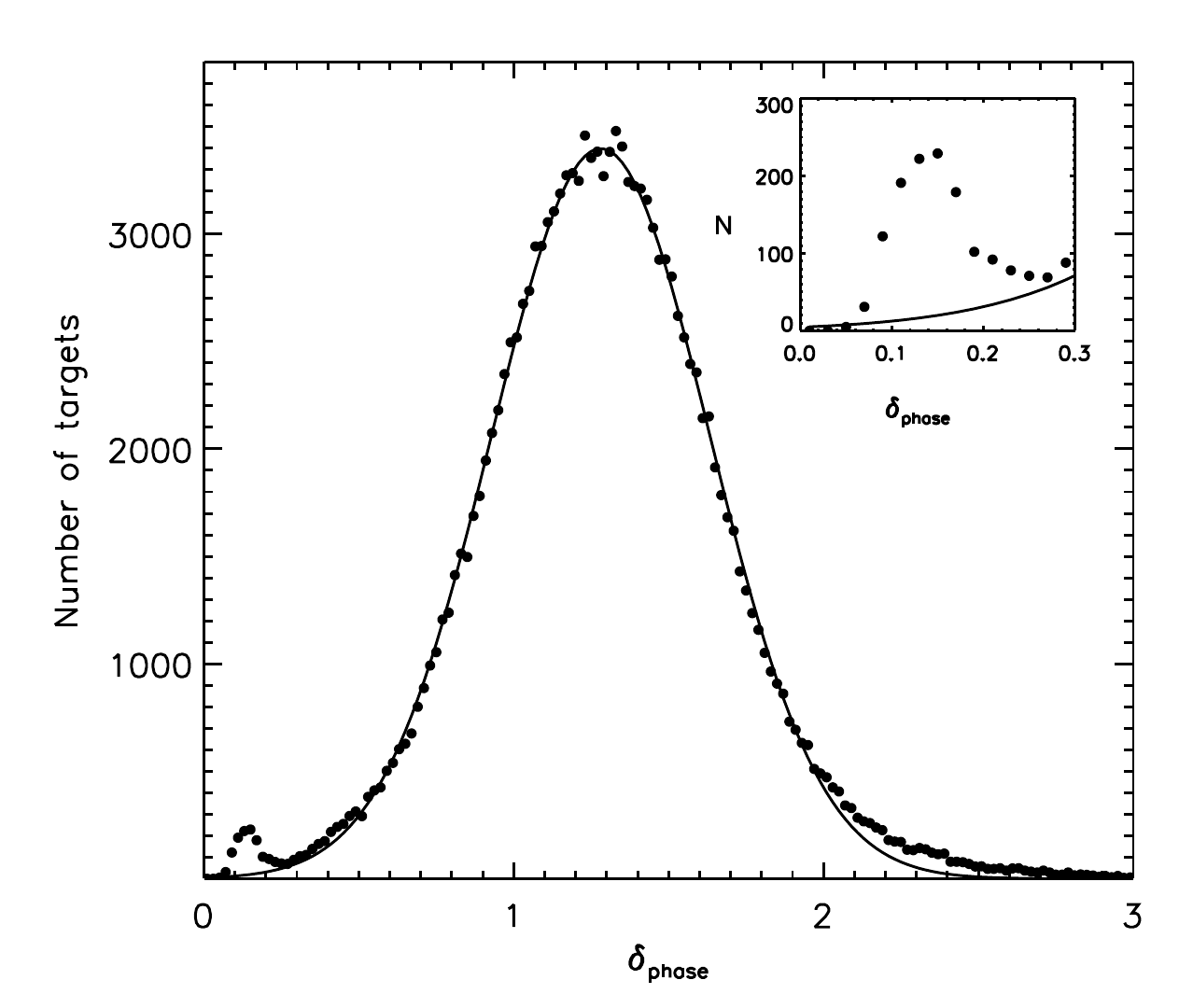}
\includegraphics[height=2.5in,angle=0]{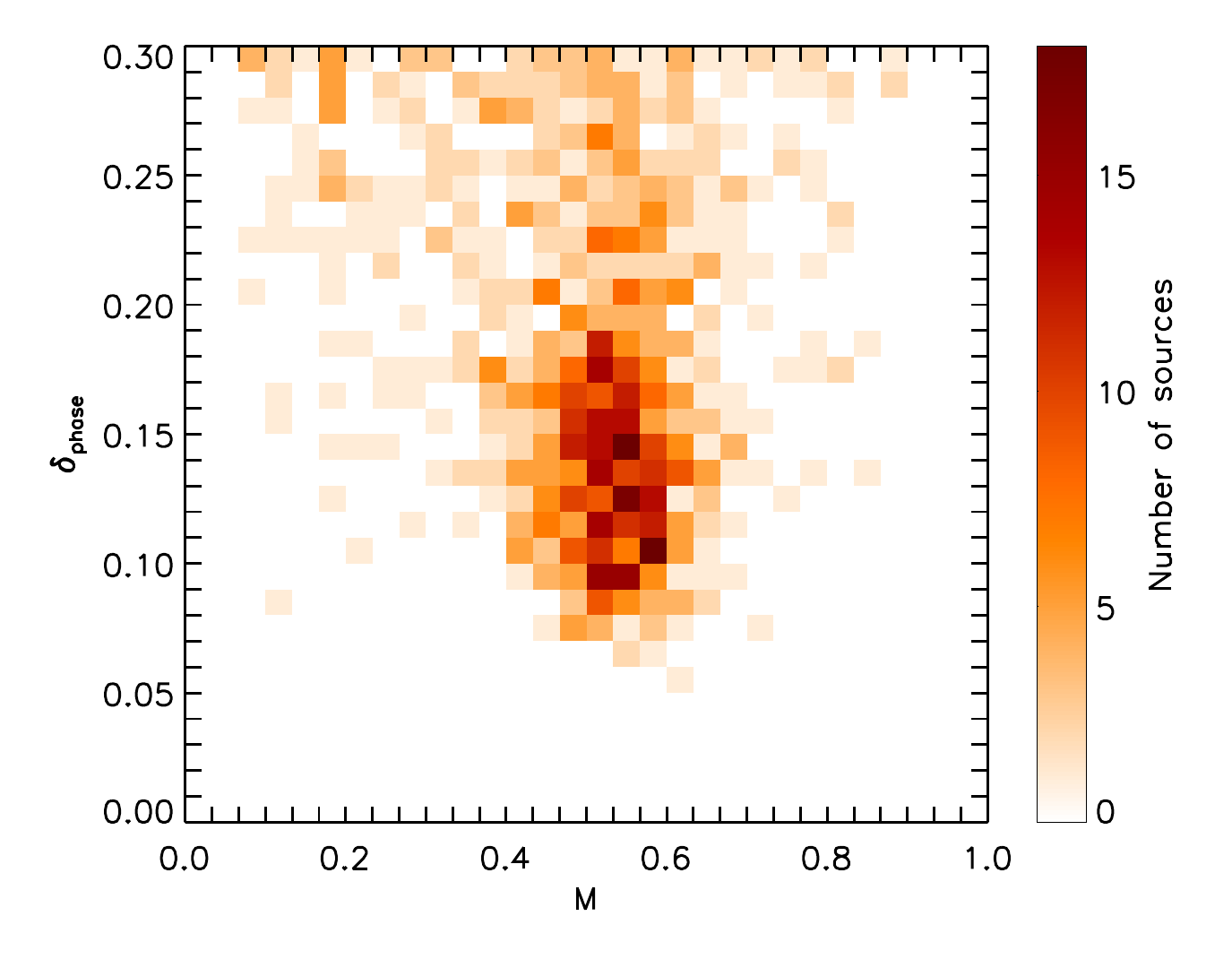}
\caption{{\it Left}: A histogram of $\delta_{\rm phase}$ value of selected VVV variable sources ($\Delta K_s > 1.5$~mag). A Gaussian distribution, shown by the solid line, is fitted to the histogram data. {\it Right}: A density map of $M$ and $\delta_{\rm phase}$ of periodic candidates.}
\label{fig:hist_delta}
\end{figure*}

\begin{figure} 
\centering
\includegraphics[width=3in,angle=0]{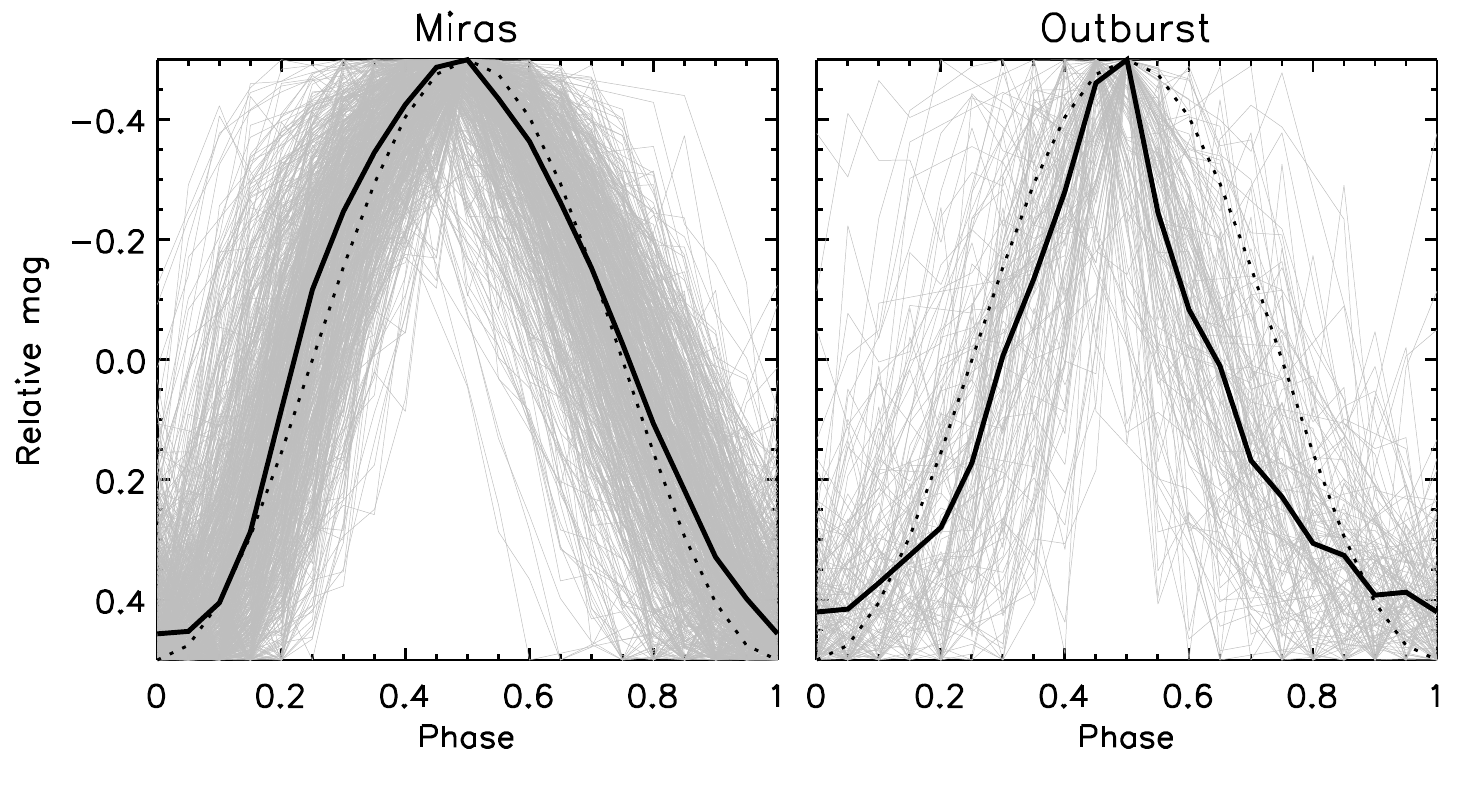}
\caption{Phase curves (grey) of individual non-saturated Miras ({\it left}) and periodic outbursting candidates ({\it right}). The median phase curves of Miras and periodic outbursting candidates are shown by solid black lines. Sinusoidal curves, shown by dotted lines, are presented as comparisons. }
\label{fig:phase_curves}
\end{figure}

\subsection{Period extraction method}

We built the period-searching pipeline based on the generalised Lomb-Scargle periodogram \citep[GLS;][]{Zechmeister2009}, with an additional linear fitting procedure to remove long-term variation.   Figure~\ref{fig:example} demonstrates our period extraction pipeline. We first applied the GLS periodogram to light curves, with fitting periods set between 10 to 1500~days. We then subtract the best sinusoidal fit provided by the GLS program from the data, and fit the residuals with a linear trend. Although polynomial or Gaussian curves may provide a better fit, the linear slope is chosen to avoid biasing the fitted period. The GLS program is then re-applied to the resulting light curve once the linear trend is removed, where the final output period is produced. A parameter $\delta_{\rm sine}$ is defined as the ratio between the standard deviation of the residual and the amplitude of the fitted sine wave (i.e. from zero to the peak).

To quantify the periodicity and the asymmetry of the time series, two parameters ($\delta_{\rm phase}$ and $M$) were defined based on the phase-folded light curves. Inspired by the phase dispersion minimisation method \citep{Stellingwerf1978}, we sampled each phase-folded light curve into ten equal phase bins, and generated a ``phase curve'' by simply linking the median magnitude within each phase-bin. The number of bins was selected as a compromise between the accuracy of the fitting and the typical morphology of the phase curve, while short-timescale variability was not considered. We then subtracted the phase curve from the phase-folded light curve, and defined $\delta_{\rm phase}$ as the ratio between the standard deviation of the residual and the photometric amplitude. The $\delta_{\rm phase}$ parameter was introduced to define the quality of the period fit in a manner that has no dependence on the analytical form of the variation. Therefore, $\delta_{\rm phase}$ was chosen in preference to $\delta_{\rm sine}$ for this reason. The distribution of $\delta_{\rm phase}$ of variable sources from the VIRAC2-$\beta$ catalogue is presented in Figure~\ref{fig:hist_delta}. In a general view, $\delta_{\rm phase}$ follows a Gaussian distribution centring at 1.3, indicating the vast majority of variable sources are aperiodic. A much smaller second peak is seen towards the lower end of the histogram ($\delta_{\rm phase}  \le 0.3$) that contains variables with periodic variations.  

The $M$-value, ranging between 0 to 1, is defined as the proportion of the phase curve that is brighter than its median value. The concept of $M$-value was originally proposed in \citet{Cody2014}, though in this work $M$ is measured on the phase curve instead of individual data point due to the sparse cadence of the VVV/VVVX time series. As an example, symmetric sinusoidal curves have $M = 0.5$, while a phase curve with a narrow dip below the typical brightness would have $M > 0.5$. Conversely, a phase curve with a narrow outburst above the typical brightness would have $M < 0.5$. This parameter is designed to separate periodic bursts from sinusoidally pulsating dusty Asympototic Giant Branch (AGB) stars (also known as Miras) and young variable stars with extinction dips \citep[e.g.][]{Bouvier2007}. 

A density map of the light curve symmetry parameter ($M$) vs. $\delta_{\rm phase}$ is presented in Figure~\ref{fig:hist_delta} for sources with $\delta_{\rm phase} \le 0.30$. A cluster of periodic sources with symmetric light curves is seen at $0.40  \le M  \le 0.65 $ and $\delta_{\rm phase}  \le 0.20$, which is the region expected to be {  mostly} occupied by dusty Miras, the most common high-amplitude periodic infrared variables. 

The number of all variables and sources with $\delta_{\rm phase} \le 0.3$ are listed in Table~\ref{tab:n_source}. Source with high level of periodicity in their light curves do not follow the same power-law distribution seen between $\Delta K_s$ and the number of all variables ($N \propto \Delta K_s^{-1.24}$). At the high-amplitude end, around 8\% of variables have $\delta_{\rm phase} \le 0.3$, and this proportion is slightly decreased to 7.2\% within the amplitude range between 2.5 and 3.0 mag. When approaching to the lower amplitude range, the fraction significantly dropped to less than 1\%, suggesting the periodic variability is either extremely rare among these variables, or are severely veiled by irregular variations, or the proportion of false positive detections rise at lower amplitude. Therefore, the threshold of $\Delta Ks \ge 1.5$~mag, in our initial periodic search, is a reasonable choice, to balance the computational budget and statistical expectations of the outcomes. Also, it becomes impractical to visually inspect larger samples. In addition, as aforementioned, this criterion also helped to eliminate contamination from post-main-sequence variables.

Visual inspection was performed on phase-folded light curves of 3396 sources that have $\delta_{\rm phase} \le 0.5$, to verify the periodicity. In total, 1520 sources were finally confirmed as periodic variables. These were then sorted into sub-groups based on period-to-period variation (i.e. $\delta_{\rm phase}$) and phase curve morphology. Most sources identified by our pipeline are high-amplitude Galactic Miras, with smooth and sinusoidal-like light curves. We note that our period extraction pipeline has limited capability to discover stellar rotation (typically with a period less than 10 days among YSOs), sources with multiple periods, and eclipsing binaries. 

\section{Result}
\label{sec:res}
\subsection{Selection of periodic outbursting YSOs}

Periodic outbursting YSO candidates are first identified by the morphology of their phase-folded light curves (i.e. $M < 0.5$) and supporting evidence for pre-main sequence status is then sought. After visual inspection of the light curves, 84 sources are selected as periodic outbursting candidates. Compared with dusty Mira variables, candidate periodic outbursting YSOs have a wide range of periods, much larger values of $\delta_{\rm sine}$ and smaller $M$ by definition. Normalised phase curves of non-saturated dusty Miras and periodic outbursting candidates are compared in Figure~\ref{fig:phase_curves}, with median phase curves computed for each category. Periodic outbursting candidates have a much narrower peak in the median phase curves than that seen in the sinusoidal light curves of Miras. Readers should note that at this initial stage, the selection of periodic outbursting candidates is purely morphological. More information is desired to confirm the accretion activity on these sources.

To identify the most likely YSOs in this set, the positions of the 84 sources were cross-matched with two catalogues of YSO candidates that were based on {\it WISE} \citep{Marton2016} and {\it Spitzer/IRAC} \citep[SPICY;][]{Kuhn2021} mid-infrared photometric colour criteria. We also visually inspected infrared images of our candidates from the {\it WISE} \citep{Wright2010} and {\it Spitzer} \citep{Werner2004} satellites to identify nearby H{\sc ii} regions, as tracers of ongoing star formation. The initial set of 84 candidate periodic outbursting YSOs was thereby refined to yield a final catalogue of 59 sources and the remaining 24 sources are regarded as lower probability YSO candidates. The 59 best YSO candidates include three that were previously spectroscopically confirmed as YSOs \citep{Guo2021}, 34 that were identified from the aforementioned YSO catalogues. Phase-folded light curves of the 59 periodic outbursting YSO candidates are shown in Figure~\ref{fig:PB1} to \ref{fig:PB3}.

The great majority of periodic outbursting sources that have mid-infrared colours from {\it WISE} (30~sources) do not match the colours of dusty Miras. We note that candidates lacking complete {\it WISE} colour data are necessarily fainter than the luminous Galactic Mira population.  We measured the gradient of the infrared spectral energy distribution \citep{Lada1987}, known as the spectral index (${\alpha_{\rm class}}$, 2 - 24 $\mu$m), for the 59 YSO candidates, using photometric data from the {\it Spitzer}/GLIMPSE survey \citep{Benjamin2003}, the {\it Spitzer}/MIPSGAL survey \citep{Carey2009} and the ALLWISE data release of the {\it WISE} survey \citep{Wright2010}. The $\alpha_{\rm class}$ of 44 sources were measured, of which 13 sources were identified as Class I protostars ($\alpha_{\rm class} \geq$ 0.3), 20 as flat-spectrum sources (-0.3 < $\alpha_{\rm class}$ < 0.3), and 11 as Class II YSOs ($\alpha_{\rm class} \leq$ -0.3). 

The 59 candidates, named with the prefix ``VVV\_PB'', are listed in Table~\ref{tab:PB}. Candidates present in either the SPICY catalogue or the \citet{Marton2016} catalogue (match radius $< 1$'') are listed as ``YSO''. Sources having at least 5 existing YSOs located within $3'$ are marked as ``near'', since they are likely located in a star-forming region. A similar YSO selection criterion was applied for highly variable VVV sources in Paper I using star-formation indicators from SIMBAD. In this work, using randomly sampled sky positions, we expect that 35\% of targets in the VVV field have more than 5 YSO candidates (from aforementioned catalogues) located with in $3'$. When applying the same method to YSOs candidates, 65\% of YSOs have more than 5 other YSO candidates located with in 3 arcmin. Sources located in or near H{\sc ii} regions (within $5'$) are marked as ``y'' or ``y?'' via visual inspections of {\it Spitzer} and {\it WISE} multi-colour images. Candidates lacking any nearby YSOs or H{\sc ii} regions were rejected from the final list.

 Using the same pipeline, in the VIRAC2-$\beta$ catalogue, we performed a second periodicity search of all non-saturated and reasonably bright sources (median $K_s$ between 11 and 16~mag) identified as YSO candidates in two {\it Spitzer}-based YSO catalogues \citep{Robitaille2008,Kuhn2021} without any amplitude cut. We found 253 sources that are very probably Miras, recognised by their characteristic phase-folded light curves, periods, {\it WISE} magnitudes and colours, despite their previous classification as YSO candidates in the SPICY catalogue. After removing the probable Miras, we found 870 stars, about 1.8\% of all YSO candidates, have stable periods confirmed after visual inspections. These are likely to be a mixture of YSOs with periodic extinction, outbursting YSOs and a small number of  Mira variables with unusual colours. At the large-amplitude end ($\Delta K_{s} > 1.5$~mag), 247 periodic sources were found, which makes up 13\% of all YSO candidates within the same amplitude range. The proportion of periodic variables steeply drops to 1.4\% among smaller-amplitude variables. This may be attributed to a mixture of small-amplitude periodic and aperiodic variability in YSOs, coupled with photometric uncertainties, therefore leading to large $\delta_{\rm phase}$ values and exclusion from our selection.

Upon visual inspection of the light curves, we found that 95 sources have outbursting morphology in the phase-folded light curves, about 10\% of all periodic YSO candidates within both the large- and small-amplitude groups. This includes 24 large-amplitude sources ($\Delta K_s > 1.5$~mag) already identified as periodic outbursting YSO candidates in our first search. The small-amplitude periodic bursting candidates found in this search resemble cases of the pulsed accretion phenomenon seen in eccentric young binaries \citep[e.g.][]{Tofflemire2017, Muzerolle2019}. The 71 extra sources found in the second search are named as ``VVV\_PB\_60'' to ``VVV\_PB\_130'', listed in Table~\ref{tab:low_amp}. The phase-folded light curves of these 71 sources are presented in the online supplementary Figures.

\begin{figure}
\includegraphics[height=2.7in,angle=0]{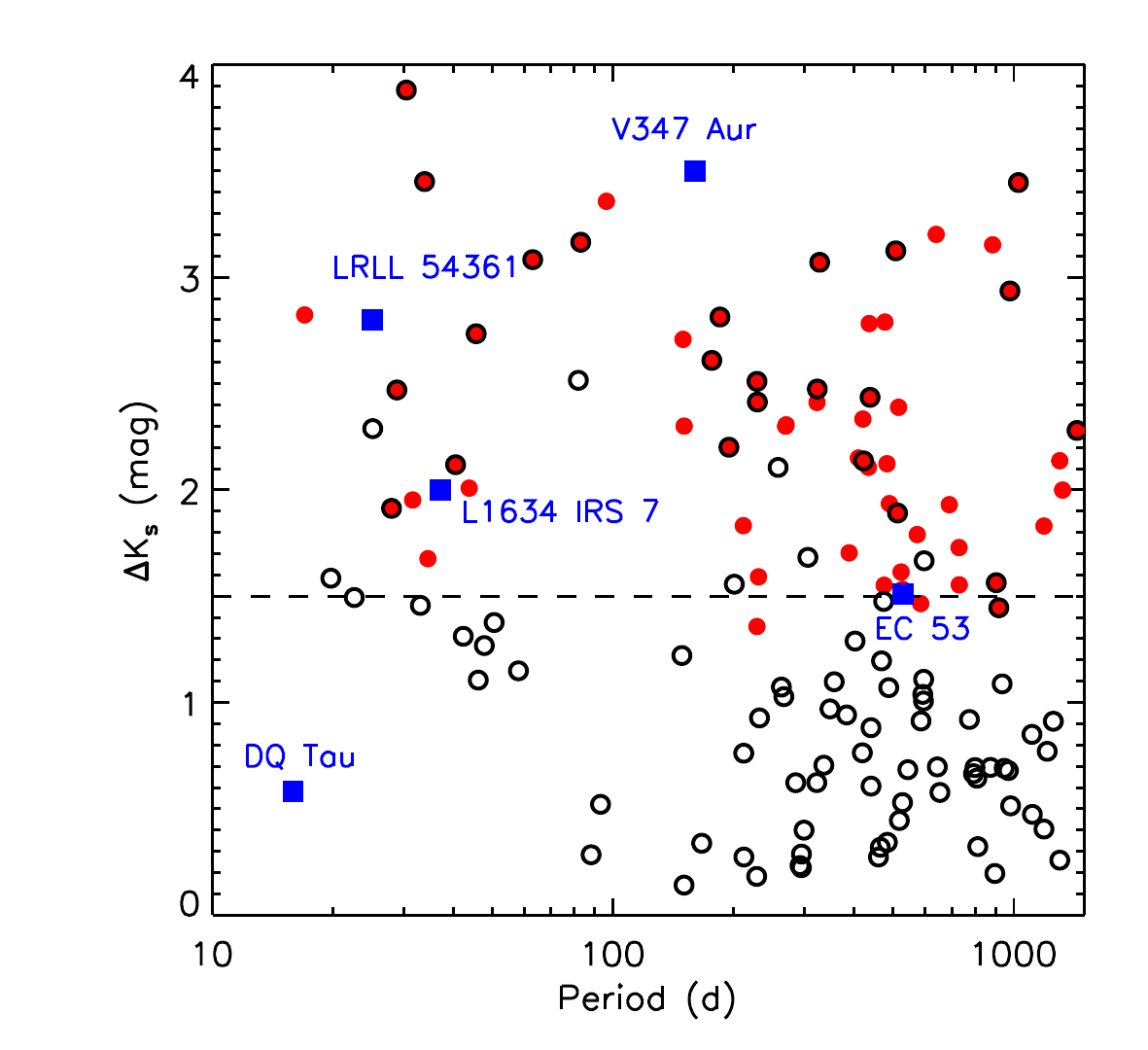}
\includegraphics[height=2.7in,angle=0]{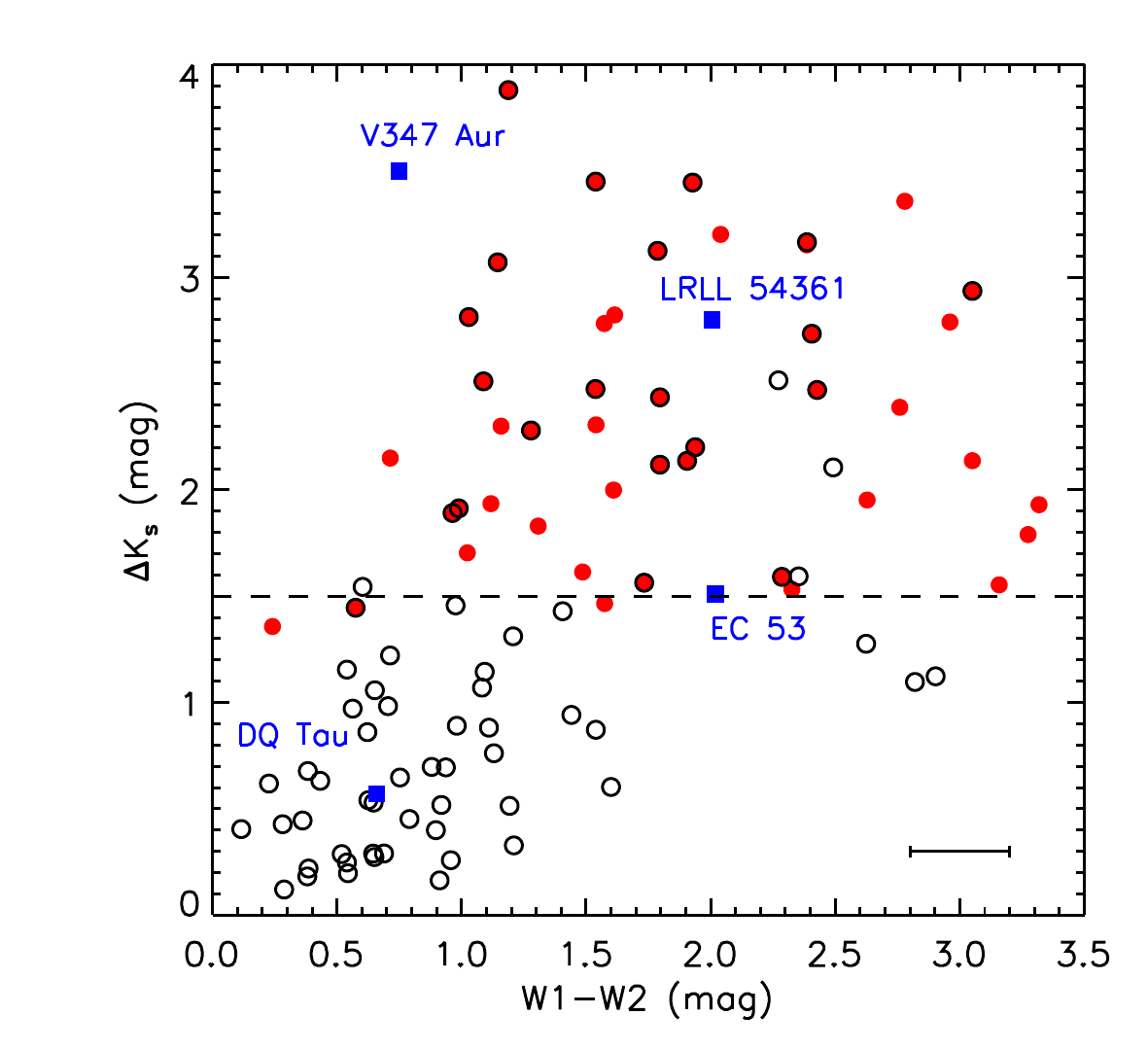}
\caption{Period vs. amplitude (upper panel) and {\it WISE} $W1-W2$ colour vs. amplitude (lower panel) of periodic outbursting YSO candidates. Candidates from archived YSO catalogues are shown by open circles, and candidates discovered by our original search are marked in red. The typical variation amplitude of $W1-W2$ colour in our sample is presented by the ``error bar'' in the lower panel.  Five previously detected periodic bursters and pulsed accretors are presented by blue squares. The amplitude of V347~Aur is measured in the $r$-band, and LLRL~54361 is measured in {\it Spitzer} 3.6 $\mu$m bandpass. }
\label{fig:buster_all}
\end{figure}
\begin{figure*}
\centering
\includegraphics[height=2.5in,angle=0]{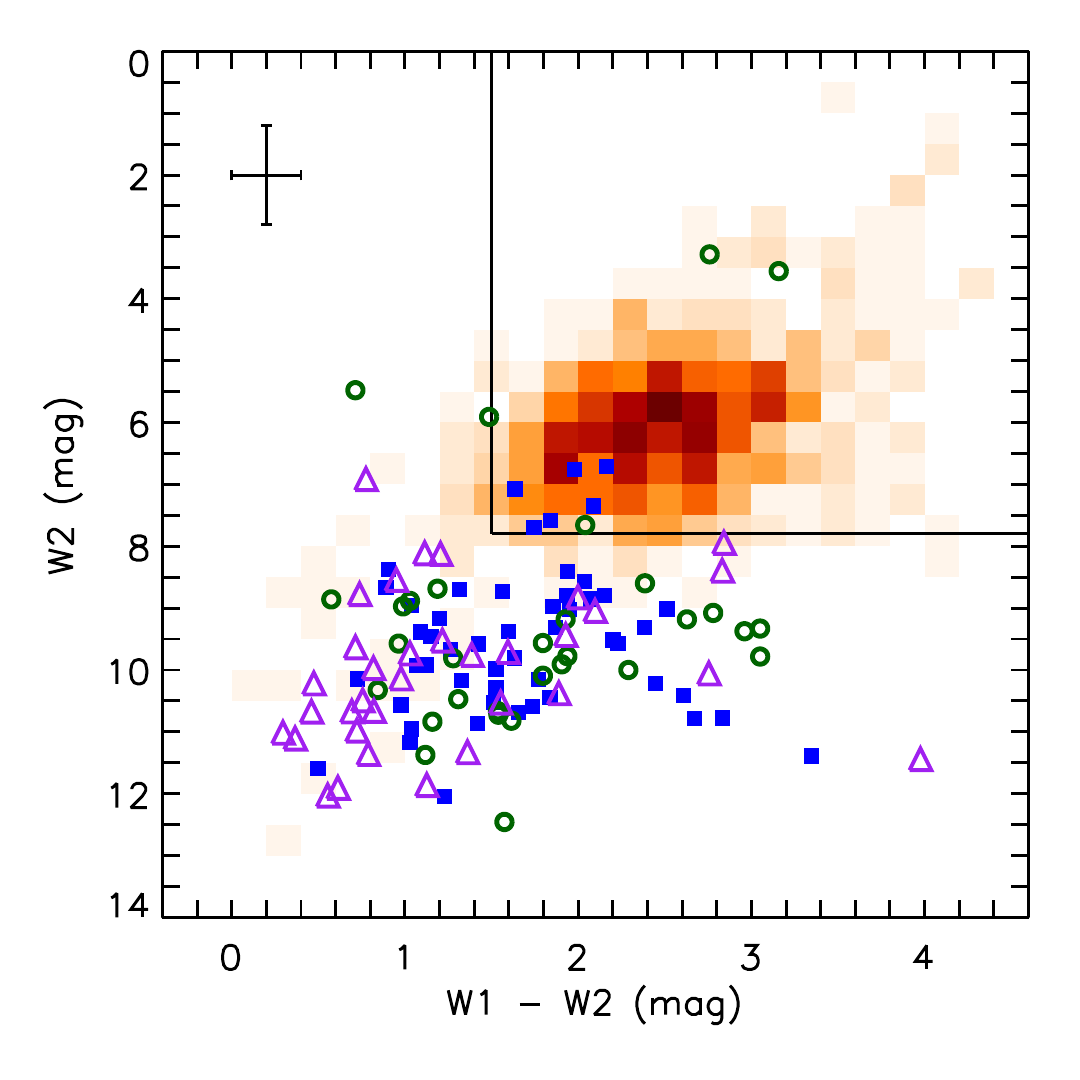}
\includegraphics[height=2.5in,angle=0]{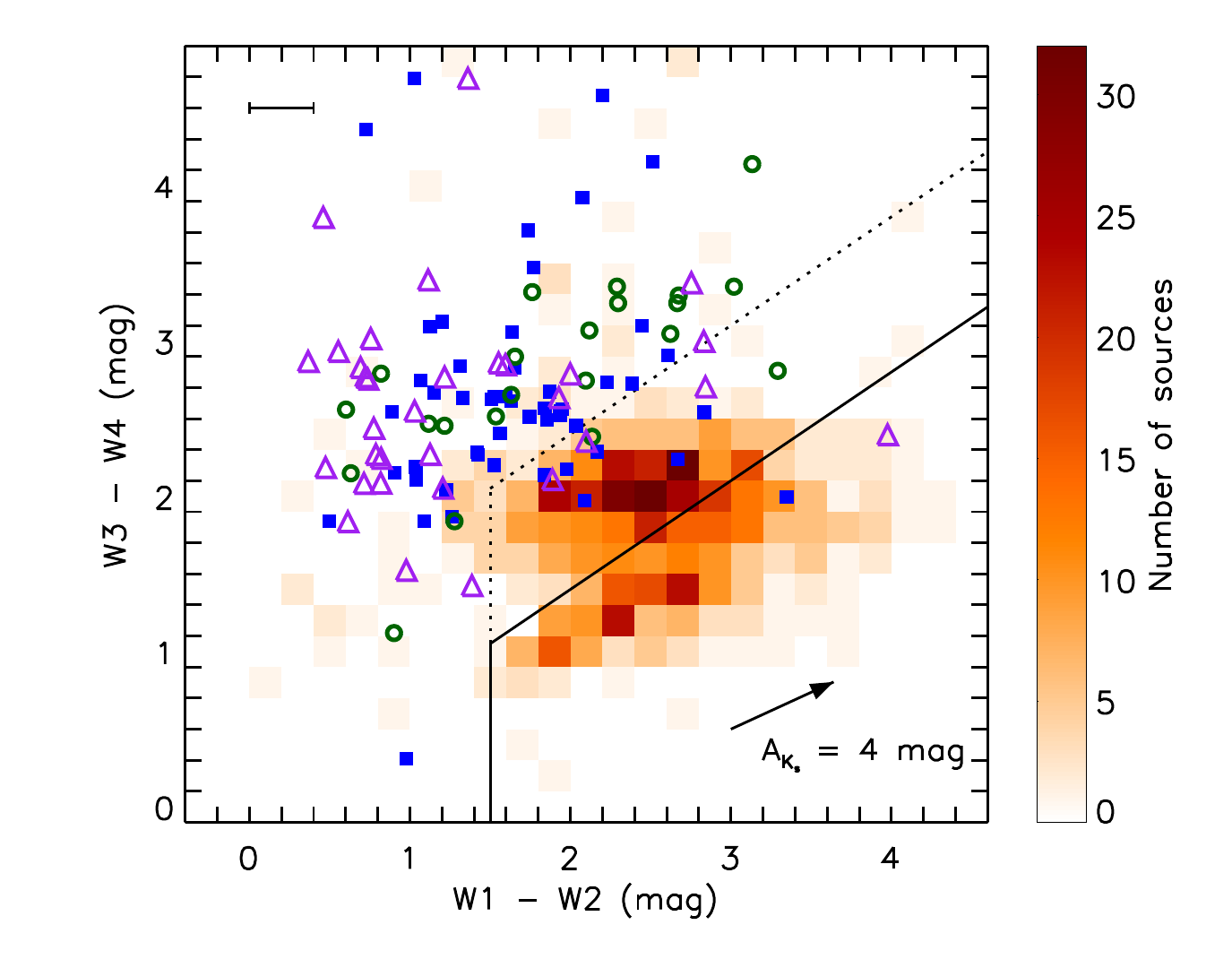}
\caption{{\it WISE} colour-magnitude and colour-colour diagrams of periodic variables. Miras are presented by density maps and candidate periodic outbursting YSOs are shown by open circles (first search) and  triangles (second search). Previously spectroscopically confirmed eruptive YSOs are shown by filled squares. The region of high-amplitude dusty AGB stars in the UKIDSS survey \citep{Lucas2017} is enclosed by solid lines. An extended region is marked by dotted lines. The typical variation amplitude in $W2$ and $W1 - W2$ is shown on the top left corner of each panel. Most photometric errors are less than 0.1 mag in all four bands, similar to the size of symbols.  The arrow in the {\it right} panel represents the extinction vector assuming $R_v = 3.1$ \citep{WangS2019}.}
\label{fig:wise_ccd}
\end{figure*}

 Periodic outbursting candidates identified from this work have a wide range of periods ($P = 17$ to $\sim$1500 days), amplitudes (up to 4~mag in $K_s$), and $W1$-$W2$ infrared colours from ALLWISE survey \citep{Wright2010}. Figure~\ref{fig:buster_all} illustrates the relations between these three parameters, along with known cases of periodic outbursts and pulsed accretion. Statistically, we found large amplitude sources ($\Delta K_s > 1.5$~mag) have redder $W1-W2$ colours than small amplitude sources, indicating that there is more hot dust around large amplitude sources. A similar trend has been previously observed among T Tauri stars with low amplitude repeated flares \citep{Cody2017}. There is a wide  scatter in the $W1-W2$ colour inside the large amplitude group, which is expected as YSOs have a broad range of $W1-W2$ colours related to their evolutionary stages and inner disc structures, and the truncation of the disc due to the existence of companions may also affect the infrared colour. No correlation is found between the $W1$-$W2$ colour and photometric period.

In the upper panel of Figure~\ref{fig:buster_all}, we see a lack of short-period candidates with small amplitudes. This might be attributed to the nature of our selection method, as intermittently periodic variables will not be identified by our pipeline. In previous works, these small-amplitude quasi-periodic flares have been detected in several young binary systems. For example, the flares in DQ~Tau ($P = 15.8$~d) only lasted for four periods in 2012 to 2013 with a small infrared amplitude, $\Delta K_s$ $\sim$0.55 mag, and the periodic flaring behaviour was intermittent in 2013 to 2014 \citep{Muzerolle2019}.

\begin{table*}
\caption{Periodic Outbursting YSO Candidates}
\centering
\begin{tabular}{c c c c c c c c r c c c c}
\hline
\hline
Name & RA (J2000) & Dec (J2000)  & Period & $\Delta K_s$ &  Med(${K_s}$) & $W1-W2$ & $W3-W4$ & $\alpha_{\rm class}$ & \multicolumn{3}{c}{references} & skewness \\
\hline
 & deg & deg  & day & mag  & mag & mag & mag & & SPICY & {\it WISE} & H{\sc ii} &\\
\hline
VVV\_PB\_1 &   177.333954 &   -61.426071 &  489.2 & 2.08 & 14.55 &  1.12 &  2.45 &  0.26 & YSO & - & y? & -0.15$\pm$0.07 \\
VVV\_PB\_2 &   187.506180 &   -62.645180 &  437.8 & 2.48 & 15.25 &  1.80 &  2.90 &  0.29 & YSO & - & y?  & -0.22$\pm$0.10 \\
VVV\_PB\_3 &   188.593475 &   -63.492653 &   45.5 & 2.53 & 16.14 &  2.41 &    -  & -0.07 & YSO & - & n & -0.03$\pm$0.27\\
VVV\_PB\_4 &   190.987946 &   -62.912498 & 1437.6 & 2.45 & 12.72 &  1.28 &  2.70 & -0.23 & YSO & near & y & 0.16$\pm$0.03 \\
VVV\_PB\_5 &   193.579849 &   -61.644066 &  885.3 & 3.13 & 14.88 &  2.39 &  3.29 &  1.07 & - & near & y &  0.14$\pm$0.04\\
VVV\_PB\_6 &   194.785309 &   -62.632748 & 1302.0 & 2.17 & 16.34 &  3.05 &  2.81 &  1.55 & near & - & n  &  0.09$\pm$0.05 \\
VVV\_PB\_7 &   199.108856 &   -63.718597 &  585.5 & 1.52 & 16.06 &  1.57 &  - &  0.55 & YSO & - & n &  0.59$\pm$0.27\\
VVV\_PB\_8 &   200.863190 &   -62.267666 &   40.5 & 2.16 & 14.65 &  1.80 &  - & -0.24 & YSO & - & y & -0.33$\pm$0.10\\
VVV\_PB\_9 &   208.341843 &   -61.599812 &  574.2 & 1.78 & 16.60 &  3.27 &    -  &    -  & near & - & y? &  0.23$\pm$0.09\\
VVV\_PB\_10 &   209.634506 &   -61.096882 &   17.0 & 2.92 & 15.69 &  1.61 &  - & -1.32 & YSO & - & y & -0.01$\pm$0.03\\
VVV\_PB\_11 &   214.888397 &   -61.670609 & 1189.0 & 1.79 & 16.18 &  1.31 &  2.65 &  0.26 & YSO & - & n & -0.58$\pm$0.18\\
VVV\_PB\_12 &   215.414703 &   -60.875877 &  176.4 & 2.63 & 14.12 &    -  &    -  &  0.61 & YSO & - & y &  0.46$\pm$0.07\\
VVV\_PB\_13 &   216.364105 &   -60.336102 &  228.5 & 2.54 & 14.16 &  1.09 &  - & -0.67 & y & - & y & -0.01$\pm$0.06\\
VVV\_PB\_14 &   221.232819 &   -60.509438 &  327.7 & 3.16 & 15.60 &  1.15 &    -  &  0.88 & YSO & - & y &  0.69$\pm$0.08 \\
VVV\_PB\_15 &   225.905167 &   -58.497616 &  230.6 & 1.54 & 16.46 &    -  &    -  &    -  & - & - & y & 0.00$\pm$0.04 \\
VVV\_PB\_16 &   226.708206 &   -58.216599 &   63.0 & 3.54 & 16.00 &  2.29 &  3.35 &  0.10 & YSO & - & n & -0.10$\pm$0.13\\
VVV\_PB\_17 &   227.398514 &   -57.589649 & 1026.7 & 3.53 & 14.10 &  1.93 &  2.38 &  0.62 & YSO & - & y? & 0.52$\pm$0.21\\
VVV\_PB\_18 &   228.723450 &   -59.338844 &  388.2 & 1.74 & 16.02 &  1.02 &    -  &    -  & - & - & y & 0.27$\pm$0.20\\
VVV\_PB\_19 &   235.625748 &   -54.862492 &   28.0 & 1.92 & 12.06 &  0.99 &  2.56 & -1.04 & YSO & - & n & -0.01$\pm$0.05\\
VVV\_PB\_20 &   237.557938 &   -54.918976 &  150.4 & 2.30 & 15.92 &    -  &    -  &    -  & near & near & y & -0.02$\pm$0.02\\
VVV\_PB\_21 &   239.374954 &   -54.053627 &  211.3 & 1.85 & 16.74 &    -  &    -  &  0.46 & near & - & y & -0.52$\pm$0.24\\
VVV\_PB\_22 &   239.866348 &   -51.965691 &   83.0 & 3.10 & 16.31 &  2.39 &    -  & -0.37 & YSO & near & y & 0.44$\pm$0.23 \\
VVV\_PB\_23 &   240.285767 &   -52.813343 &   34.5 & 1.68 & 15.63 &    -  &    -  & -0.13 & YSO & - & y & -0.51$\pm$0.23 \\
VVV\_PB\_24 &   243.751526 &   -50.745075 &   28.9 & 2.40 & 15.92 &  2.43 &    -  &  0.45 & YSO & near & y &  0.17$\pm$0.16 \\
VVV\_PB\_25 &   244.162247 &   -50.836494 &  420.0 & 2.43 & 16.17 &    -  &    -  & -0.33 & near & near & y & 0.00$\pm$0.08\\
VVV\_PB\_26 &   245.053955 &   -49.465725 &   43.7 & 2.01 & 15.45 &    -  &    -  &    -  & - & - & y &  0.00$\pm$0.09 \\
VVV\_PB\_27 &   245.056427 &   -50.268200 &  476.8 & 2.51 & 15.69 &  2.96 &  3.24 &  0.41 & YSO & - & y & 0.31$\pm$0.06\\
VVV\_PB\_28 &   245.667404 &   -49.107346 &   30.5 & 3.09 & 12.71 &  1.19 &  1.84 & -0.64 & YSO & - & n & 0.31$\pm$0.28 \\
VVV\_PB\_29 &   247.648468 &   -47.766544 &  507.5 & 3.15 & 14.76 &  1.79 &    -  &  0.29 & YSO & near & y & 0.62$\pm$0.21\\
VVV\_PB\_30 &   250.300003 &   -47.131538 &  149.4 & 2.46 & 15.94 &    -  &    -  &    -  & near & - & y & -0.08$\pm$0.25\\
VVV\_PB\_31 &   250.942032 &   -46.763126 &  512.6 & 1.76 & 14.19 &  0.96 &  2.79 &  0.05 & YSO & near & y & -0.03$\pm$0.26\\
VVV\_PB\_32 &   251.085663 &   -46.762188 &   33.9 & 3.47 & 16.32 &  1.54 &    -  &  0.11 & YSO & - & y & 0.62$\pm$0.09\\
VVV\_PB\_33 &   252.817123 &   -41.385250 &  433.5 & 2.05 & 15.66 &    -  &    -  &    -  & - & - & n & -0.41$\pm$0.21\\
VVV\_PB\_34 &   254.949860 &   -42.219059 &  270.3 & 2.25 & 16.13 &  1.54 &  2.47 & -0.33 & near & near & y & 0.35$\pm$0.13\\
VVV\_PB\_35 &   255.927750 &   -42.612728 &  322.7 & 2.12 & 16.47 &    -  &    -  &  0.38 & - & near & y & -0.43$\pm$0.18\\
VVV\_PB\_36 &   256.013977 &   -41.768227 &  474.7 & 1.58 & 16.84 &    -  &    -  & -0.06 & near & near & n & -0.14$\pm$0.24 \\
VVV\_PB\_37 &   257.410919 &   -41.647724 &  640.1 & 2.70 & 14.46 &  2.04 &  3.07 &  0.47 & YSO & - & y & 0.32$\pm$0.09 \\
VVV\_PB\_38 &   257.691284 &   -39.875633 &  323.3 & 1.86 & 13.70 &  1.54 &  - & -0.22 & YSO & near & n & 0.25$\pm$0.07 \\
VVV\_PB\_39 &   259.517212 &   -38.872913 &  184.8 & 2.51 & 13.48 &  1.03 &  - &  0.24 & YSO & near & y & 0.29$\pm$0.19\\
VVV\_PB\_40 &   262.260071 &   -34.010117 &  978.0 & 2.80 & 14.96 &  3.05 &  3.35 &  1.66 & near & near & y & 0.41$\pm$0.14\\
VVV\_PB\_41 &   262.307404 &   -34.538891 &   96.2 & 3.38 & 15.01 &  2.78 &  - &  0.24 & near & - & y & 0.80$\pm$0.09 \\
VVV\_PB\_42 &   262.317566 &   -34.686844 & 1323.4 & 2.02 & 15.30 &  1.61 &    -  & -0.82 & near & - & y & 0.02$\pm$0.10\\
VVV\_PB\_43 &   262.549194 &   -34.786137 &  229.1 & 2.47 & 15.73 &    -  &    -  &  0.17 & YSO & near & y & 0.00$\pm$0.02\\
VVV\_PB\_44 &   263.398834 &   -32.876404 &  515.7 & 1.93 & 16.37 &  2.76 &  1.32 &    -  & near & YSO & y & -0.32$\pm$0.14\\
VVV\_PB\_45 &   264.494568 &   -31.369286 &  482.5 & 2.31 & 16.11 &    -  &    -  & -0.97 & near & - & n & 0.59$\pm$0.17 \\
VVV\_PB\_46 &   264.650757 &   -23.374863 &  435.7 & 2.29 & 14.98 &  1.57 &    -  &    -  & - & YSO & n & 0.17$\pm$0.19\\
VVV\_PB\_47 &   265.289276 &   -30.641739 &  228.5 & 1.68 & 16.15 &  0.24 &    -  & -2.36 & near & near & y & -0.73$\pm$0.17\\
VVV\_PB\_48 &   267.106201 &   -29.205429 &  729.8 & 1.65 & 13.48 &    -  &    -  &    -  & near & - & y & -0.42$\pm$0.04\\
VVV\_PB\_49 &   267.114075 &   -28.489040 &  690.1 & 2.10 & 16.24 &  3.32 &    -  &    -  & near & near & y? & 0.34$\pm$0.10\\
VVV\_PB\_50 &   268.253082 &   -28.641108 &  523.2 & 1.67 & 15.46 &  1.49 &  1.49 &    -  & - & YSO & n & 0.47$\pm$0.14\\
VVV\_PB\_51 &   268.372894 &   -25.294584 &  527.8 & 1.68 & 15.51 &  2.33 &    -  &    -  & near & YSO & y? & -0.03$\pm$0.12\\
VVV\_PB\_52 &   269.130432 &   -25.593458 &  421.7 & 2.43 & 13.42 &  1.91 &  3.32 & -0.21 & YSO & near & n & 0.59$\pm$0.12\\
VVV\_PB\_53 &   269.674896 &   -22.533661 &  194.5 & 2.24 & 13.77 &  1.94 &  - & -0.25 & YSO & - & y & 0.26$\pm$0.08 \\
VVV\_PB\_54 &   269.974121 &   -19.974720 &  269.4 & 2.29 & 13.83 &  1.16 &  1.12 & -0.80 & YSO & - & n & -0.05$\pm$0.07 \\
VVV\_PB\_55 &   269.998688 &   -22.440241 &  409.4 & 2.28 & 14.05 &  0.71 &  1.12 &    -  & - & YSO & n & 0.02$\pm$0.18\\
VVV\_PB\_56 &   270.174805 &   -23.789545 &  904.0 & 1.51 & 14.66 &  1.73 &    -  &  0.06 & YSO & near & y & 0.49$\pm$0.28\\
VVV\_PB\_57 &   270.539398 &   -22.787409 &   31.6 & 1.93 & 14.56 &  2.63 &  3.05 & -0.33 & YSO & - & y & -0.20$\pm$0.23\\
VVV\_PB\_58 &   270.950897 &   -22.821819 &  917.6 & 1.56 & 12.32 &  0.57 &  2.15 &  0.30 & YSO & - & n & -0.45$\pm$0.06 \\
VVV\_PB\_59 &   271.995789 &   -20.458395 &  730.9 & 1.69 & 16.66 &  3.16 &  1.98 &    -  & near & near & y & 0.48$\pm$0.29 \\
\hline
\end{tabular}
\flushleft{references: criteria to identify YSO candidates, including catalogues based on {\it Spitzer/IRAC} \citep{Fazio2004} and {\it WISE} photometry \citep{Wright2010}, and nearby H{\sc ii} regions seen from mid-infrared images.}
\label{tab:PB}
\end{table*}

\subsection{Other periodic YSOs}

 A further 148 sources, neither Miras nor candidate periodic outbursting YSOs, are identified as periodic or quasi-periodic variables from our pipeline. Based on the phase curve morphologies, these variables were classified as ``burster'', ``LPV'' (Long Period Variable),  ``dipper'', ``STV'' (Short Timescale Variable), ``QP'' (Quasi-Periodic) and other periodic variables, following the categories introduced in Paper I.  The ``bursters'' are sources not identified as YSO candidates due to either having Mira-like {\it WISE} colours or not located near to any star-forming regions, although some outbursting morphologies are seen in their phase-folded light curves. ``Dippers'' have larger $M$ values than periodic outbursting sources, and a light curve morphology resembling a single dip below an approximately constant level of brightness.
 
 We cross matched the location of these periodic variables with archived {\it Spitzer} or {\it WISE} colour-based YSO catalogues, and 54 sources were identified as YSO candidates from existing catalogues. Besides these, another 69 sources have at least 5 YSOs from archived catalogues located within 3', and one source is located next to a Galactic H{\sc ii} region. Periodic YSO candidates exhibit a wide range of phase curve morphologies including LPV, dipper and QP as the three most frequently seen categories. The physical mechanisms of these sources are beyond the scope of this paper, but they are most likely related to the star-disc interaction or a tilted disc warp structure. We list some basic information of these periodic/quasi-periodic sources in the Appendix.

\begin{figure*}
\centering
\includegraphics[height=2.6in,angle=0]{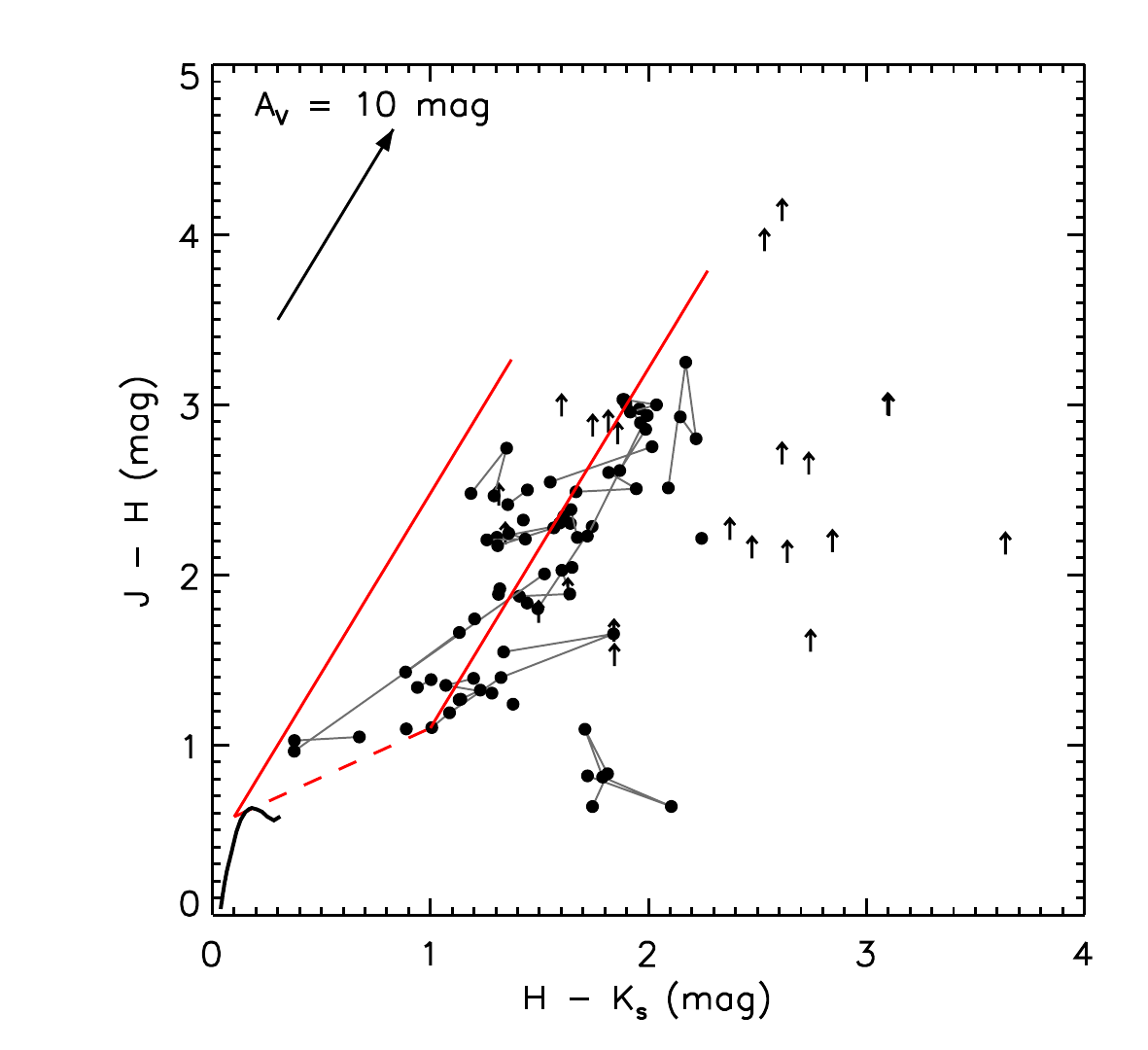}
\includegraphics[height=2.6in,angle=0]{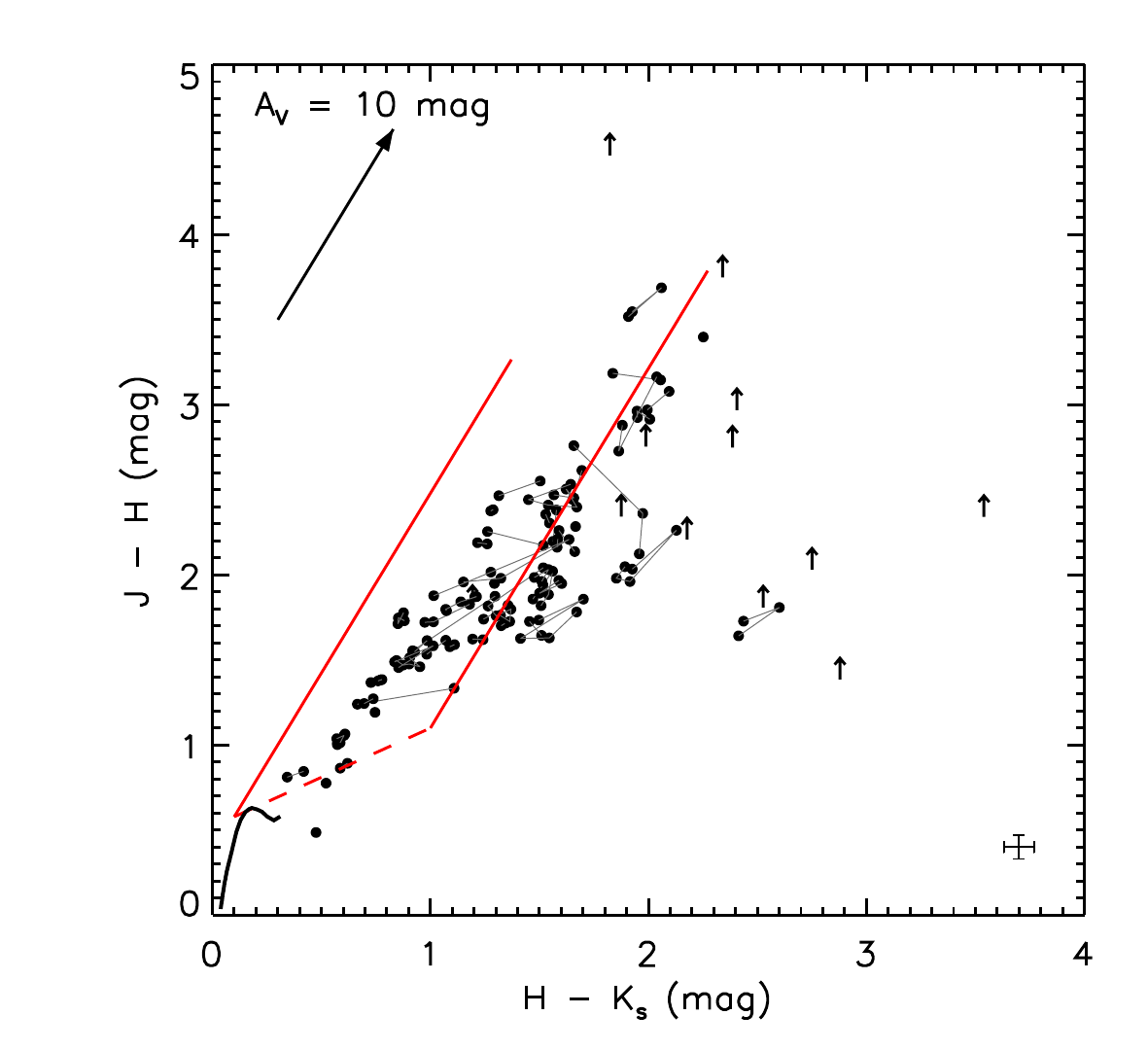}
\includegraphics[height=2.6in,angle=0]{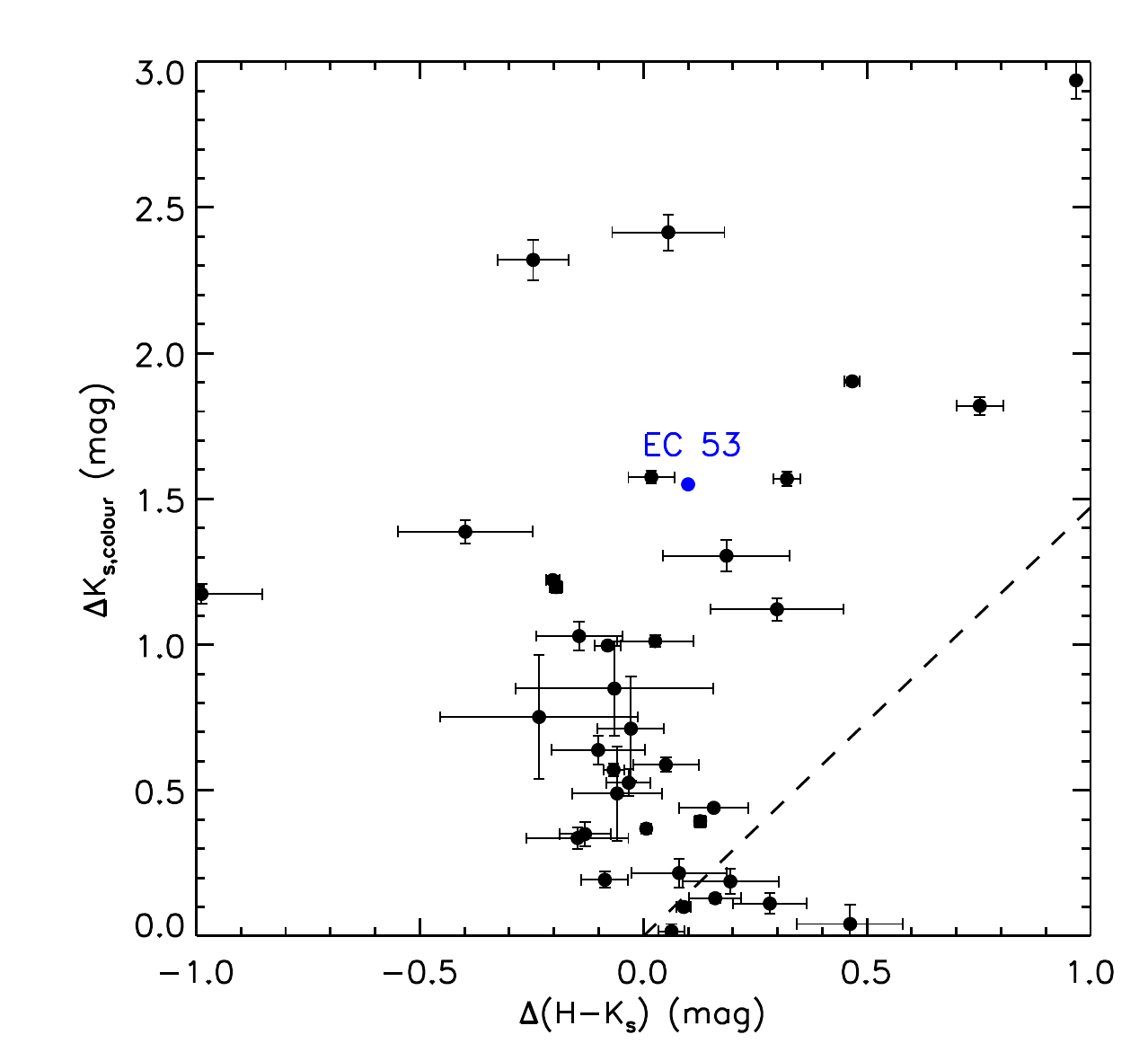}
\includegraphics[height=2.6in,angle=0]{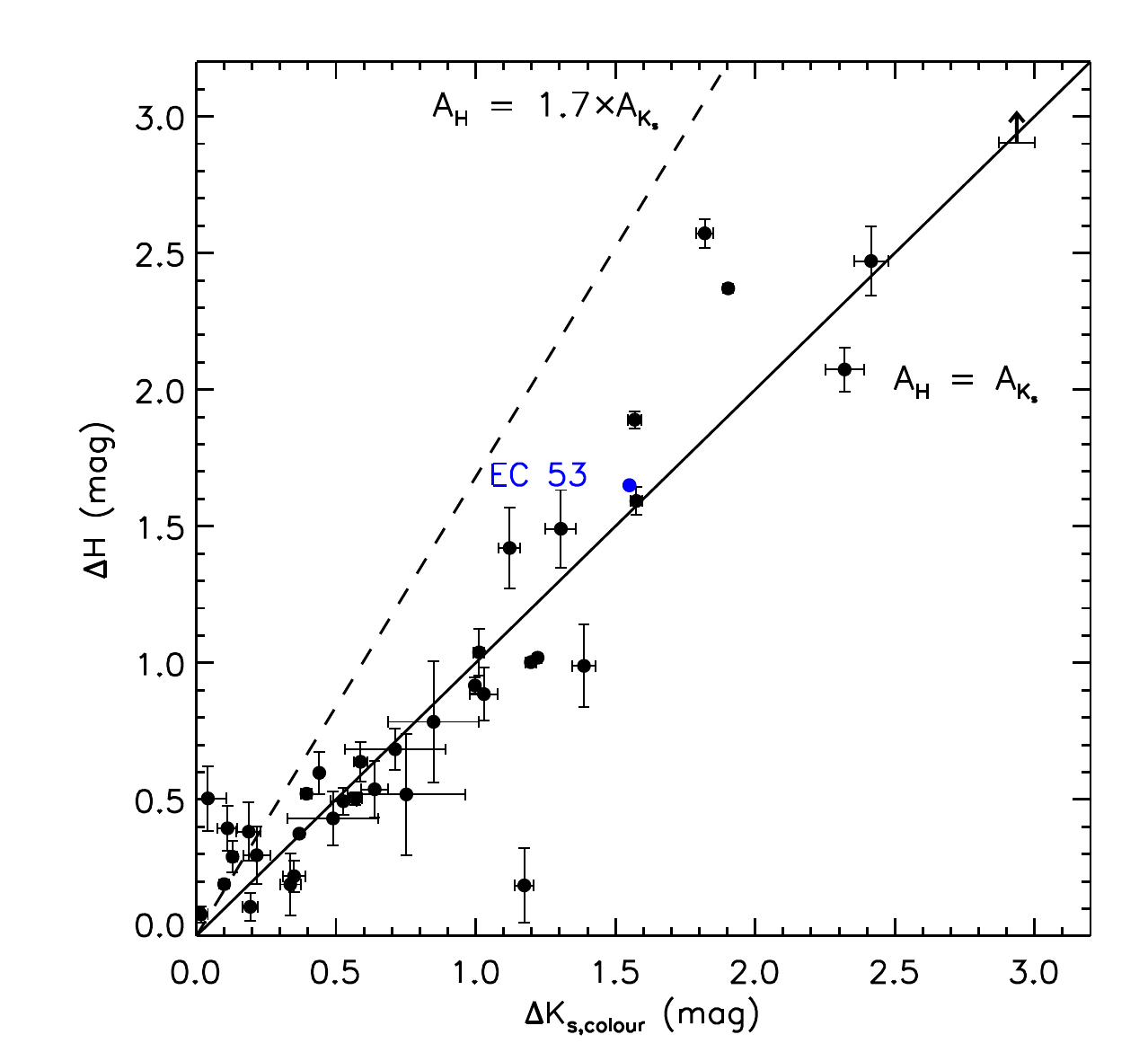}
\caption{{\it Upper Panels:} Colour-colour diagram of periodic outbursting YSO candidates identified from our first ({\it left}) and second ({\it right}) searches.  $J$ is artificially set to 20~mag and marked by upward arrows when $J$ detections are lacking. Multi-epoch detections of the same source are linked by grey lines. The dashed red line represents the intrinsic locus of Class II objects \citep{Meyer1997}. The black solid curve marks the empirical colour of main sequence stars \citep{Pecaut2013}, and red solid straight lines are reddening vectors. Most sources have photometric errors less than 0.07 mag, shown as the error bar in the {\it upper right} panel. {\it Lower Panels:} Contemporaneous $K_s$-band amplitude ($\Delta{K_{\rm s, colour}}$), $H$-band amplitude ($\Delta{H}$), and variation of $H - K_s$ colour of periodic outbursting YSO candidates identified in the first search. One source having no $H$ detection during photometric minimum is marked by the upward arrow assuming $H \ge 19$~mag. The $H$ and $K_s$ amplitude of the periodic burster EC53 are shown in blue. The near-infrared extinction law $A_{H} = 1.7 \times A_{K_s}$ assuming $R_v = 3.1$ \citep{WangS2019} is shown by dashed lines in two lower panels.}
\label{fig:JHK_ccd}
\end{figure*}

\section{Discussion}
\label{sec:dis}
\subsection{WISE colours of Periodic Sources}

We obtained NEOWISE $W1$- and $W2$-band light curves \citep{Mainzer2014} since 2014 and ALLWISE $W1$ to $W4$ photometry from 2010 for periodic sources identified in this work. The NEOWISE light curve data were averaged in 1 day bins, after excluding outliers more than 5$\sigma$ from the mean of the bin. ALLWISE images were visually inspected to eliminate suspicious detections due to line of sight nebulosity and saturated sources. In total, 1182 candidate Miras and 20 candidate periodic outbursting YSOs from our first search have reliable measurements across all four {\it WISE} bands. The {\it WISE} colour-magnitude and colour-colour diagrams of periodic variable sources are shown in Figure~\ref{fig:wise_ccd}, where $W2$ and $W1-W2$ are median values from the NEOWISE light curves, and $W3-W4$ is from the AllWISE survey. Spectroscopically confirmed eruptive YSOs from the VVV survey \citep{Contreras2017b, Guo2021}, and low-amplitude periodic outbursting YSO candidates from our second search, are presented as comparisons. In addition, with solid lines, we mark the regions defined in {\it WISE} colour and magnitude space for candidate dusty high-amplitude asymptotic giant branch (AGB) stars from the UKIDSS survey \citep{Lucas2017}.

In the colour-magnitude diagram, the regions occupied by candidate Miras and spectroscopically confirmed YSOs are largely separate, with only a $\sim$10\% overlap. A bimodal distribution is seen for candidate Miras in the colour-colour diagram, which is similar to the result from \citet{Suh2021}. A large number of sources with redder $W3 - W4$ colour are well beyond the region identified in \citet{Lucas2017}, which could not be simply explained by the interstellar extinction. Therefore, a new set of {\it WISE} colour identification for high-amplitude dusty Miras in VVV is proposed as
\begin{equation}
\begin{split}
W1 - W2 > 1.5, \\
W3- W4 < 0.7(W1-W2)+1.0,
\end{split}
\label{eq:mira}
\end{equation}
where 91\% of candidate Miras are located in this new region.

We see in \ref{fig:wise_ccd} that $\sim$90\% of the periodic outbursting YSO candidates found by our two searches lie outside the new, enlarged colour space of the Mira variables. Regarding the small amount of overlap in colour, the blending of infrared colours between YSOs and AGB stars is not unusual. Some periodic AGB stars are known to have similar colours to YSOs, such as D-type symbiotic stars containing a Mira-type object and a lower-mass companion \citep{Corradi2008}. In our recent spectroscopic observation of VVV variable sources \citep{Guo2021}, two symbiotic systems were identified by $\Delta v$=3 $^{12}$CO absorption in the $H$ bandpass, which have  {\it WISE} colours located outside the location of Miras defined in Equation~\ref{eq:mira}. On the other hand, a recent study found that some previously identified YSOs in nearby star forming regions are actually AGB stars, using evidence from {\it WISE} colours and the associated SiO maser \citep{LeeJE2021}. However, the total number of these ``unclear'' or possibly ``misclassified'' sources is small and therefore only a minor statistical impact is expected. Some YSOs may have sinusoidal light curves attributable to variable line-of-sight extinction by a warped inner disc \citep{Chiang2004}.

\begin{figure*}
\centering
\includegraphics[height=2.7in,angle=0]{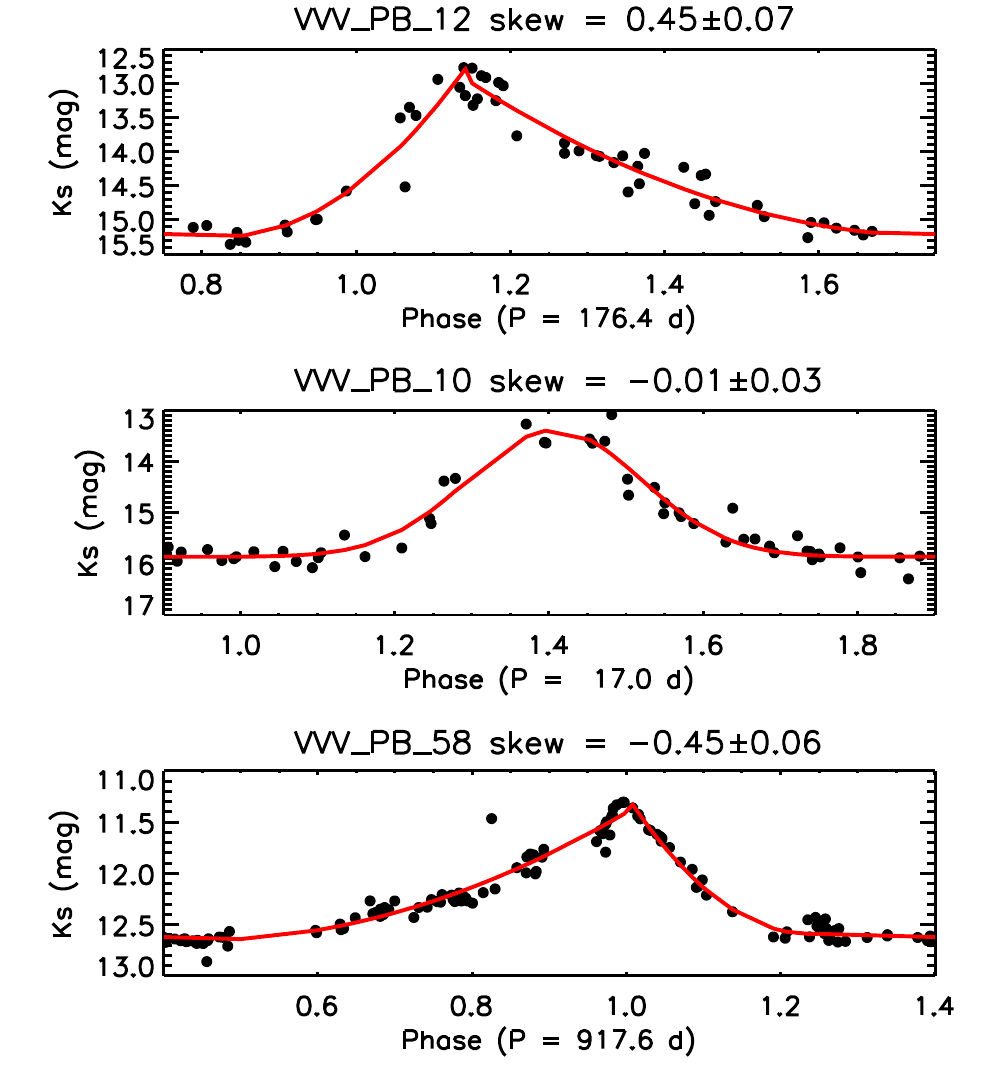}
\includegraphics[height=2.7in,angle=0]{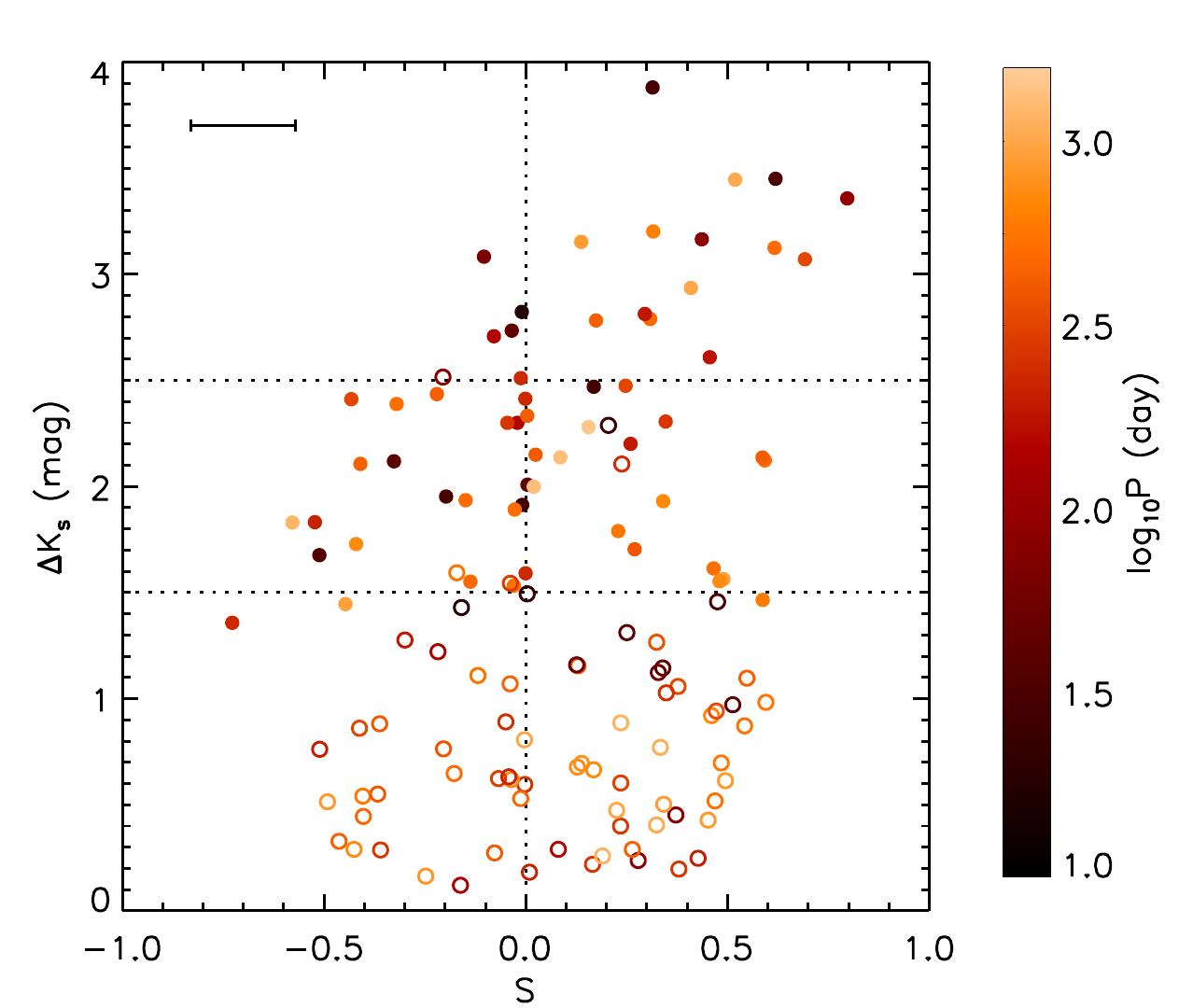}
\caption{{\it Left}: Examples of analytical fittings of phase-folded light curves and the measured skewness. {\it Right}: Skewness ($S$) and amplitude ($\Delta K_s$) of candidate periodic outbursting YSOs. As in Figure~1, two groups of candidates are shown by open and filled circles, respectively, and colour-coded by period. Dotted lines mark out $\Delta K_s = 1.5$ and 2.5 mag. The median uncertainty of the measurement of skewness is shown as the error bar on the {\it right} panel.}
\label{fig:skew}
\end{figure*}

\subsection{Near-infrared colour of periodic outbursting YSOs}

The infrared excess of disc-bearing YSOs with respect to photospheric templates has been widely used to identify their evolutionary stages. In Figure~\ref{fig:JHK_ccd}, we present the colour-colour diagrams of periodic outbursting YSO candidates that have at least one  contemporaneous $H$ and $K_s$ detection in VIRAC2-$\beta$. For most regions in the VVV/VVVX survey, there are two to three multi-colour epochs.  Sources with more than one epoch of $J, H, K_s$ detections are presented as tracks on both diagrams. 

Almost all candidates are located to the upper-right side of the main-sequence locus, owing to the existence of circumstellar and interstellar dust. About half of candidates (that have VVV $J, H, K_s$ detections) from the first search are located between the red lines, known as the region occupied by disc-bearing Class II YSOs. The rest lie in the Class I region, with redder $H-K_s$ colours. The proportion of candidates located in the Class II region is slightly increased for the second search group. Some variables, with tracks on the colour-colour diagram, move across the boundary between Class I and Class II objects, which would alter the classification of the evolutionary stage when using single epoch observations. Most variation tracks of large-amplitude sources do not follow the infrared extinction vector \citep{WangS2019}, indicating that the variation of these periodic variables is not dominated by changing of line-of-sight extinction. 

The departures from extinction-related trends are clearer in the lower panels of Figure~\ref{fig:JHK_ccd}, where the $H - K_s$ colour variations are presented. The $\Delta K_{\rm s, colour}$ here represents the maximum variation in $K_s$ between contemporaneous $J, H, K_s$ epochs, which is different to the $\Delta K_s$ defined above. Most sources have small $H-K_s$ colour variation despite large photometric amplitude, opposite to the interstellar extinction slope. The relatively constant $H - K_s$ colour favours the accretion-burst scenario \citep{Kospal2011}, though we note that grey extinction is occasionally reported in some edge-on systems \citep[e.g.][]{Scholz2015a}, attributed to large dust grains. The existence/disappearance of warm inner disc structure \citep{Covey2021} may also affect the observed colour. Similar colour variation is seen between $J$- and $H$-band detections.

We found large-amplitude periodic outbursting YSO candidates have similar colour variations to previously confirmed periodic bursters \citep[e.g. EC~53;][]{YHLee2020}. This indicates that higher amplitude variations most likely result from accretion outbursts. However, we are aware that the contemporaneous $H$ and $K_s$-band amplitudes were measured using a small number of VVV multi-colour epochs having very limited phase coverage. Therefore, they often do not represent the full variability picture, especially in cases where the amplitude between two epochs are small.

\subsection{Analytical phase curves and skewness of periodic outbursting YSOs}
\label{sec:fit}

 The timescales of the rising and descending stages of an outburst provide information on the physical mechanism and the viscosity of the circumstellar disc. The periodic bursts seen in EC~53 and the cyclical flares of DQ~Tau \citep{Kospal2018} share a similar morphology: a rapid rise followed by a slower decline attributable to viscous spread. In the case of young binary star systems, the light curve is likely dominated by the emission from the accretion disc surrounding the more massive component. The accumulation of warm dust in the inner disc, before the outburst, may explain the slow rise seen in some of the light curves as a slow rising phenomenon, as the instability propagates inwards. 
 
 To parameterise the outbursting YSO light curve profiles, we fitted each phase-folded light curve with simple analytical functions, either linear, Gaussian or second order polynomial. The long-term linear trends measured by our pipeline were removed before the fitting.  For individual candidates, the phase-folded light curve is divided into quiescent, rising and descending stages, where analytical curves are fit to the rising and descending stages separately.  The best fitting results, chosen from the above three types of analytical functions via the $\chi^2$ minimization method, are then linked together to form the analytical phase curve. These analytical phase curves are smoother and less sensitive to outliers or gaps in the phase coverage than the interpolation-based phase curves. Three examples are shown in the left panel of Figure~2 in the main article, and more fitting results are presented in Figure~\ref{fig:PB1} to \ref{fig:PB3}. 
 
 We apply the statistical concept of skewness ($S$), the third moment of a distribution, to parameterise the degree of asymmetry of analytical phase curve functions that we fitted to the phase-folded light curves (see examples Figure~\ref{fig:skew}). A positive skewness describes an outburst where most of the event extends to the right of the peak (i.e. fast-rise/slow-decay), while a negative skewness means that most of the event extends to the left side.  We adopted the statistical skewness function from the {\sc IDL} software package to measure the asymmetry of the analytical phase curves as
\begin{equation}
    S = \frac{1}{N}\sum_{i=1}^{N}\left(\frac{X_i - \bar{X}}{\sigma}\right)^3,
\end{equation}
where $N$ is the number of test particles, $\sigma$ is the standard deviation of $X$, and $\bar{X}$ is the mean value of the distribution. The skewness of a phase curve was calculated numerically by the following steps. First, an ``outbursting window'' is defined for each analytical phase curve, as lasting from the beginning of the rising stage to the end of the descending stage. Then the analytical phase curve in the outbursting window is re-sampled into 100 equal bins on the phase space, and a synthetic distribution of 10000 test particles is generated following the distribution of the curve in each bin, where the skewness is measured using the equation provided above. We applied a bootstrapping method to estimate the error introduced from fitting the analytical phase curves, where each phase-folded light curve is randomly re-sampled by 200 attempts. Some rarely seen outliers that could not be fitted by certain analytical functions were rejected from the group and were replaced by additional random attempts. The error bar is then defined as the standard deviation of skewness measured in these 200 attempts. 
 
 The skewness of all 130 periodic outbursting YSO candidates are shown in the right panel of Figure~\ref{fig:skew}, plotted against outburst amplitude in $K_s$. The 18 sources with the highest amplitudes all have skewness $>$ -0.2, including 14 sources having positive skewness,  consistent with viscous evolution of an accretion burst that clears out a finite mass reservoir in the inner disc. However, the skewness of low- to intermediate-amplitude sources ($\Delta K_s \leq 2.5$~mag) is symmetrically and evenly distributed from negative to positive. The statistical result is clear: fast-rise/slow-decay bursts predominate in the sub-sample with the highest amplitude, even though the measurement of skewness may have some uncertainties in individual sources, especially the ones with large period to period variation. The skewness values of all periodic outbursting YSO candidates are listed in Table~\ref{tab:PB} and Table~\ref{tab:low_amp} with the uncertainties. No correlation is found between the skewness and the period.

\begin{figure} 
\centering
\includegraphics[width=3.0in,angle=0]{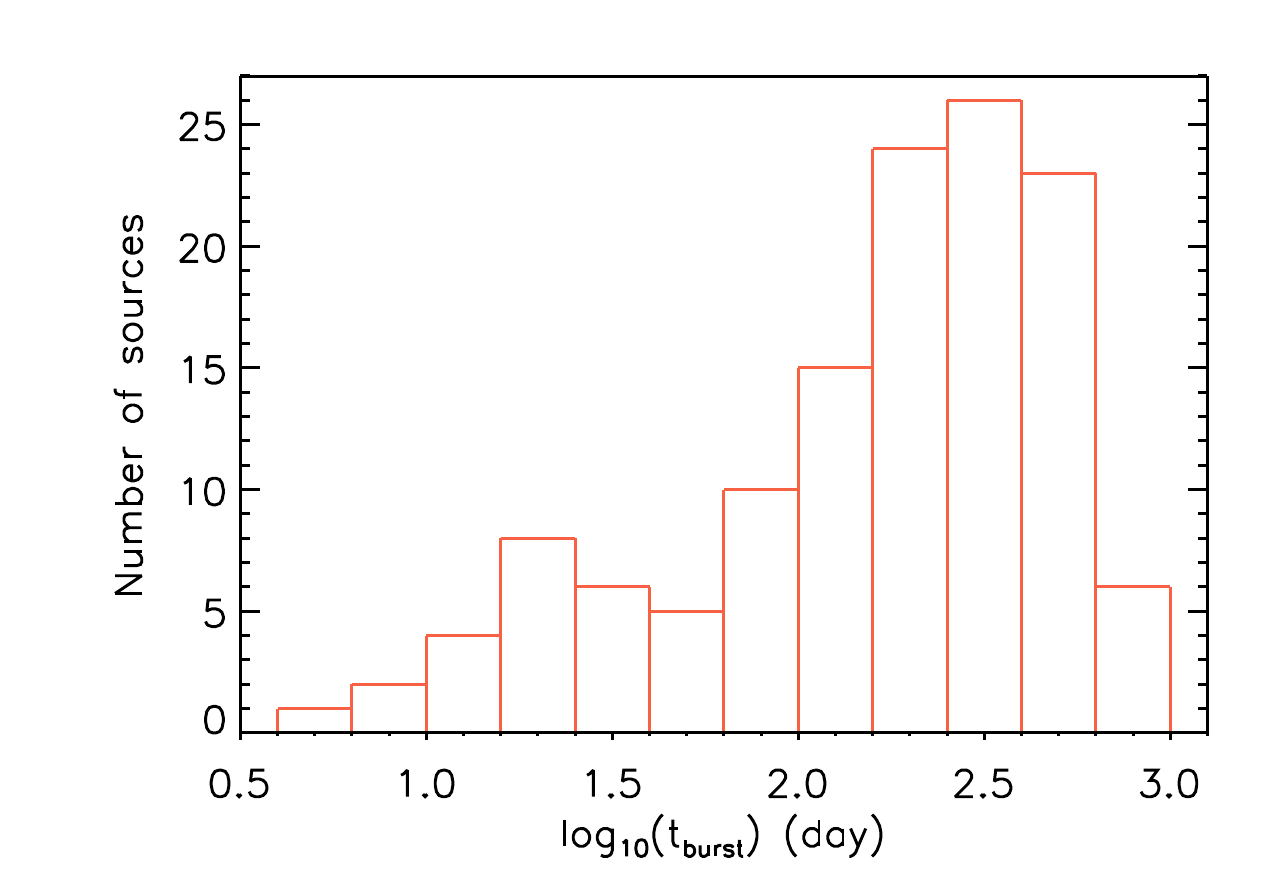}
\caption{Histogram of the duration of outbursts ($t_{\rm burst}$) measured from the analytical phase curves.}
\label{fig:timescale}
\end{figure}

The duration of the outbursting phase ($t_{\rm burst}$) is also measured using the analytical curves. To be consistent with our definition of the $M$ factor, $t_{\rm burst}$ is defined as the timescale when the source is brighter than its median brightness on the analytical phase-curve. The distribution of $t_{\rm burst}$ is presented in Figure~\ref{fig:timescale}.  Similar to the variation period, we observe a wide range of $t_{\rm burst}$ among outbursting candidates, from less than 10 days (consistent with the stellar rotation period) to a few hundred days which is comparable to episodic accretion bursts seen among YSOs \citep{Audard2014}. We do not observe any obvious correlations among $t_{\rm burst}$, $\Delta K_s$ and the skewness, suggesting complicated physical mechanisms.

Short duration outbursts (i.e. $t_{\rm burst} < 30$~day) resemble the scenario of accretion flares on young binaries \citep[e.g. DQ~Tau;][]{Kospal2018, Muzerolle2019}, as consequences of dynamical perturbation or other mechanisms at or close to the inner edge of the accretion disc. Notably, some short timescale outbursts have reached exceptional amplitudes ($\Delta K_s$ > 3~mag) in less than 30 days, which are higher than the periodic outbursts detected on LRLL~54361 \citep{Muzerolle2013}. Similar rapid changing of $K_s$-band brightness was observed in the Class I YSO DR4\_30 \citep{Guo2021}, though no periodicity was detected on that source. The majority of periodic outbursting candidates have $t_{\rm burst}$ between 100 and 1000 days, which is comparable to some EXor-type outbursts seen in literature \citep[e.g.][]{Herbig2008, Lorenzetti2009}. If the periodic outbursts result from dynamical perturbations, the Keplerian orbit of the perturbation source should be around 0.5~au to several au, with accretion bursts being triggered during the periastron passage. By simply assuming that the enhancement of $K_s$-band brightness is solely attributable to, and linearly proportional to, the increase in accretion luminosity, the mass accretion rate should be cyclically modulated by at least one order of magnitude to reach $\Delta K_s$ = 2.5~mag. 

Simultaneous multi-wavelength photometric monitoring would help to understand the origin of the outbursts. For example, if the burst originates far away from the inner edge of the circumstellar disc and then propagates inward, a phase delay is expected between the mid-infrared and near-infrared light curves. Since most of our candidates are optically faint, we explored the possibility of measuring the phase delay between the VVV $K_s$ and the NEOWISE $W2$-band time series. However, the potential for such measurements is severely limited by the sparse cadence of NEOWISE light curves. We conclude that there is no general phase delay (exceeding 10\% of the phase) between the $K_s$ and $W2$-band light curves. However, we found different outburst morphologies between the near- and mid-infrared phase-folded light curves (see the case of VVV\_PB\_41). 

\begin{figure*} 
\centering
\includegraphics[width=6.5in,angle=0]{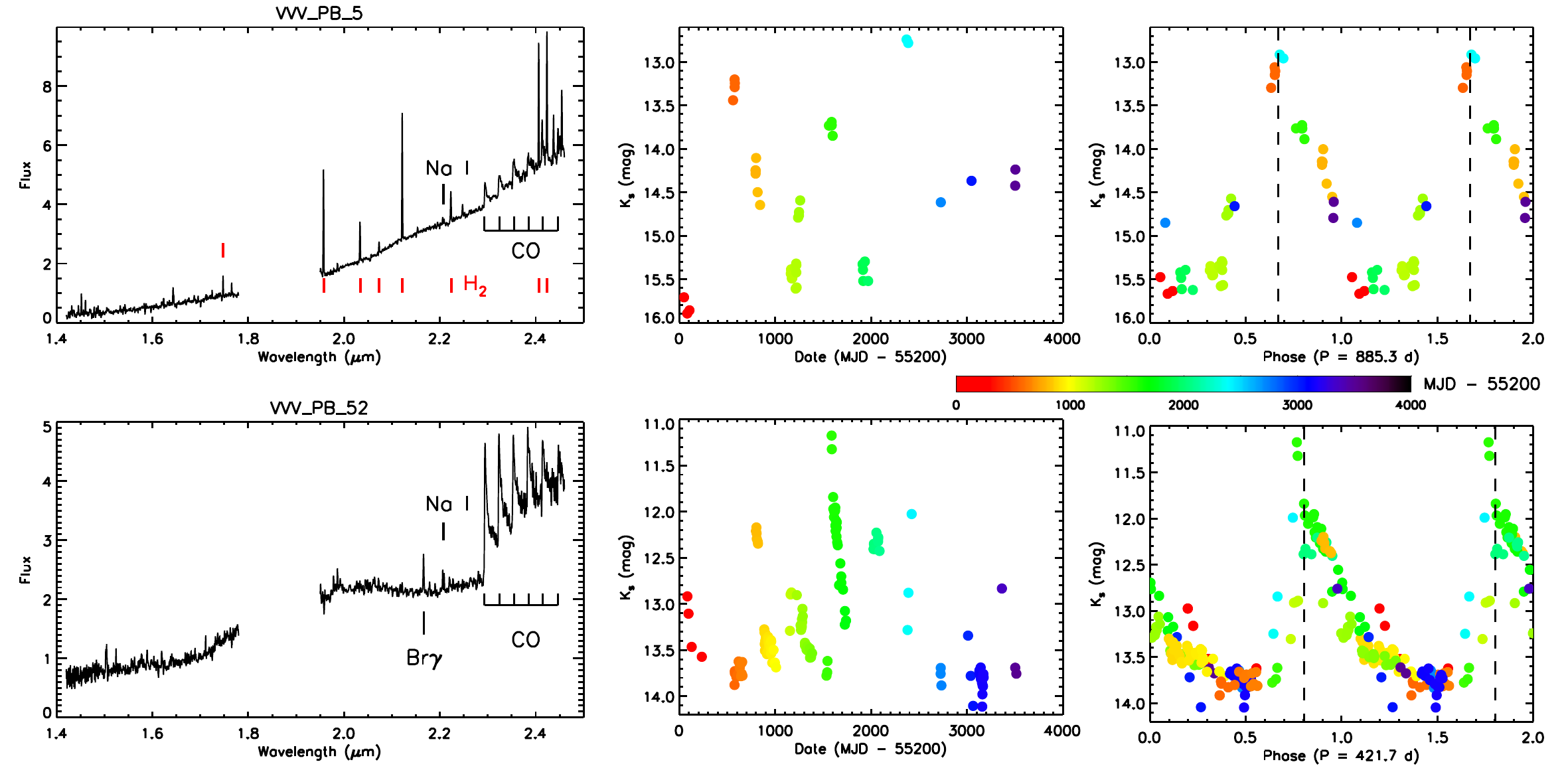}
\caption{  SOAR near-infrared spectra ({\it left}), VVV/VVVX $K_s$-band light curves ({\it middle}) and phase-folded light curves ({\it right}, long term trend removed) of two periodic outbursting sources ({\it top}: VVV\_PB\_5; {\it bottom}: VVV\_PB\_53). The flux unit of the spectra is erg/s/cm$^2$/\AA. The light curves are colour-coded by the observation time (Modified Julian Date - 55200). Photometric phases when SOAR spectra were taken are marked by dashed lines.}
\label{fig:soar}
\end{figure*}

\subsection{Individual objects}

In this section, we highlight YSOs for which we have information on the variability mechanisms, either from near-infrared spectra or sufficient phase coverage from multi-wavelength light curves.

\subsubsection{Spectroscopically confirmed YSOs in previous works}
\label{sec:spec_confirm_yso}

 The nature of the candidates as accreting YSOs is confirmed in five cases by near-infrared spectroscopy. Three of these were published previously \citep{Guo2021},  designated here as VVV\_PB\_4, VVV\_PB\_17 and VVV\_PB\_40, or VVVv32, DR4\_v17, and DR4\_v55 in the previous work, respectively. The multi-wavelength phase-folded light curves of three periodic sources were presented in Figure~4 of \citep{Guo2021}. VVV\_PB\_4 and VVV\_PB\_40 have colour variations indicating temperature changes in the inner disc, while the colour variation of VVV\_PB\_17 is close to the infrared extinction law. Indicators of high magnetospheric accretion rates, such as H{\sc i} lines and CO overtone bandheads, were detected in VVV\_PB\_4 and VVV\_PB\_17. Meanwhile, the spectrum of VVV\_PB\_40, observed at minimum brightness, is dominated by H$_2$ emission that originates in a stellar wind or outflow. However, the nine quasi-periodic YSOs in our previous work are not recovered in this work due to large $\delta_{\rm phase}$ values, or a very long period ($P = 1840$~d) in the case of VVVv467.

\subsubsection{SOAR/TripleSpec~4.1 Spectroscopy}

 Here, we present the near-infrared spectra of two candidates (VVV\_PB\_5 and VVV\_PB\_52) taken during their photometric maxima (see Figure~\ref{fig:soar}), obtained by the TripleSpec~4.1 spectrograph on the 4-m Southern Astrophysical Research (SOAR) Telescope. The spectra were obtained during an engineering night on April 30$^{\rm th}$, 2021. TripleSpec~4.1 provides intermediate-resolution ($R = 3500$) long-slit spectra within the wavelength range of 0.97 to 2.50 $\mu$m \citep{Herter2020}. Targets were observed in an {\it ABBA} nodding pattern with three-minute single exposures. Total exposure times are 42~min for VVV\_PB\_5 and 24~min for VVV\_PB\_52. The extraction of spectra and telluric correction were performed with the data reduction pipeline written by K. Allers, as a modified version of the {\sc Spextool} package \citep{Vacca2003, Cushing2004}.

 Both spectra contain several strong emission features including indicators of a high mass accretion rate, such as CO ($\Delta\nu = 2$) overtone bands beyond 2.29 $\mu$m, the Na{\sc i} doublet at 2.2 $\mu$m (marginally detected in VVV\_PB\_5), and Br$\gamma$ emission (only detected in VVV\_PB\_52 with 5.8 \AA\,  equivalent width).  The absence of Br$\gamma$ in VVV\_PB\_5 might relate to a low disc inclination angle, where the line emission from the inner accretion disc is either veiled or obscured by the outer disc. ${\rm H}_2$ emission lines, tracers of stellar winds and outflows from young stars, are observed in VVV\_PB\_5 with blue-shifts of around -140 km/s departed from the stellar radial velocity. The spectroscopic observations confirm that VVV\_PB\_5 and VVV\_PB\_52 are YSOs undergoing strong mass accretion during photometric maxima. In addition, both sources have near-infrared colour variations between VVV multi-colour epochs that do not follow the extinction law. Here, we conclude that current evidence supports the hypothesis that the variability of both sources is due to a periodic accretion process, even though spectra at the photometric minima were lack.

\subsubsection{VVV\_PB\_10}
Located in a {\it Spitzer} dark cloud \citep[SDC~G310.948+0.723][]{Peretto2009}, VVV\_PB\_10 has $\Delta K_{s}$~=~2.92~mag and a period of 16.99~d. Contemporaneous $H$ to $W3$ band photometry around MJD~55240 yields a spectral index, $\alpha_{\rm class} = -1.09$, suggesting VVV\_PB\_10 is a Class II object. A similar value is given in the SPICY catalogue based on {\it Spitzer}/IRAC photometry. There were no detections in the {\it WISE} $W4$ and {\it Spitzer}/MIPS 24~$\mu$m bands. This source appears to suffer high extinction, given that there was no VVV $J$-band detection during the photometric maximum (hence $J - H > 4.1$~mag, see the left panel of \ref{fig:JHK_ccd}). 

\begin{figure*}
\centering
\includegraphics[width=2.6in,angle=0]{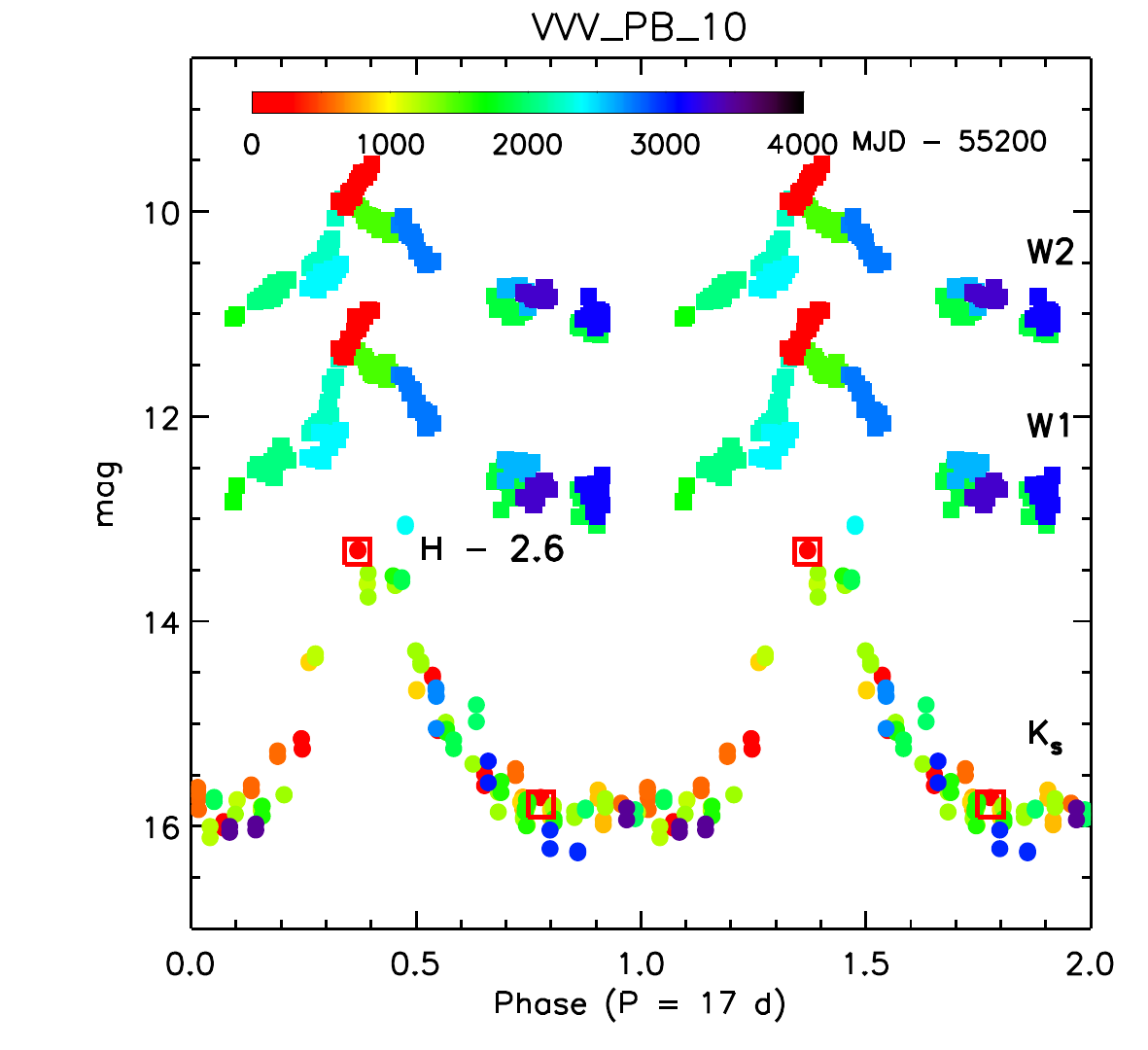}
\includegraphics[width=2.6in,angle=0]{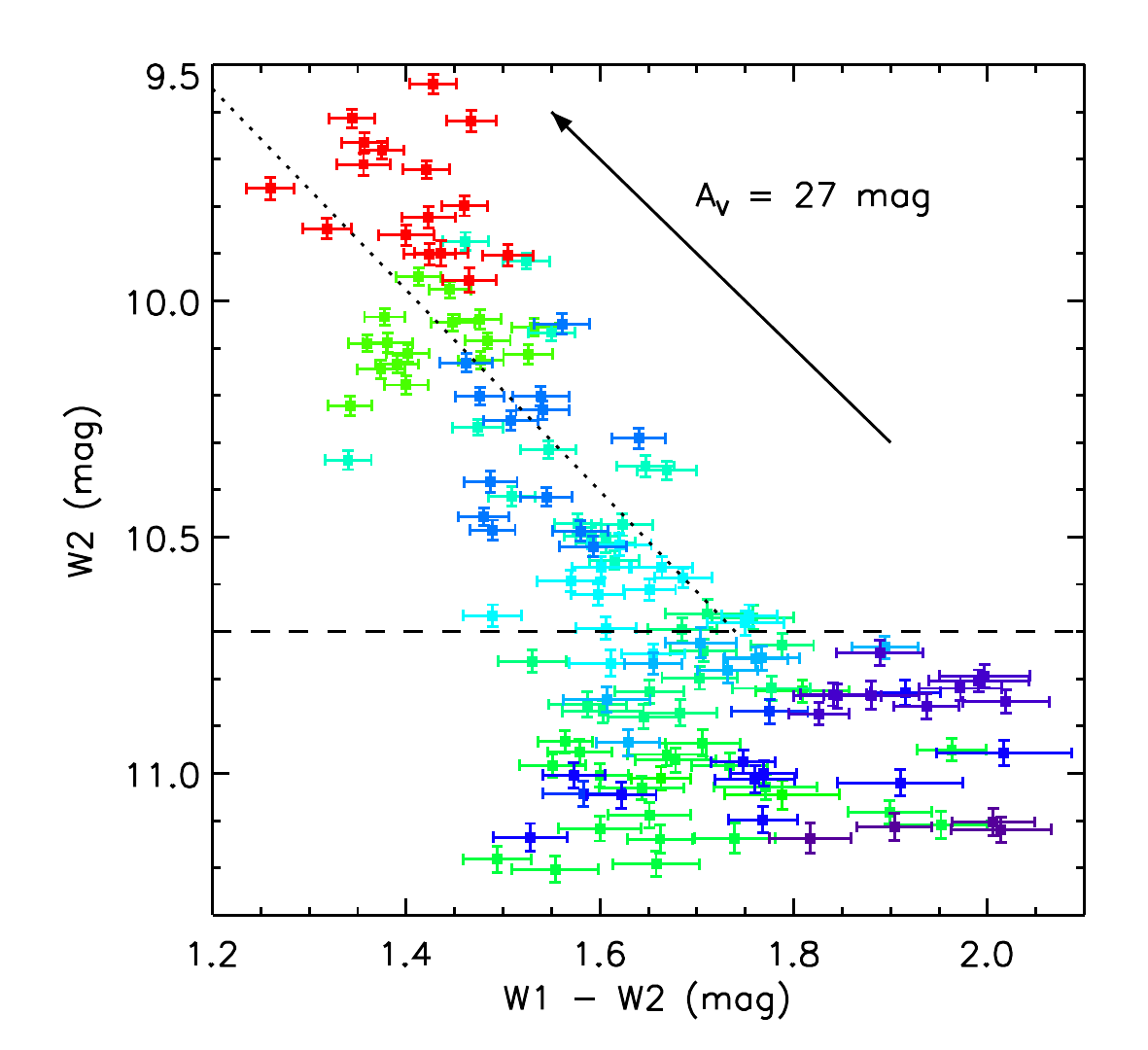}
\caption{{\it Left}: Phase-folded $K_s$, $W1$ and $W2$ light curves of VVV\_PB\_10, under a 17-day period. Two VVV $H$-band detections are shown by open squares and are shifted by -2.6 mag. Epochs are colour-coded by the observation time. {\it Right}:  $W1 - W2$ and $W2$ colour magnitude diagram of VVV\_PB\_10, with the same colour-code as the {\it left} panel. The dotted line is the linear fitting result of detections with $W2 \le 10.7$~mag. The extinction vector is also presented assuming $R_V=3.1$ \citep{WangS2019}. The typical error bars in $W1$ and $W2$ bands are less than 0.1 mag.}
\label{fig:PB10}
\end{figure*}

This source is highlighted due to its short period and well-sampled $K_s$ to $W2$-band phase-folded light curves (shown in Figure~\ref{fig:PB10}). The $W1$ and $W2$-band light curves are obtained from the {\it WISE} All-Sky Single Exposure catalogue and the NEOWISE catalogue. VVV\_PB\_10 has a symmetric analytical phase curve ($S$=-0.01, see Figure~\ref{fig:skew}), and has similar values of $\Delta H$ and $\Delta K_s$. Different amplitudes are measured in the near and mid-infrared bands: $\Delta K_s = 2.76$~mag, $\Delta W1 = 2.17$~mag, and $\Delta W2 = 1.66$~mag. No obvious long-term linear variation trends were observed on $K_s$, $W1$ and $W2$ light curves. Some period-to-period variation is seen in the {\it WISE} light curves, especially during the photometric maximum. The {\it WISE} colour-magnitude diagram is shown in the right panel of Figure~\ref{fig:PB10}. The $W1 - W2$ colour variation is consistent with the extinction vector when $W2$ is brighter than 10.7~mag, although some colour variations are also seen within individual epochs. The colour-magnitude variation was not studied during the faint stage ($W2$ > 10.7~mag) where scatter amongst the individual scans within each epoch predominates. 

The photometric variation of VVV\_PB\_10 is summarised as follows: 1) the morphologies of $K_s$ to $W2$ phase-folded light curves suggest periodic accretion bursts; 2) the relatively constant $H - K_s$ colour between photometric minimum and maximum is consistent with a variable mass accretion process; 3) the {\it WISE} colour variation is parallel to the extinction vector. However, the declining amplitude in the mid-infrared could also indicate that the variation involves hot matter, expected to lie in the innermost disc, and the outbursts have less impact on the SED at longer wavelengths, where cooler matter located further out in the disc dominates the SED. The stellar rotation period of a Class II source is somewhere between 1 to 20 days according to recent photometric surveys \citep{Rebull2020}. Therefore, the 16.99~d period of VVV\_PB\_10 is most likely associated with a dynamical perturber or an obscuration source (optically thick in the near-infrared) on an orbit that lies close to the inner edge of the circumstellar disc. Several YSOs with periods around 20 days were previously detected. For example, the 25~d periodic protostar LRLL~54361 \citep{Muzerolle2013} has a symmetric outbursting morphology like VVV\_PB\_10, but with larger infrared amplitudes ($\Delta m_{[3.6]} \sim 3.0$~mag and $\Delta m_{[4.5]} \sim 2.8$~mag). Dynamical perturbation from a stellar or sub-stellar companion was proposed as the variation mechanism of LRLL~54361.

\begin{figure}
\centering
\includegraphics[width=2.6in,angle=0]{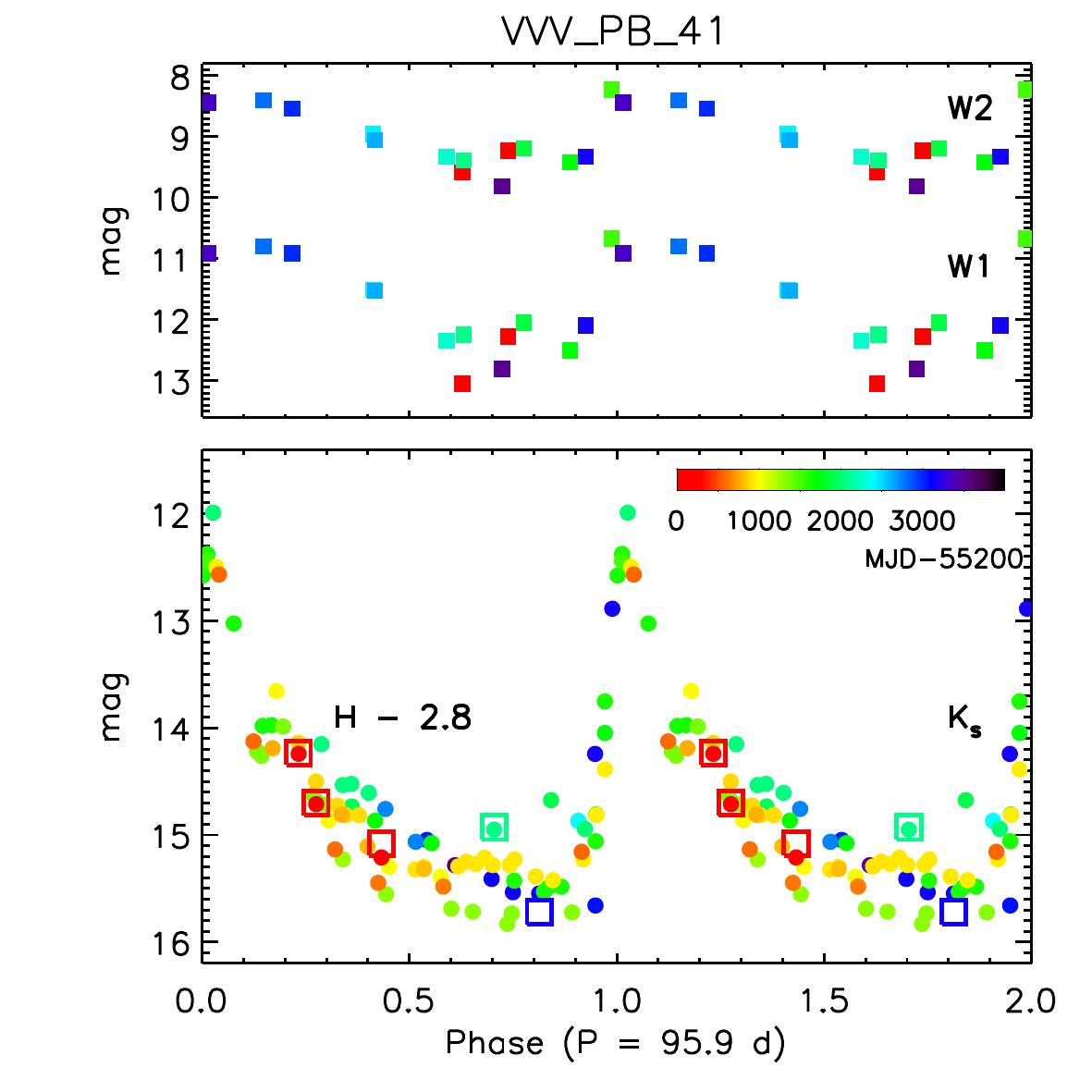}
\caption{Phase-folded $K_s$, $W1$ and $W2$ light curves of VVV\_PB\_41 under a 95.9~d period. $H$-band detections are shown by open squares and are shifted by -2.8 mag to match simultaneous $K_s$ detections. Epochs are colour-coded by the observation time (see the colour bar in the lower panel). Photometric error bars are smaller than the symbols.}
\label{fig:pb41}
\end{figure}
\subsubsection{VVV\_PB\_41}

The large-amplitude near-infrared variability of VVV\_PB\_41 has been previously reported \citep{Teixeira2018}. Associated with an infrared outflow, VVV\_PB\_41, also known as G353.40-0.07, was identified as an extended green object (EGO) from the {\it Spitzer} GLIMPSE~II survey \citep{ChenXi2013}. Estimated by infrared SED (1.2 to 70 $\mu$m) fitting \citep{Robitaille2007, Teixeira2018}, VVV\_PB\_41 is an embedded YSO with $M \sim 2.8 {\rm\, M}_\odot$, though this is uncertain because the distance is not independently constrained.

In this work, a 95.9~d period is extracted from the VVV/VIRAC2-$\beta$ $K_s$ light curve with $\Delta K_{s} = 3.38$~mag. No robust period was reported in \citet{Teixeira2018} due to the lower precision of the standard VISTA aperture photometry pipeline used in that work and the shorter time series available then. Periods extracted from the $W1$ and $W2$ light curves are 96.1~d and 96.2~d, respectively. The differences of periods extracted among three bands are negligible considering the sparse cadences in both time series after 2015. The phase-folded light curves (adopting $P = 95.9$~d) of VVV\_PB\_41 in the $H$, $K_s$, $W1$, and $W2$ bands are presented in Figure~\ref{fig:pb41}. An outbursting morphology is seen in the $K_s$ phase-folded light curve, with an asymmetric shape including a rapid brightening stage ($\sim$12.5~d) and a more gradual decay afterward ($\sim$38.5~d). Five $H$-band detections were taken during the decaying stage with relatively consistent $H - K_s$ colours. The $W1$ and $W2$ phase-folded light curves appear to have different shapes than the $K_s$ light curve, being less sharp during the outburst peak, though some caution is needed due to the limited sampling around the photometric maxima.

Similar periodic variable behaviour was recently observed in a flat spectrum YSO, V347~Aur \citep{Dahm2020}, with the same fast-rising morphology. Adopting the estimated mass of 2.83 M$_\odot$ for VVV\_PB\_41,  a 96~d Keplerian orbit corresponds to a semi-major axis, $a=0.58$~AU. If the perturbing source is a giant planet on an eccentric orbit, it will provide some unique perspective to study the planet forming and migration history. The challenge of this scenario is whether the perturbation from a low-mass companion at a wide orbit is able to significantly modulate the stellar mass accretion, then efficiently heat the inner edge of the circumstellar disc, leading to a $\sim$3.5 mag variation in $K_s$. In the case of young binary systems, numerical simulations suggested binaries on eccentric orbits can periodically modulate the mass accretion rate on the entire system by one order of magnitude \citep{Munoz2016, Kuruwita2020}. However, so far, there is no clear evidence that VVV\_PB\_41 is in a binary system.

\subsection{Variation Mechanisms}

We note that the $K_s$-band emission, mainly from the inner edge of the circumstellar disc, presents a wide range of variability periods, from the stellar rotation period of a few days (the dynamical timescale of the magnetosphere) to the Keplerian timescale of several AU ($\sim$1000~d). We do not assume that all periodic outbursting sources are attributed to dynamical perturbations, since some other mechanisms, perhaps associated with the stellar magnetic field, might be able to produce repeated accretion bursts on YSOs. For example, the ``magnetic gating'' mechanism was proposed to explain EXor-type outbursts \citep{DAngelo2010, DAngelo2012}, where the inner disc material is trapped outside the star-disc co-rotation radius, until the pressure overcomes the magnetic barrier, causing an accretion burst. However, some outstanding theoretical challenges remain. It is unknown whether this mechanism can produce periodic variations on YSOs that last nearly a decade and it can hardly explain outbursts with periods much longer than the dynamical timescale of the magnetosphere. Moreover, models with simple viscous evolution of the burst struggle to explain the shorter period systems because an extremely high value of the viscosity parameter, $\alpha$, is required \citep{YHLee2020}. An alternative mechanism to consider is perturbation by cyclical reversal of the stellar magnetic field \citep{Clarke1995, Armitage2016}. The amplitude and period of the variability would then depend on the characteristics of the stellar activity cycle. However, this mechanism is plausible only for periods of years so it is unlikely to be responsible for all the observed cases.

Numerical simulations have predicted that young binaries in eccentric orbits can modulate the accretion flow in a periodic way \citep{Dunhill2015, Munoz2016}. These hydrodynamic simulations of pulsed accretion have found that the mass accretion on the primary component of an eccentric binary system can be periodically modulated by one order of magnitude. A similar result has been reported in magneto-hydrodynamic simulations of wide binaries \citep[P > 20 yr;][]{Kuruwita2020}. A recent numerical simulation work also explored the scenario that close-in hot Jupiter planets on eccentric orbits can significantly modulate the mass accretion rate of the parent star \citep{Teyssandier2020}. Given the observational evidence for pulsed accretion in eccentric stellar binaries, we conclude that at present dynamical perturbation from a stellar or planetary companion is probably the most promising model to produce variability on a wide range of timescales.

\section{Summary}
\label{sec: con}
Periodic outbursts and pulsed accretion are rarely observed phenomena in YSOs. We performed two Lomb-Scargle based period searches on the 10-year-long VVV/VVVX $K_s$ light curves provided by the VIRAC2-$\beta$ catalogue. Approximately 130 new periodically outbursting YSO candidates were identified by the morphology of their phase-folded light curves, with periods ranging from 10 to 1500 days. Some highlights of our findings are listed below. 

\begin{itemize}

\item Our first search among large-amplitude VVV variables ($\Delta K_s > 1.5$~mag) discovered 59 periodic outbursting YSO candidates, initially selected using their light curves properties. These sources either were identified as YSO candidates in the literature, or are identified by association with H{\sc ii} regions or close proximity to other YSO candidates. We measured the spectral index of 44 outbursting YSO candidates: 15 sources are Class I YSOs, 17 are flat-spectrum sources, and 12 are disc-bearing Class II sources. Five sources have been spectroscopically confirmed as accreting YSOs. 

\item The second search in this paper identified 71 lower amplitude periodic outbursting YSO candidates from existing Galactic YSO catalogues, which resemble pulsed accretors except for their exceptionally long and regular periods, as opposed to the intermittent bursts seen previously. In general, about 10\% of periodic YSO candidates have outbursting morphology in the $K_s$ light curves. 

\item Most periodic outbursting YSO candidates have similar values of $\Delta H$ and $\Delta K_s$, supporting the hypothesis that the photometric variability is due to a cyclical accretion process rather than changing extinction. Variable extinction that is grey or optically thick cannot be formally ruled out but this explanation appears unlikely in view of the sharply peaked light curve morphology.

\item Statistically, we found large-amplitude sources ($\Delta K_s > 1.5$~mag) have redder $W1-W2$ colours than small-amplitude sources, indicating the existence of more hot dust around large-amplitude sources. 

\item Symmetric bursts, fast-rise/slow-decay bursts and slow-rise/fast-decay bursts are seen amongst the YSO candidates. However, fast-rise bursts predominate at the highest amplitude end ($\Delta K_{s} > 2.5$~mag).

\item No correlation is found between the period and variation amplitude of periodic outbursting candidates. However, there is lack of low-amplitude and short-period sources, which agrees with the quasi-periodic nature of pulsed accretors seen among young binary systems. 

\end{itemize}

Our results provide evidence of the dynamical impact of stellar or planetary companions on the mass accretion process in young stellar systems. They also provide valuable laboratories to study how accretion discs respond to perturbations, which has relevance to some of the disc instability theories proposed to explain the unpredictable episodic accretion events seen in some YSOs in particular and planet formation in general. Future multi-wavelength photometric and spectroscopic follow-up of these periodic outbursting YSO candidates are crucial to confirm ongoing mass accretion behaviour, therefore to understand their variable nature. Specifically, the physical conditions of the YSO and its circumstellar disc, before, during and after the outburst, will be further revealed by well-organised observations with sufficient phase coverage.

\section*{Acknowledgements}

Authors thank for the helpful comments and suggestions from the anonymous referee. ZG, PWL, CJM and NM acknowledge support by STFC Consolidated Grants ST/R00905/1, ST/M001008/1 and ST/J001333/1 and the STFC PATT-linked grant ST/L001403/1. AB, DM and MC gratefully acknowledge support by the ANID BASAL project FB210003. JB, RK, JA-G., and MC thank support from the Ministry for the Economy, Development and Tourism, Programa Iniciativa Cientifica Milenio grant IC120009, awarded to the Millennium Institute of Astrophysics (MAS). JA-G. acknowledges support from FONDECYT Regular 1201490. AB acknowledges support by ANID's Millennium Science Initiative Program NCN19\_171 and support from FONDECYT grant 1190748. FN is grateful for financial support by Proyecto Gemini CONICYT grants \#32130013 and \#32140036, and by P. Universidad Cat\'olica de Chile's Vicerrector\'{i}a de Investigaci\'on (VRI).

This work used the high-performance computing facility of University of Hertfordshire. We gratefully acknowledge data from the ESO Public Survey program ID 179.B-2002 taken with the VISTA telescope, and products from the Cambridge Astronomical Survey Unit (CASU). Based on observations obtained at the Southern Astrophysical Research (SOAR) telescope, which is a joint project of the Minist\'{e}rio da Ci\^{e}ncia, Tecnologia e Inova\c{c}\~{o}es (MCTI/LNA) do Brasil, the US National Science Foundation’s NOIRLab, the University of North Carolina at Chapel Hill (UNC), and Michigan State University (MSU). This research has made use of the NASA/IPAC Infrared Science Archive, which is funded by the National Aeronautics and Space Administration and operated by the California Institute of Technology. This publication makes use of data products from the Wide-field Infrared Survey Explorer, which is a joint project of the University of California, Los Angeles, and the Jet Propulsion Laboratory/California Institute of Technology, funded by the National Aeronautics and Space Administration. This publication also makes use of data products from NEOWISE, which is a project of the Jet Propulsion Laboratory/California Institute of Technology, funded by the Planetary Science Division of the National Aeronautics and Space Administration.

\section*{Data Availability}

The {\it WISE} and {\it Spitzer} data underlying this article are publicly available at \url{https://irsa.ipac.caltech.edu/Missions/wise.html} and \url{https://irsa.ipac.caltech.edu/Missions/spitzer.html}, respectively.
The VVV and VVVX data are publicly available at the ESO archive \url{http://archive.eso.org/cms.html}. The relevant reduced products for the most recent VVVX epochs have not yet been publicly released but are available on request to the first author. The SOAR spectra are also available on request to the first author.

\bibliographystyle{mnras}
\bibliography{reference}

\appendix

\section{Figures and Tables}

\begin{figure*}
\centering
\includegraphics[width=6.3in,angle=0]{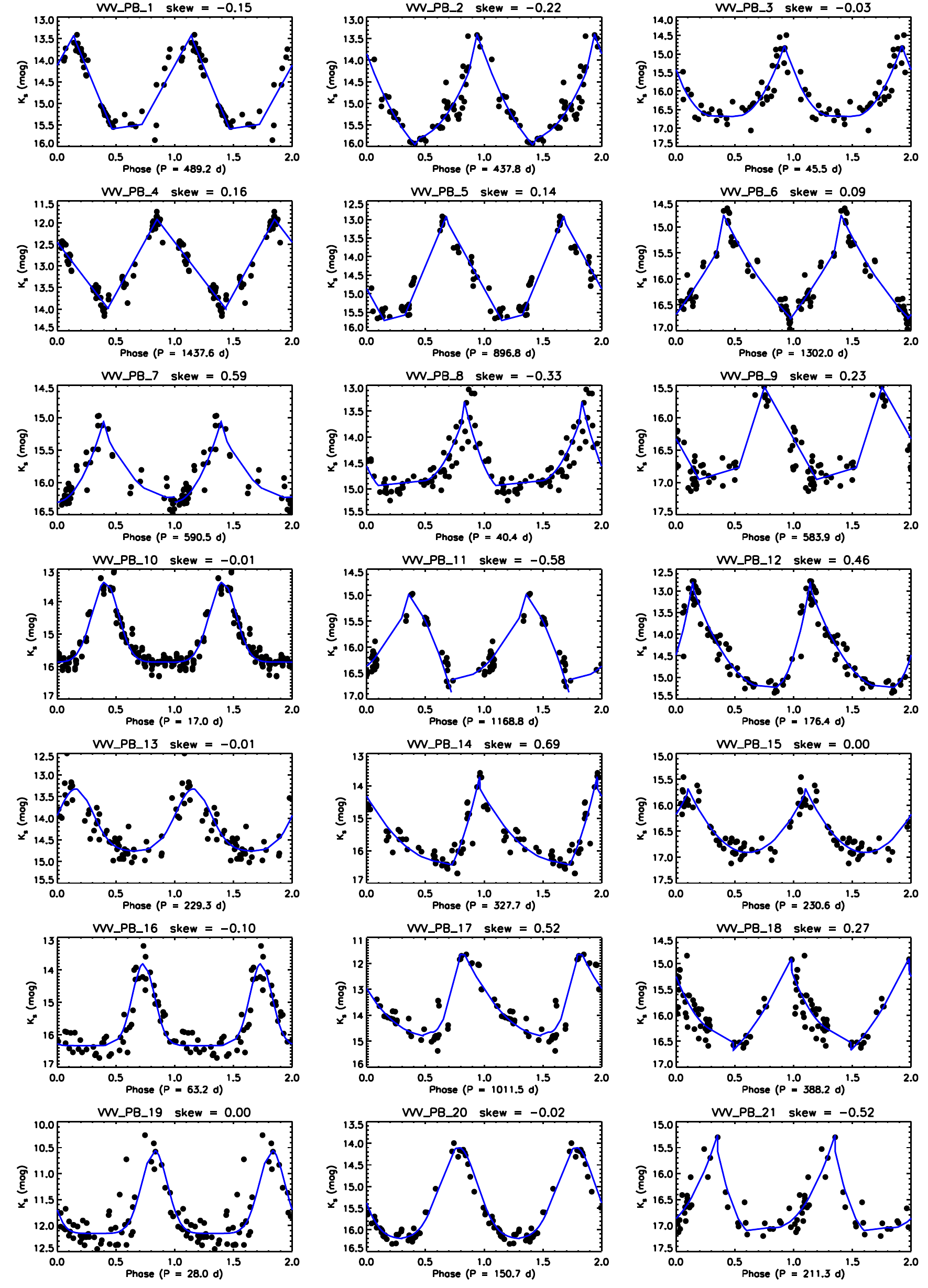}
\caption{  VVV/VVVX $K_s$-band phase-folded light curves of periodic outbursting YSO candidates discovered in the first search. Long-term variation is removed before the phase-folding. Analytical fittings are shown by the blue curves. Unfolded light curves are presented in Figure 1 to Figure 3 of online supplementary files.}
\label{fig:PB1}
\end{figure*}

\begin{figure*}
\centering
\includegraphics[width=6.5in,angle=0]{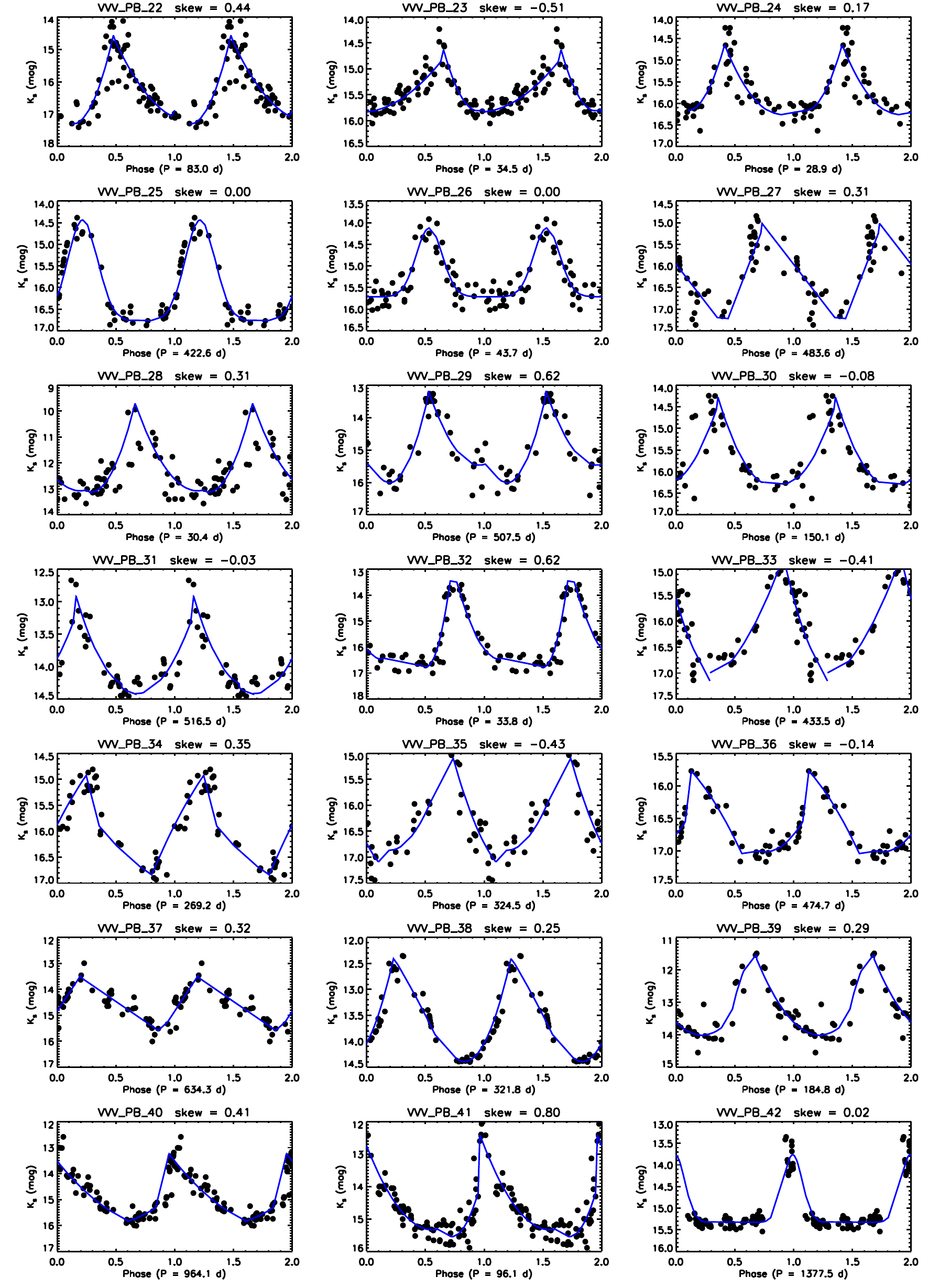}
\caption{\ref{fig:PB1} continued.}
\label{fig:PB2}
\end{figure*}

\begin{figure*}
\centering
\includegraphics[width=6.5in,angle=0]{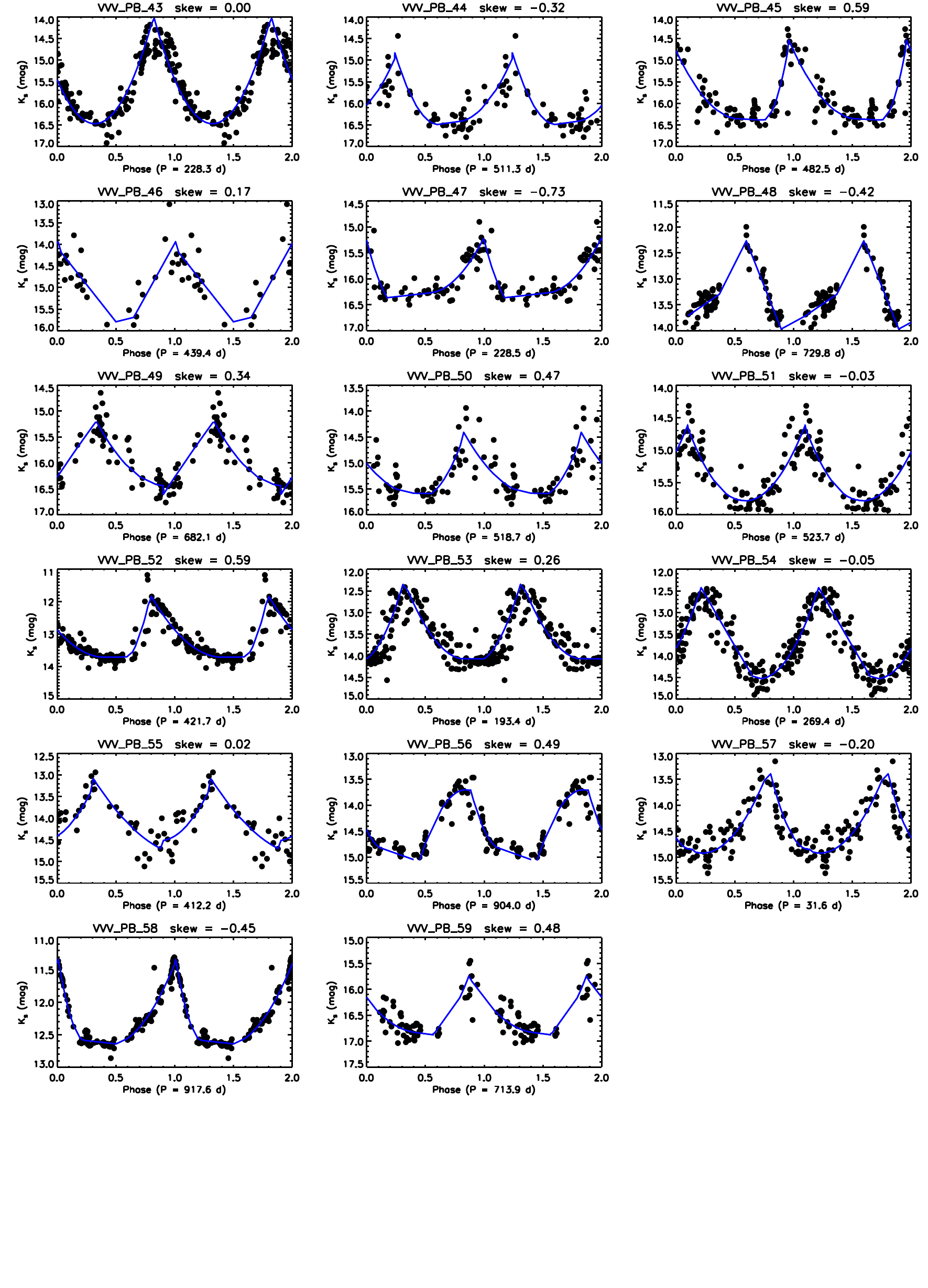}
\caption{\ref{fig:PB1} continued.}
\label{fig:PB3}
\end{figure*}

The VVV/VVVX $K_S$ phase-folded light curves of periodic outbursting YSO candidates identified in the first search are presented in Figure~\ref{fig:PB1} to Figure~\ref{fig:PB3} with analytical phase curves. The unfolded light curves and the phase-folded light curves of sources identified in our second search are presented in the online supplementary files. The information of periodic YSO candidates identified in our second search and other highlighted periodic sources discovered by our pipeline are presented in Table~\ref{tab:low_amp} and Table~\ref{tab:periodic_others}. Some variables that have outbursting morphology but failed our YSO candidate selection are listed as ``burster'' in Table~\ref{tab:periodic_others}. For reference, sources having at least 5 published YSO candidates located within 5' are marked as ``near''. YSO candidates identified in the SPICY catalogue are marked as ``SPICY'' and sources associated with Galactic H{\sc ii} regions are listed as H{\sc ii}. Sources that failed the YSO selection are listed as ``non-YSO''.

\begin{table*} 
\caption{Periodic outbursting YSO candidates discovered in our 2nd search}
\centering
\begin{tabular}{c c c c c c c r r}
\hline
\hline
Name & RA (J2000) & Dec (J2000)  & Period & $\Delta K_s$ &  Med$({K_s})$ & $W1-W2$ & skewness\\
\hline
 & deg & deg  & day & mag  & mag & mag & \\
\hline
VVV\_PB\_60 &   175.992783 &   -63.127666 &    88.1 & 0.3 &  14.09 &   -  & 0.28$\pm$0.14\\
VVV\_PB\_61 &   180.034622 &   -62.951687 &   150.4 & 0.1 &  12.37 &  0.3&-0.16$\pm$0.11\\
VVV\_PB\_62 &   187.290604 &   -62.825199 &    19.8 & 1.6 &  14.41 &  1.4 &-0.16$\pm$0.08\\
VVV\_PB\_63 &   187.526474 &   -62.888618 &    47.7 & 1.3 &  13.70 &  1.1 & 0.34$\pm$0.20\\
VVV\_PB\_64 &   187.535919 &   -62.431339 &   291.3 & 0.2 &  13.51 &  0.5 & 0.38$\pm$0.21\\
VVV\_PB\_65 &   194.954193 &   -63.759903 &   595.6 & 1.1 &  16.78 &   -  &-0.12$\pm$0.13\\
\hline
\end{tabular}
\flushleft{The full table is available in the online supplementary data.}
\label{tab:low_amp}
\end{table*}

\begin{table*} 
\caption{Other periodic sources}
\centering
\begin{tabular}{c c c c c c c c c c}
\hline
\hline
RA (J2000)  & Dec (J2000) & Period & $\Delta K_{s}$ &  Median$({K_s})$ & category & YSO reference \\
\hline
deg & deg & day & mag & mag & & \\
\hline
  179.873962 &   -61.118557 &  306.89 &    1.74 & 16.44 & dipper & non-YSO\\
  179.998535 &   -62.436974 &  412.38 &    1.52 & 13.43 & QP & SPICY\\
  188.922180 &   -61.876019 &  433.13 &    2.24 & 13.57 & LPV & SPICY\\
  194.129242 &   -61.647305 & 1395.47 &    2.58 & 15.63 & LPV & non-YSO\\
  196.643051 &   -61.483536 &  286.00 &    3.03 & 12.88 & LPV & H{\sc ii}\\
  201.818939 &   -61.960693 &  306.74 &    1.71 & 16.32 & burster & non-YSO\\
  \hline
\hline
\end{tabular}
\flushleft{QP: quasi-periodic; LPV: Long-term periodic; STV: short-time variable. YSO reference: see descriptions in Appendix. The full table is available in the online supplementary data.}

\label{tab:periodic_others}
\end{table*} 

\section{Mira Variables}
\label{sec:mira}

\begin{figure*}
\centering
\includegraphics[height=2.7in,angle=0]{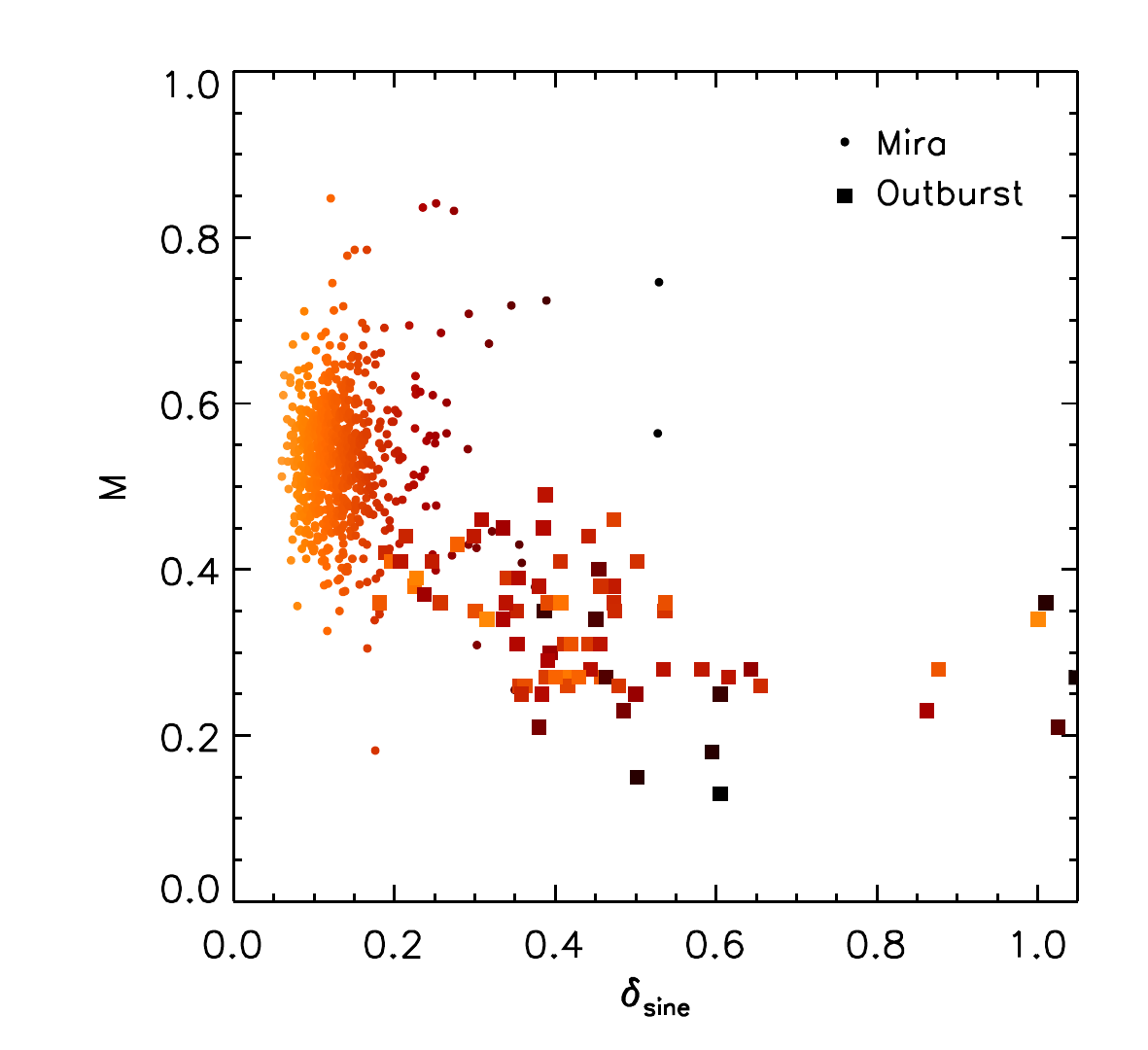}
\includegraphics[height=2.6in,angle=0]{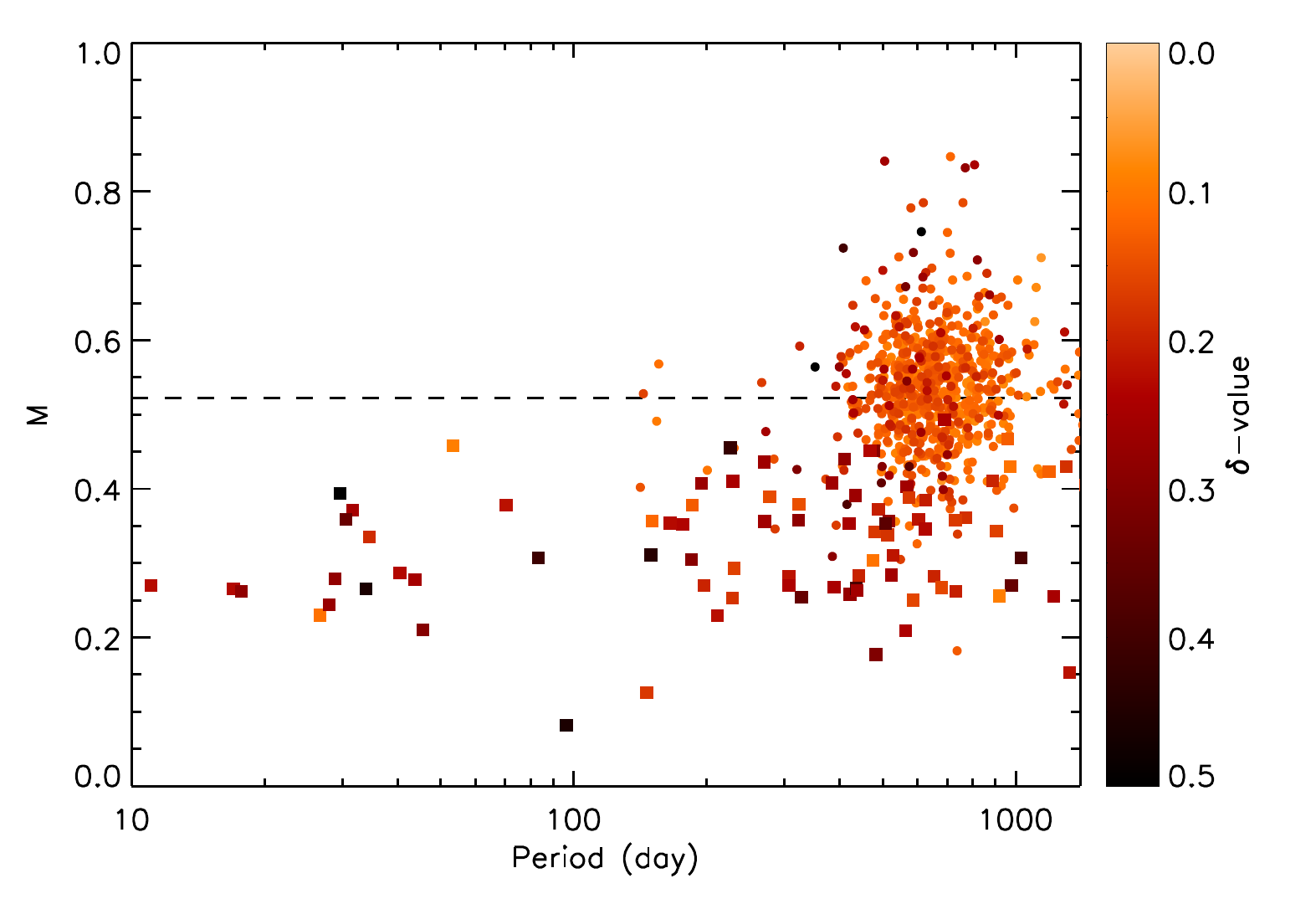}
\includegraphics[height=3.0in,angle=0]{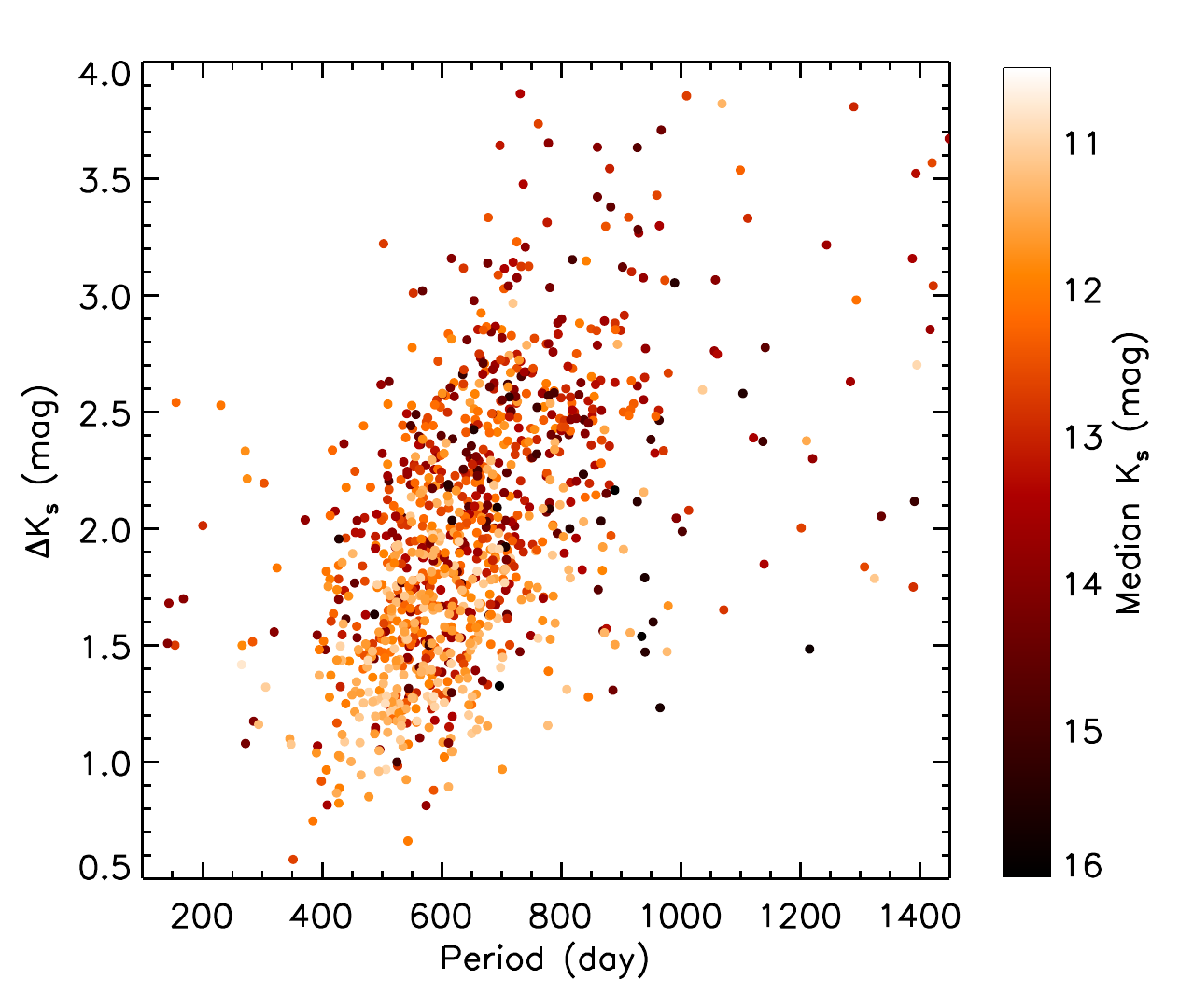}
\caption{{\it Top Panels:} $M$, $\delta_{\rm sine}$ and period of Miras (dots) and periodic outbursting candidates (squares). Individual sources are colour coded by $\delta_{\rm phase}$. The dashed line presents the median $M$ for Miras. {\it Bottom:} Period and $K_s$-band amplitude (long-term trend removed) of Mira candidates, colour-coded by the median $K_s$ magnitude. }
\label{fig:mira_stat}
\end{figure*}

\begin{figure*}
\centering
\includegraphics[height=2.25in,angle=0]{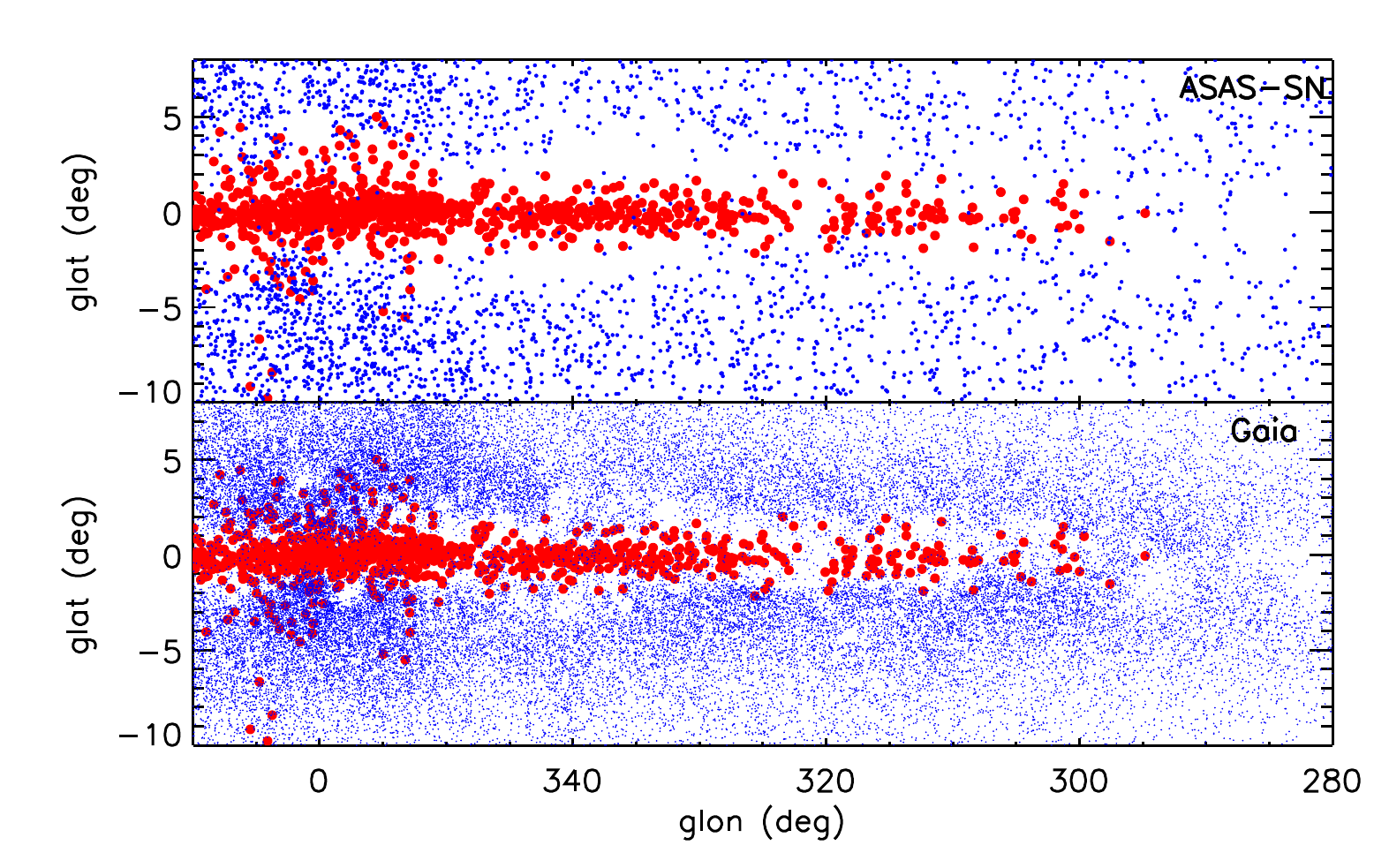}
\includegraphics[height=2.3in,angle=0]{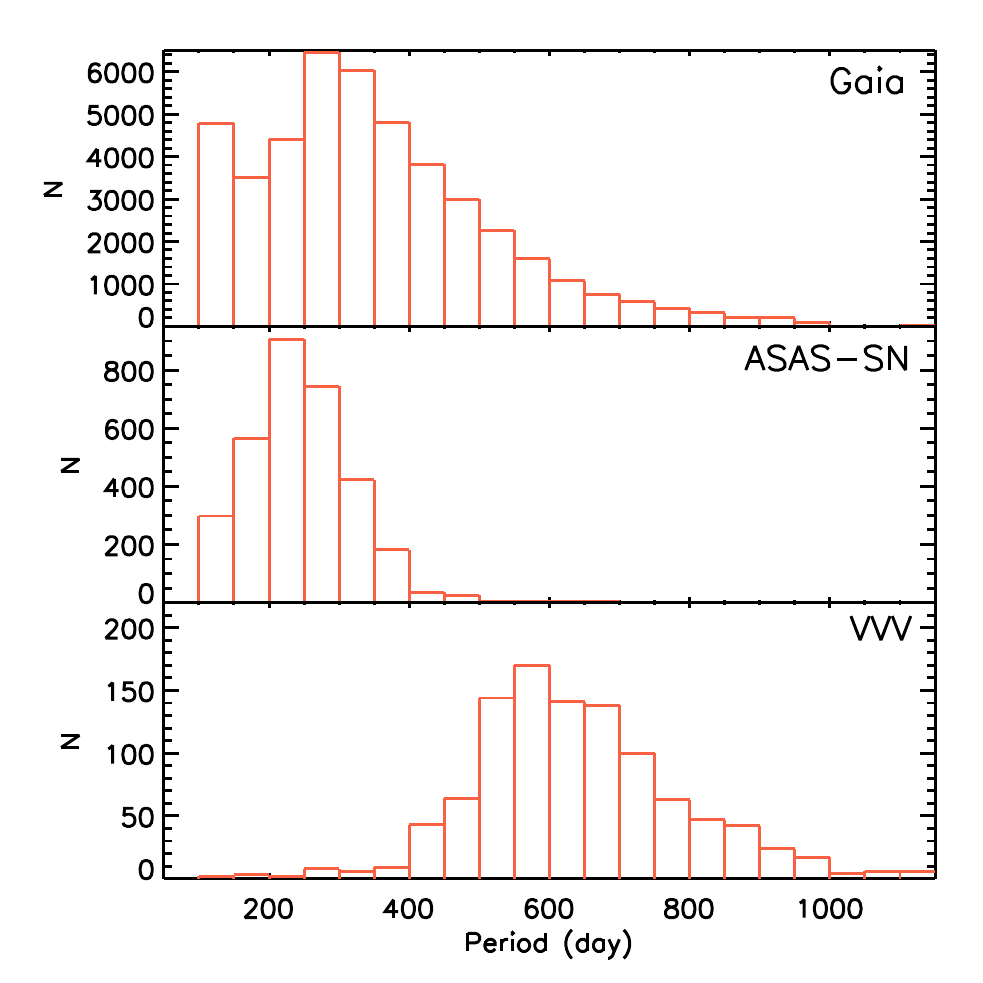}
\caption{{\it Left}: Galactic location of Mira candidates/variables discovered in this work (red) and from  {\it ASAS-SN} survey. LPVs identified from {\it Gaia} around the Galactic bulge and the inner disc are also presented. {\it Right}: Histograms of the period of Miras / LPVs discovered in each survey located in the frame of the {\it left} panels.}
\label{fig:VVV_Gaia_ASASSN}
\end{figure*}

The most abundant long period variables in the VVV survey are dusty pulsating AGB stars, also known as Miras, especially among the brighter end of the variable sample \citep{Catelan2013}. Miras are bright sources ($-10 < M_{K_s} < - 6$~mag) with periods from 10$^2$ to $\sim$10$^3$ days generated by the pulsating photosphere \citep{Whitelock2008}. During the stellar pulsation, a Mira variable undergoes rapid loss of mass \citep{Hofner2018} which forms a dust shell at a few stellar radii \citep{Fleischer1992}. 

In this work, 1313 sources are identified as Mira-candidates due to the morphology of phase-folded light curves, which meet the following criteria. First, the measured periods should be larger than 100 days, consistent with past results \citep{Whitelock2008}. Limited cycle-to-cycle variation is expected from the pulsating nature, as confirmed by the visual inspection. Symmetric phase-folded light curves are expected in most Miras according to previous observations \citep{Matsunaga2009}, although semi-regular and irregular AGB stars may be excluded from this search. We conclude that the region of peak density in the $M$-$\delta_{\rm phase}$ distribution (Figure~\ref{fig:hist_delta}) is mainly populated by Miras. A group of bright sources (median $K_s \le 12$~mag) experienced saturation during their photometric maxima, resulting in asymmetric phase-folded light curves, therefore slightly larger $\delta_{\rm phase}$, which were excluded from further comparisons between Miras and periodic outbursting candidates.

The distributions of period, $\Delta{K_s}$, median $K_s$, and $M$ factor of Mira- and periodic outbursting YSO candidates are presented in Figure~\ref{fig:mira_stat}. A loose trend is seen between $\Delta{K_s}$ and period, consistent with previous results \citep{Molina2019}, as short-period Miras have relatively smaller amplitude. Most Mira-candidates have periods between 400 to 1000 days with unusually higher-amplitude comparing to other Galactic Miras \citep{Mowlavi2018}, which indicate a relatively young Mira population based on the period-age relationship \citep{Feast2007}.

Thousands of Galactic LPVs, including Miras, were detected by optical time-serie surveys, such as ASAS-SN \citep{Jayasinghe2018, Jayasinghe2019}, OGLE \citep{Groenewegen2005} and {\it Gaia} \citep{Mowlavi2018, Gaia2018dr2catalog}. In Figure~\ref{fig:VVV_Gaia_ASASSN}, we compared Galactic locations and periods of Miras identified in this work (VVV-Miras) with previous detections from ASAS-SN (ASAS-SN-Miras) and {\it Gaia}~DR2. VVV-Miras are located much closer to the Galactic plane, compared to those in both optical surveys, as the $K_s$ band is less affected by the high interstellar extinction towards low Galactic latitude regions. According to the period-luminosity relationship \citep[][]{Yuan2017}, Miras with 600~d periods should have $M_{K_s} \sim -8.5$~mag. Hence, Miras in the inner Galaxy found from the VIRAC2-$\beta$ catalogue ($K_s < 10$~mag) should typically experience high circumstellar or foreground interstellar extinction. VVV-Miras have periods significantly longer than ASAS-SN-Miras, which arises from the amplitude selection of our targets and the detection limits of both surveys.  Heavily extincted by the circumstellar dust shell, Miras are generally brighter in the infrared than in the optical. Following the amplitude-period relationship (also seen in Figure~\ref{fig:mira_stat}), short-period Miras, often with old age, do not have large enough photometric amplitudes to be detected by our pipeline. On the other hand, long-period Miras were not discovered by  ASAS-SN due to the limited time span of the survey. 

In another recent study using the VVV photometry, 130 Mira candidates were identified from 10 VVV tiles in the Galactic bulge area \citep{Nikzat2022}, in which both short- ($<$200~d) and long-period ($>$600~d) Miras are detected. Trends between period, amplitude and age of Mira-type variables are found in this work, as short-period Miras have relatively small amplitude, which is smaller than the selection criterion in our work. Therefore, our work only provides good completeness for long-period, high-amplitude Miras and most short-period Miras were missed.

Inter-period variations have been detected on top of the regular pulsation signal of Miras, especially in carbon-rich Miras, due to the dust formation as a consequence of mass-loss \citep{Whitelock1997, Olivier2001}. The long-term variation is measured by subtracting the sinusoidal fits from the $K_s$ light curves (see Figure~\ref{fig:example}). We found that 
  most long-term trends are linear slopes, while a few Gaussian dips and higher order polynomial curves are seen indicating temporary or asymmetric structures in the dust shell. About 30\% of Miras have long-term variation amplitude ($\Delta K_{s, \rm long}$) with median value of 0.58~mag, comparable to the sinusoidal amplitude.

\section{Completeness of this work}

In Paper~I, 816 VVV variables were discovered using $K_s$ light curves covering the interval between early 2010 and late 2014, including five periodic outbursting YSO candidates. The definition of periodic source in this work is narrower than previous works \citep{Guo2021}, as quasi-periodic sources were not included.  There were 28 variables classified as ``LPV-YSO'' in Paper I with $K_s > 1.5$~mag, but only seven were identified as periodic sources in this work, including three sources re-classified as Miras. Also, there is one periodic outbursting source that was previously mis-classified as an ``EB'' in Paper I. The light curves of the remaining ``LPV-YSOs'' from Paper I were visually inspected, and some level of periodicity was seen but with $\delta_{\rm phase}$ greater than 0.5. A substantial number of periodic sources from Paper I were not included in this work since their amplitudes are less than 1.5 mag.  Applying the same period extraction pipeline, periods were measured in 20 lower amplitude ``LPV-YSOs'' from Paper I but only one of these has outbursting morphology in the phase-folded light curve.

There are 51 targets with projected locations in star forming regions that were provisionally categorised as ``LPV-Mira'' and have $\Delta K_s \ge 1.5$~mag in VIRAC2-$\beta$. In this work, 41 out of these 51 ``LPV-Miras'' are identified as Mira candidates. Among the remaining ten ``LPV-Miras'', eight sources have $<$100 detections in $K_s$,  one source (VVVv436) is re-classified as a YSO candidate and one source is not identified as a periodic source. A further 81 Mira candidates discovered in this work were also previously listed in Paper~I, which comprise 75 sources labeled in Paper I as ``LPV'' (the designation given there to Mira candidates that were not projected in star formation regions), one source labelled as ``EB'' (as eclipsing binary), and five sources labelled as ``LPV-YSO''. Separately, several Miras located in the Galactic centre area were discovered using VVV light curves \citep{Molina2019}, among which all five high-amplitude Miras were successfully recovered in this work. 

In a recent publication \citep{Nikzat2022}, 130 Mira-type variables in the Galactic bulge area are discovered by applying point-spread function photometry to ten VVV tiles (2010 -- 2016). Among these Mira variables, 71 sources have median $K_s \ge 11$~mag and $\Delta K_s \ge $~1.5~mag (provided via private communication). We cross-matched these variables with Mira candidates identified in this work, and 47 of them appear in our catalogue. The remaining 24 sources without a match in our catalogue are either due to saturation around photometric maxima (2/24), lack of enough epochs in VIRAC2-$\beta$ (8/24) or do not have small enough $\delta_{\rm phase}$ to be classified as periodic variables by our pipeline.

In conclusion, this paper has a high level of completeness for large-amplitude and unsaturated LPVs with sufficient VVV $K_s$ detections. Some small-amplitude variables were missed due to the nature of our selection criteria, but our second search among YSO candidates in existing catalogues has made up this gap. Therefore, this work provides a relatively complete catalogue of large-amplitude periodic outbursting YSO candidates, serving to increase the total number of periodic outbursting YSOs by an order of magnitude.

\section{Online supplementary figures}

\begin{figure*}
\centering
\includegraphics[width=6.3in,angle=0]{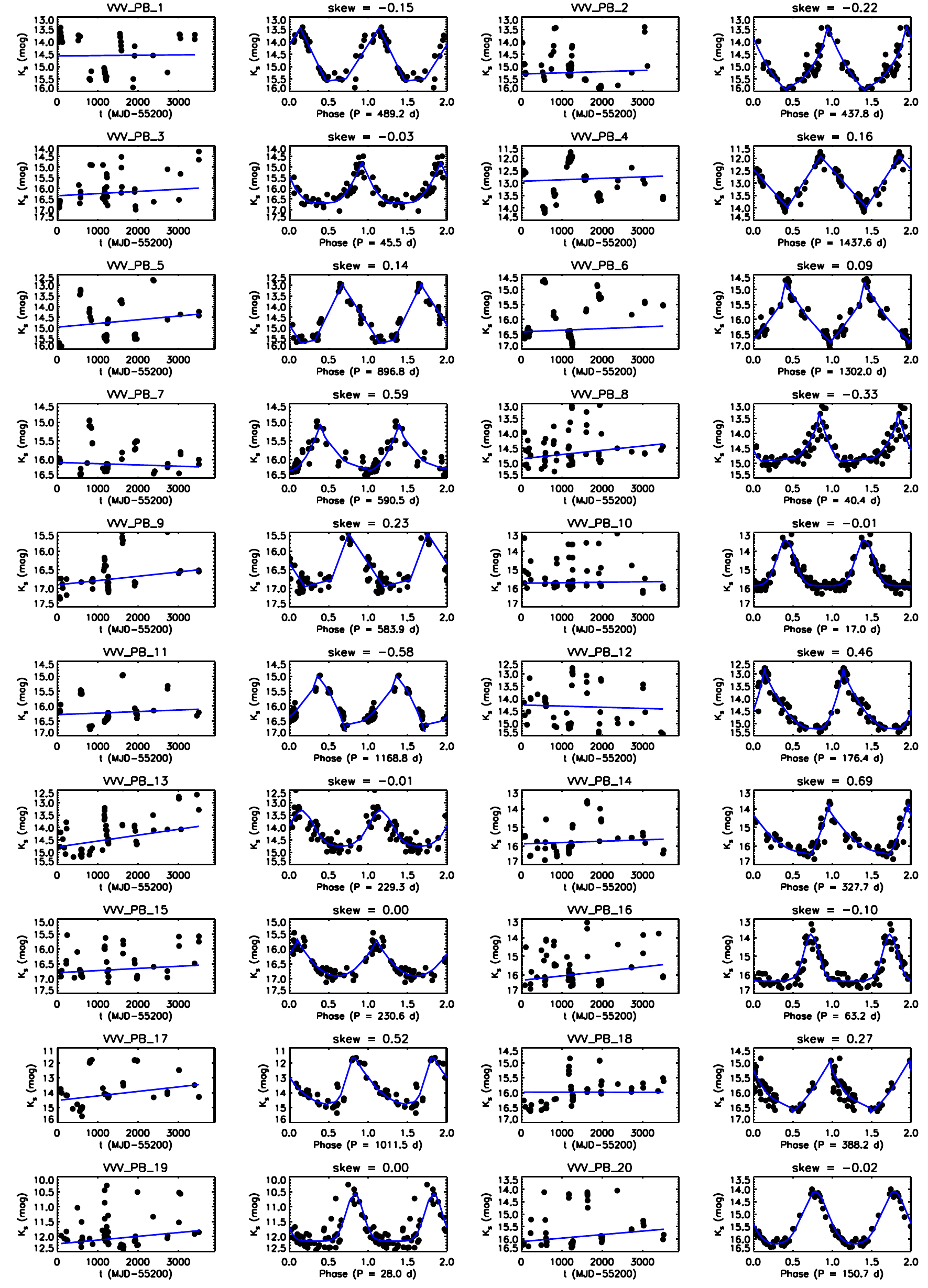}
\caption{VVV/VVVX $K_s$-band light curves and phase-folded light curves of periodic outbursting YSO candidates discovered in the first search. Long-term variation is shown by the blue solid line on light curves, which is removed before the phase-folding. Analytical fittings are shown by the blue curves on the phase-folded light curves.}
\label{fig:PB_online_1}
\end{figure*}

\begin{figure*}
\centering
\includegraphics[width=6.5in,angle=0]{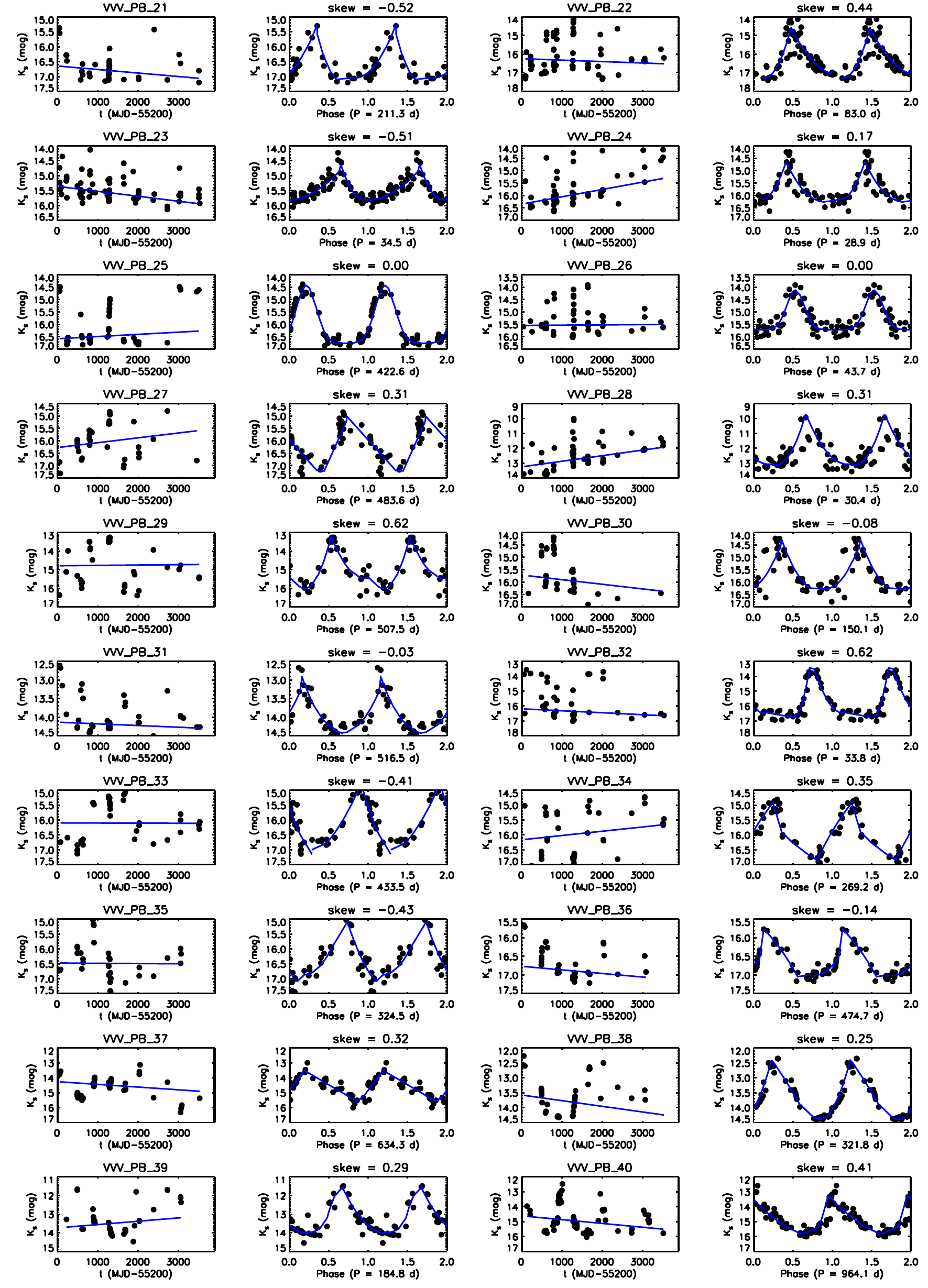}
\caption{Figure~\ref{fig:PB_online_1} continued.}
\end{figure*}

\begin{figure*}
\centering
\includegraphics[width=6.5in,angle=0]{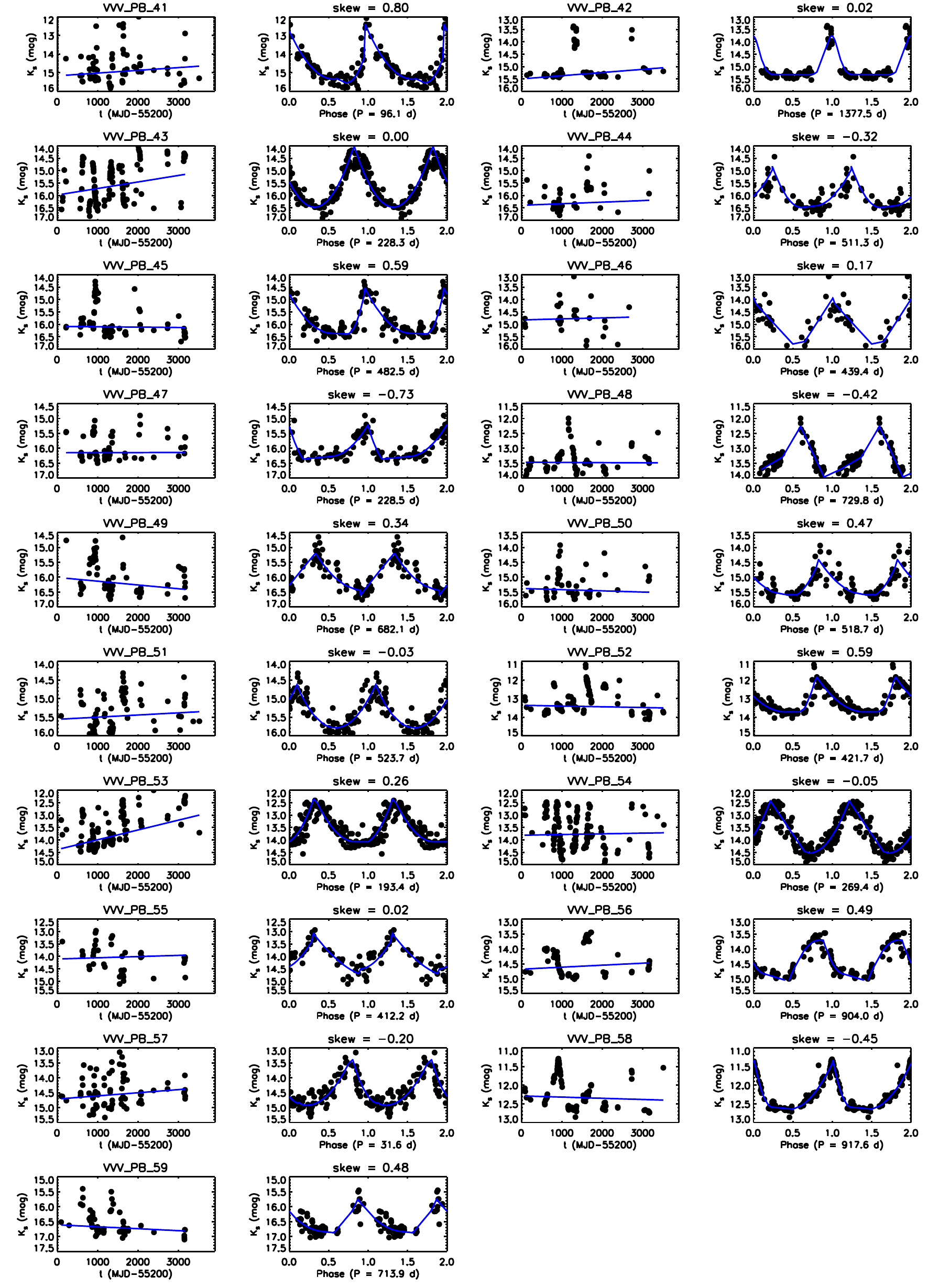}
\caption{Figure~\ref{fig:PB_online_1} continued.}
\end{figure*}

\begin{figure*}
\centering
\includegraphics[width=6.5in,angle=0]{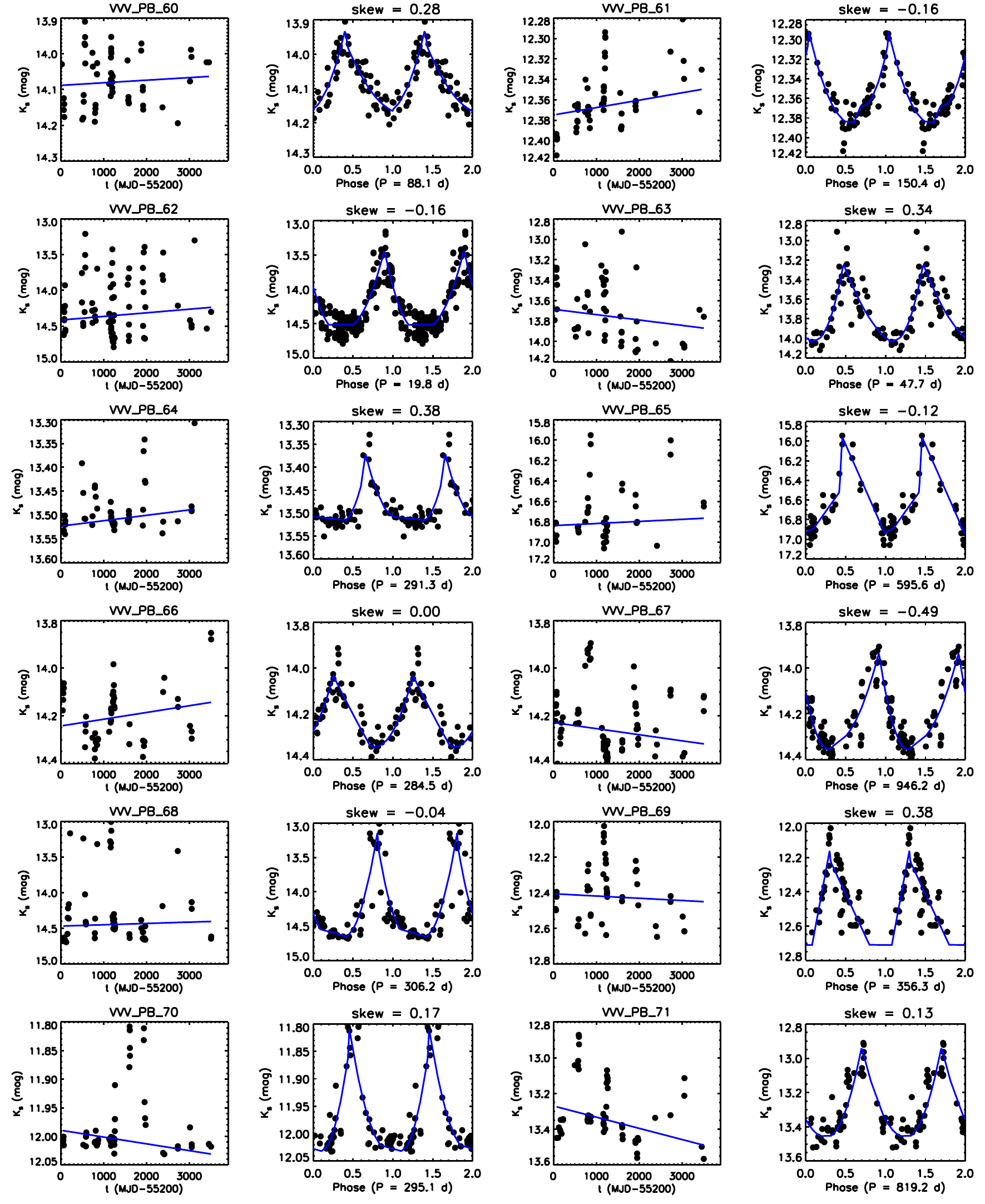}
\caption{VVV/VVVX $K_s$-band light curves and phase-folded light curves of periodic outbursting YSO candidates discovered in the second search. Long-term variation is shown by the blue solid line on light curves, which is removed before the phase-folding. Analytical fittings are shown by the blue curves on the phase-folded light curves.
}
\label{fig:PB_low_1}
\end{figure*}

\begin{figure*}
\centering
\includegraphics[width=6.5in,angle=0]{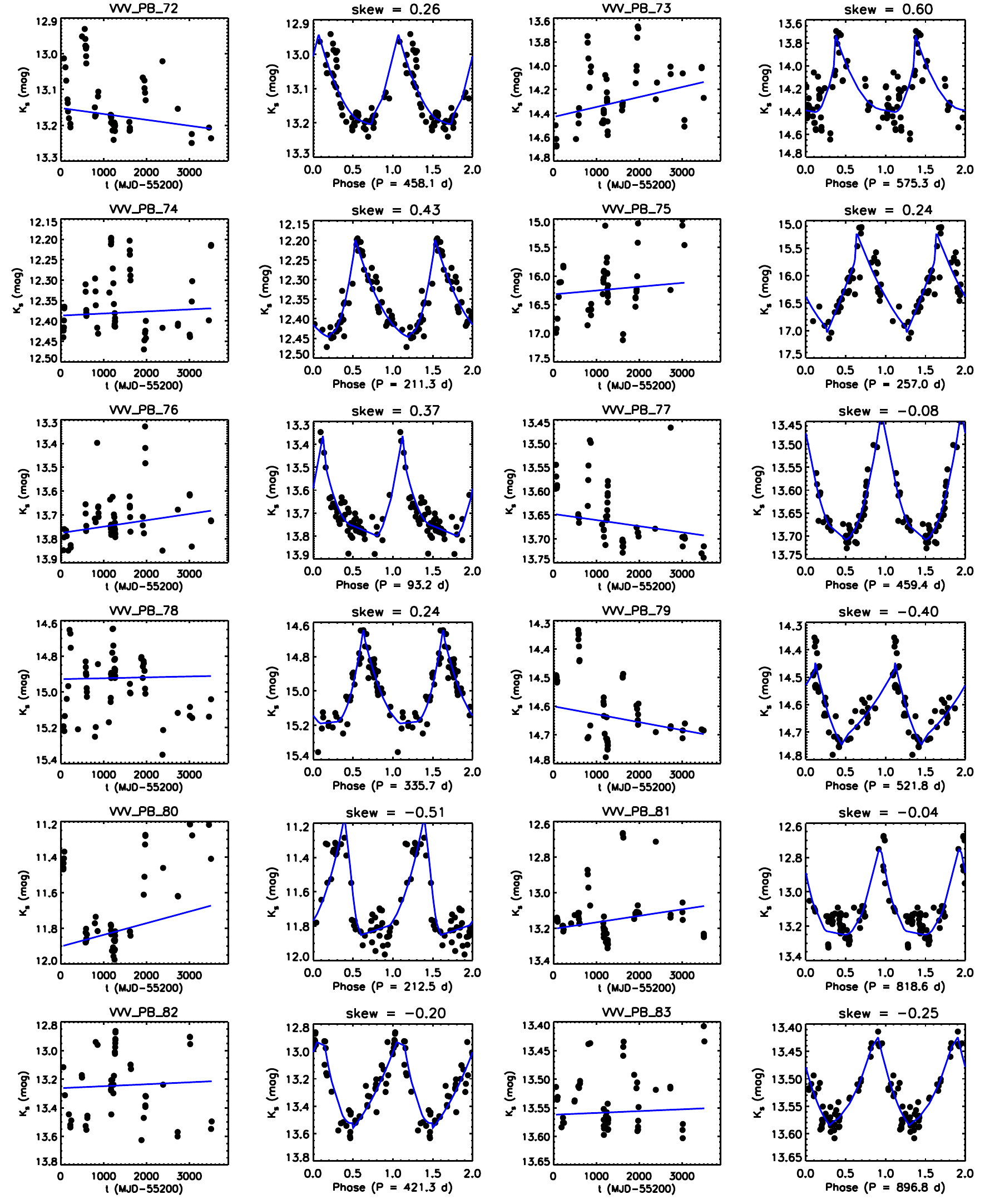}
\caption{Figure~\ref{fig:PB_low_1} continued.}
\label{fig:PB_low_2}
\end{figure*}

\begin{figure*}
\centering
\includegraphics[width=6.5in,angle=0]{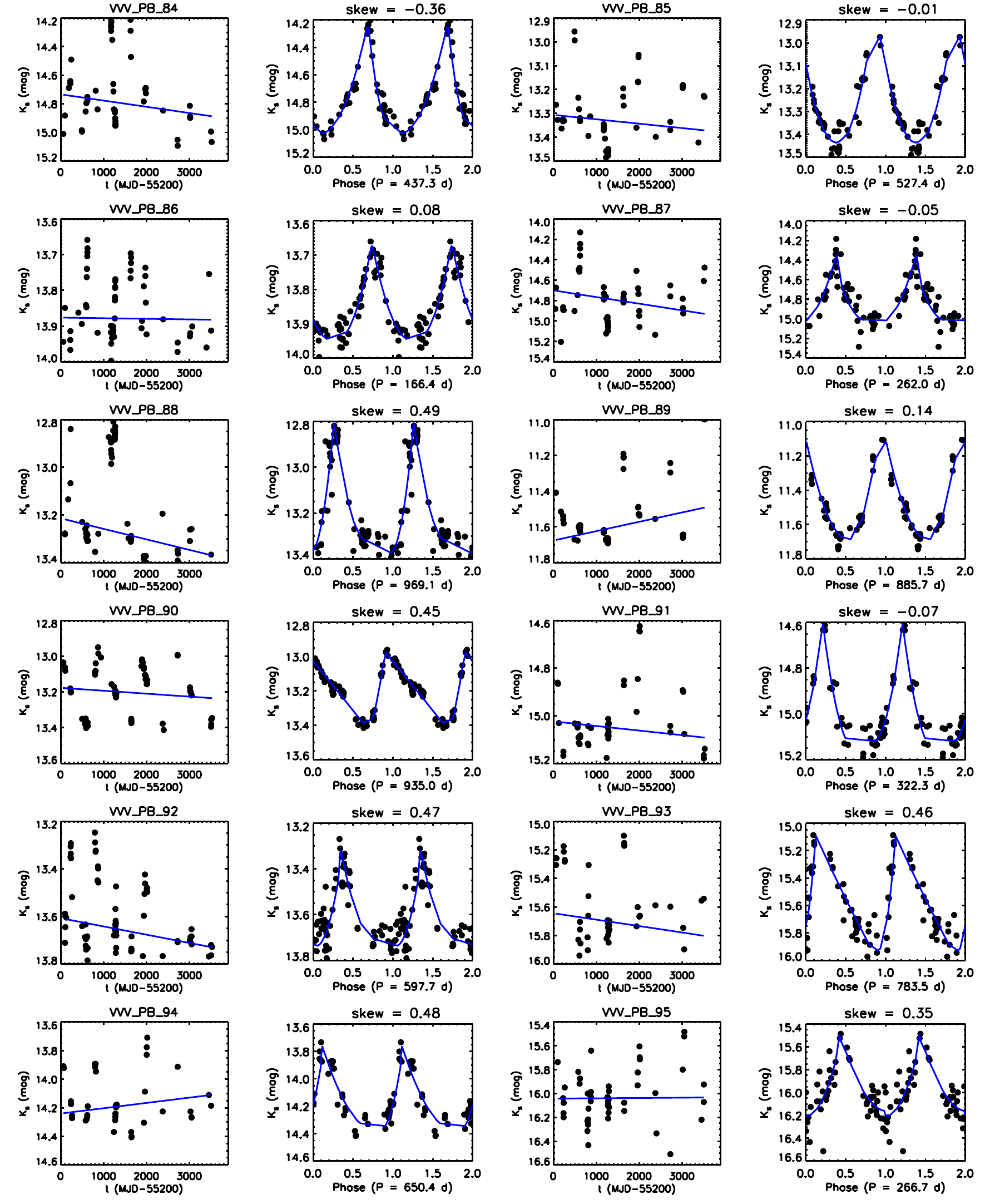}
\caption{Figure~\ref{fig:PB_low_1} continued.}
\label{fig:PB_low_3}
\end{figure*}

\begin{figure*}
\centering
\includegraphics[width=6.5in,angle=0]{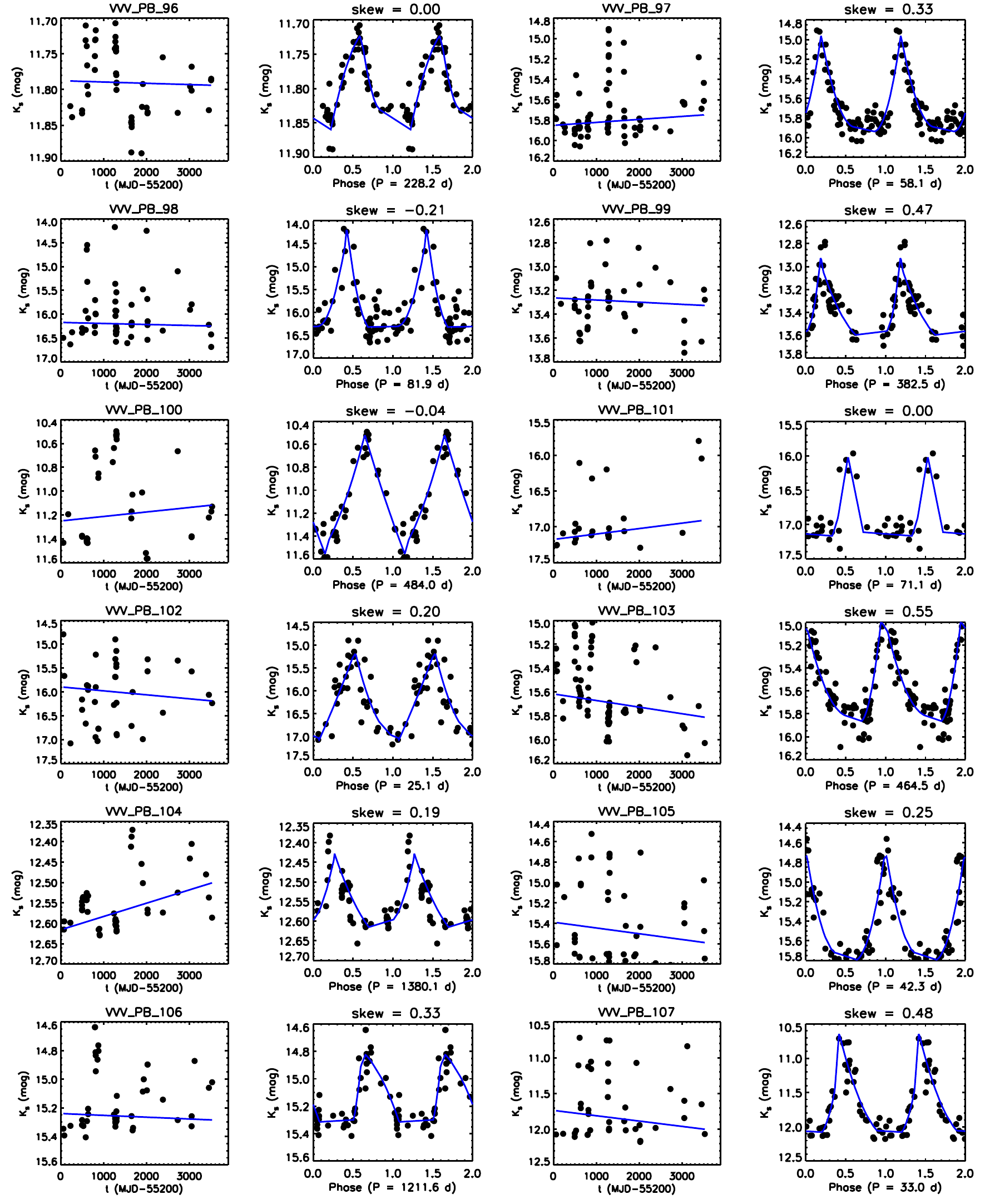}
\caption{Figure~\ref{fig:PB_low_1} continued.}
\label{fig:PB_low_4}
\end{figure*}

\begin{figure*}
\centering
\includegraphics[width=6.5in,angle=0]{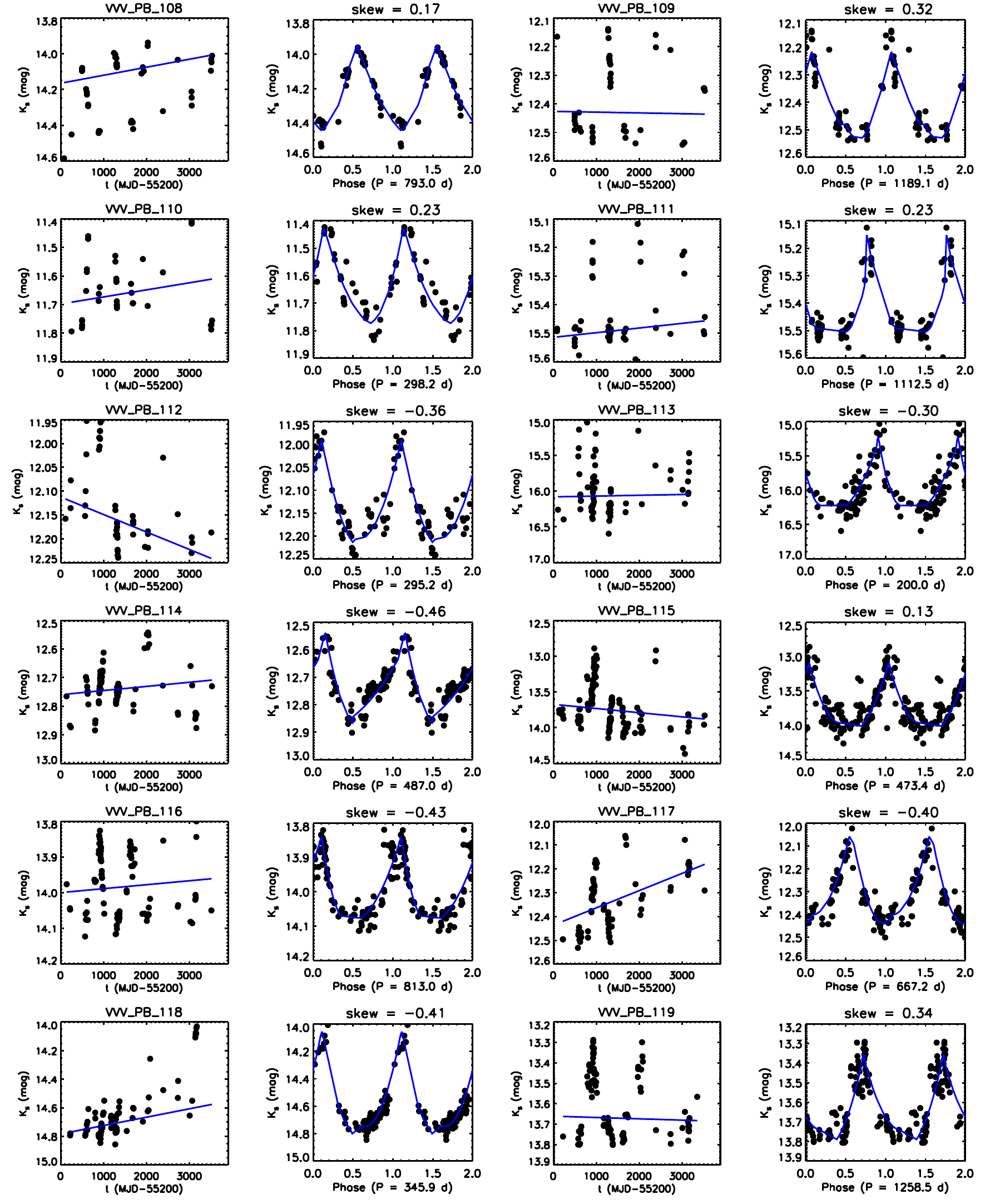}
\caption{Figure~\ref{fig:PB_low_1} continued.}
\end{figure*}

\begin{figure*}
\centering
\includegraphics[width=6.5in,angle=0]{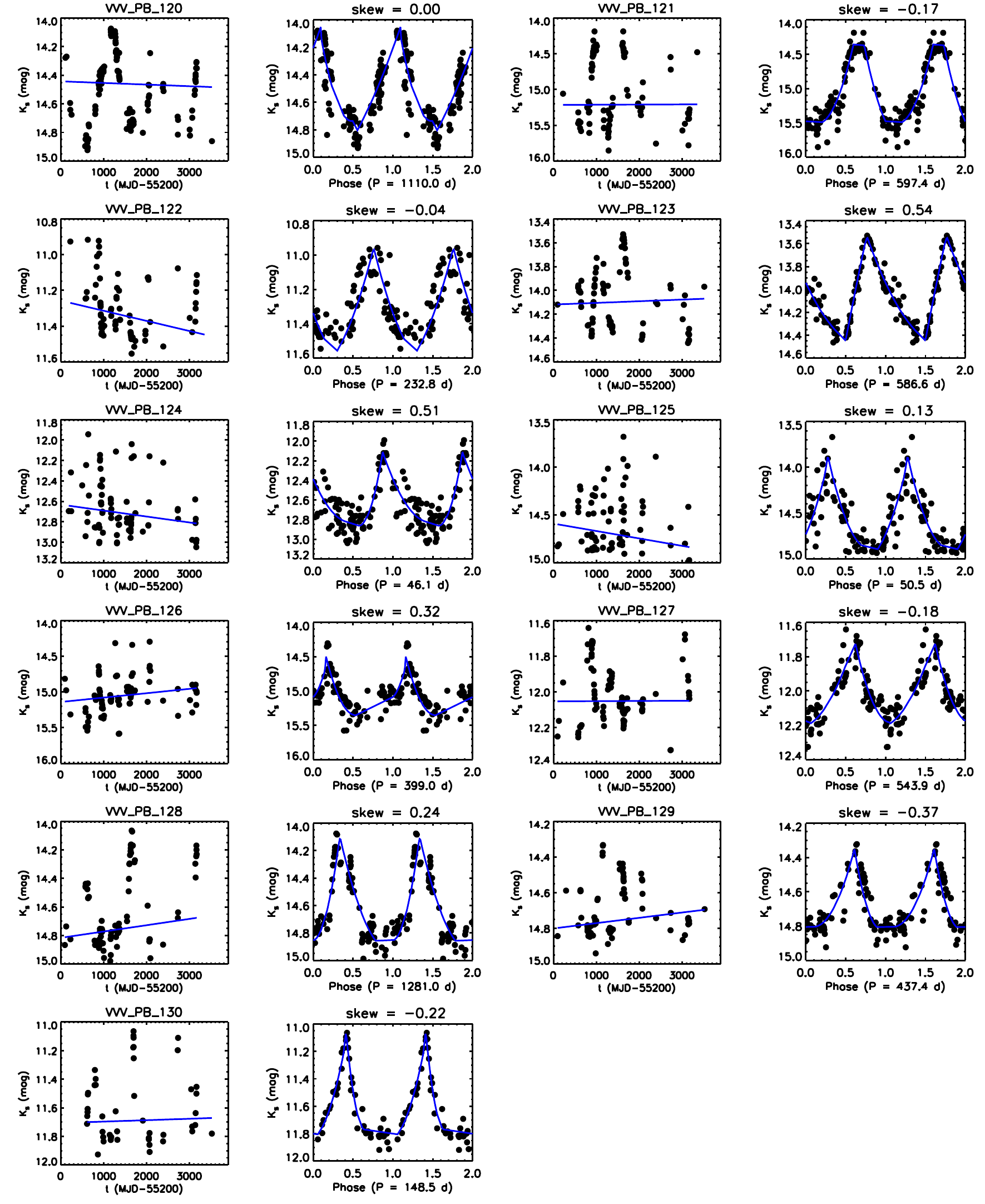}
\caption{Figure~\ref{fig:PB_low_1} continued.}
\end{figure*}

\end{document}


\begin{figure*}
\centering
\includegraphics[width=6.3in,angle=0]{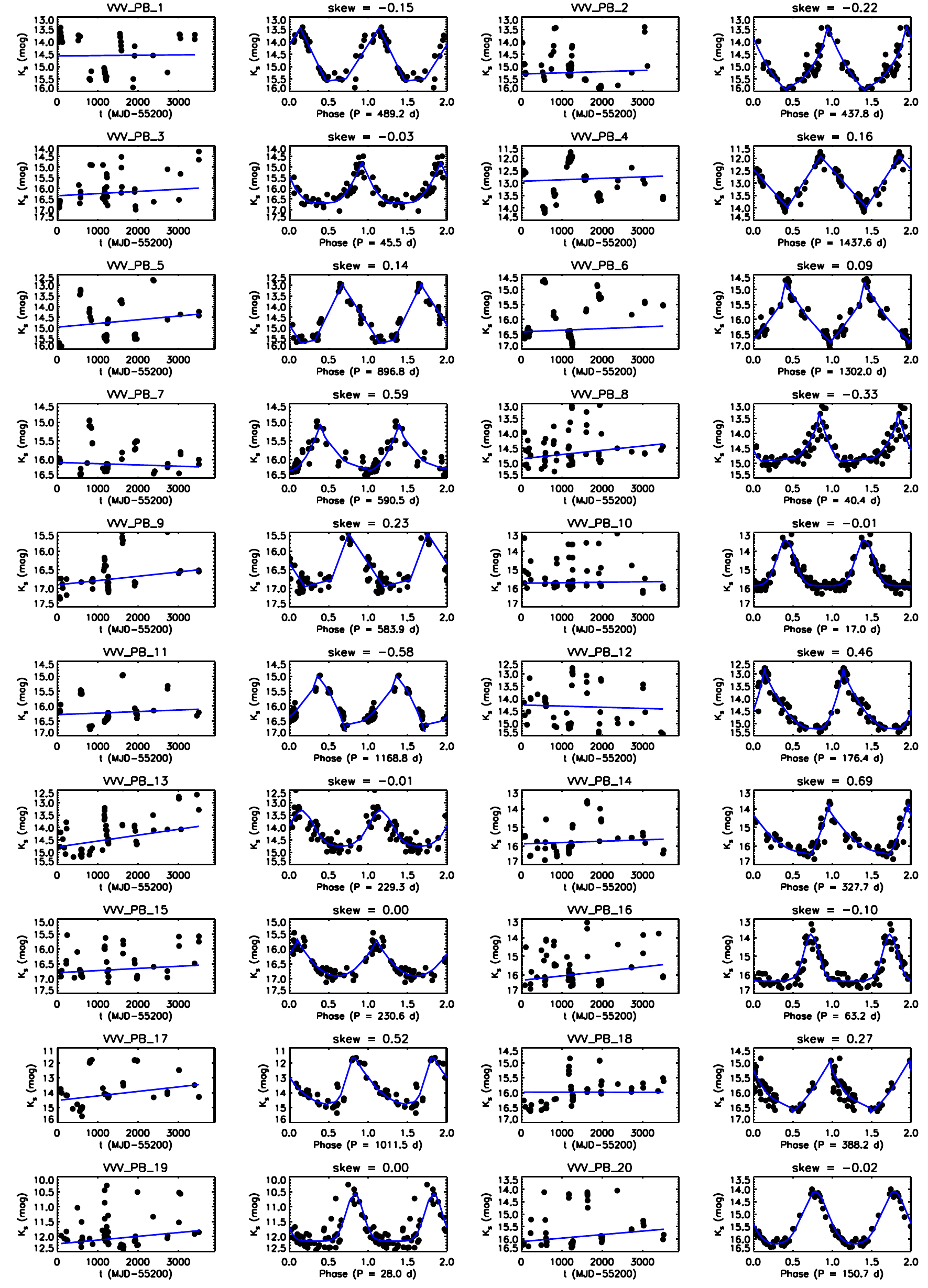}
\caption{VVV/VVVX $K_s$-band light curves and phase-folded light curves of periodic outbursting YSO candidates discovered in the first search. Long-term variation is shown by the blue solid line on light curves, which is removed before the phase-folding. Analytical fittings are shown by the blue curves on the phase-folded light curves.}
\label{fig:PB_1}
\end{figure*}

\begin{figure*}
\centering
\includegraphics[width=6.5in,angle=0]{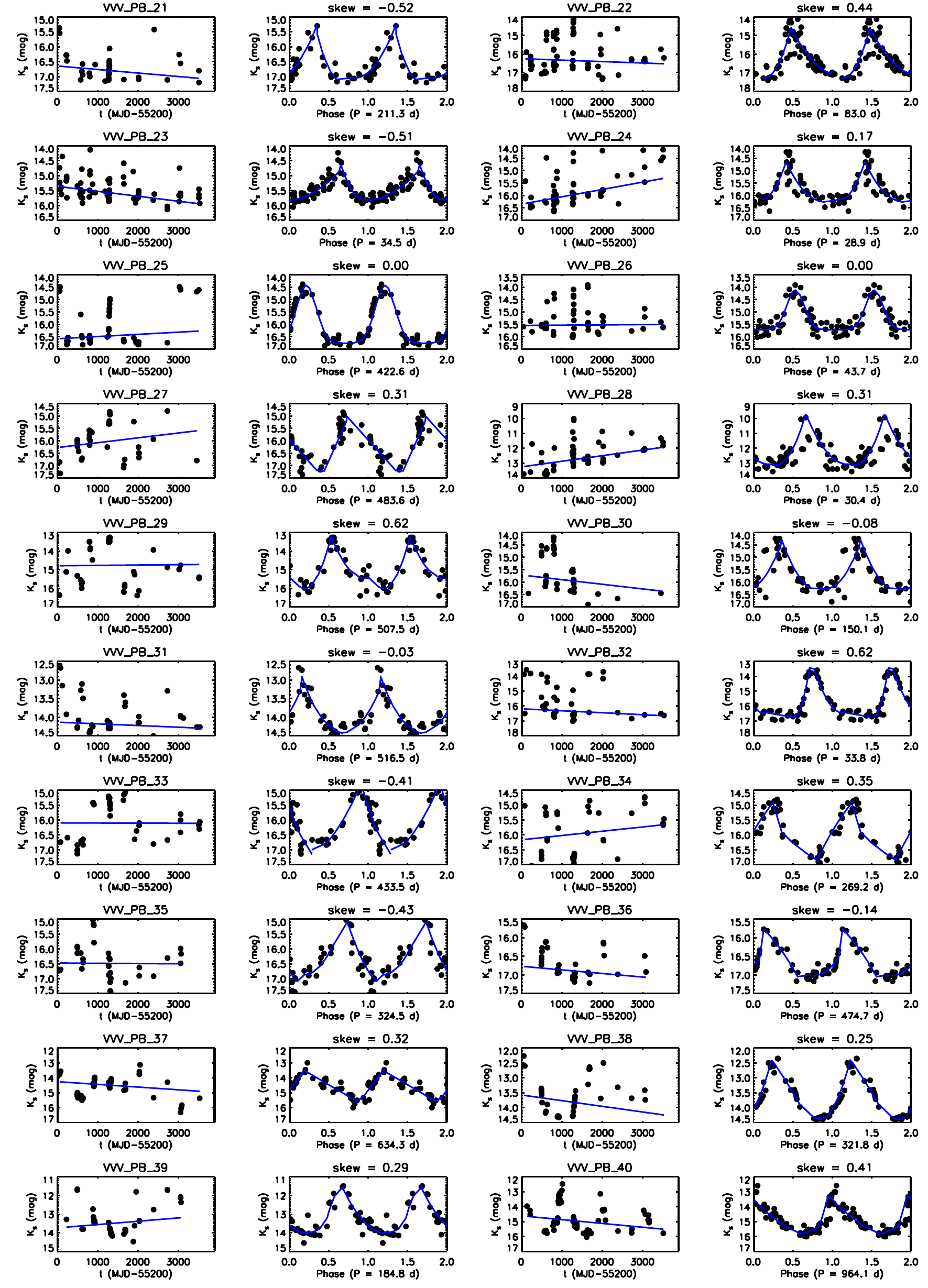}
\caption{Figure~\ref{fig:PB_1} continued.}
\end{figure*}

\begin{figure*}
\centering
\includegraphics[width=6.5in,angle=0]{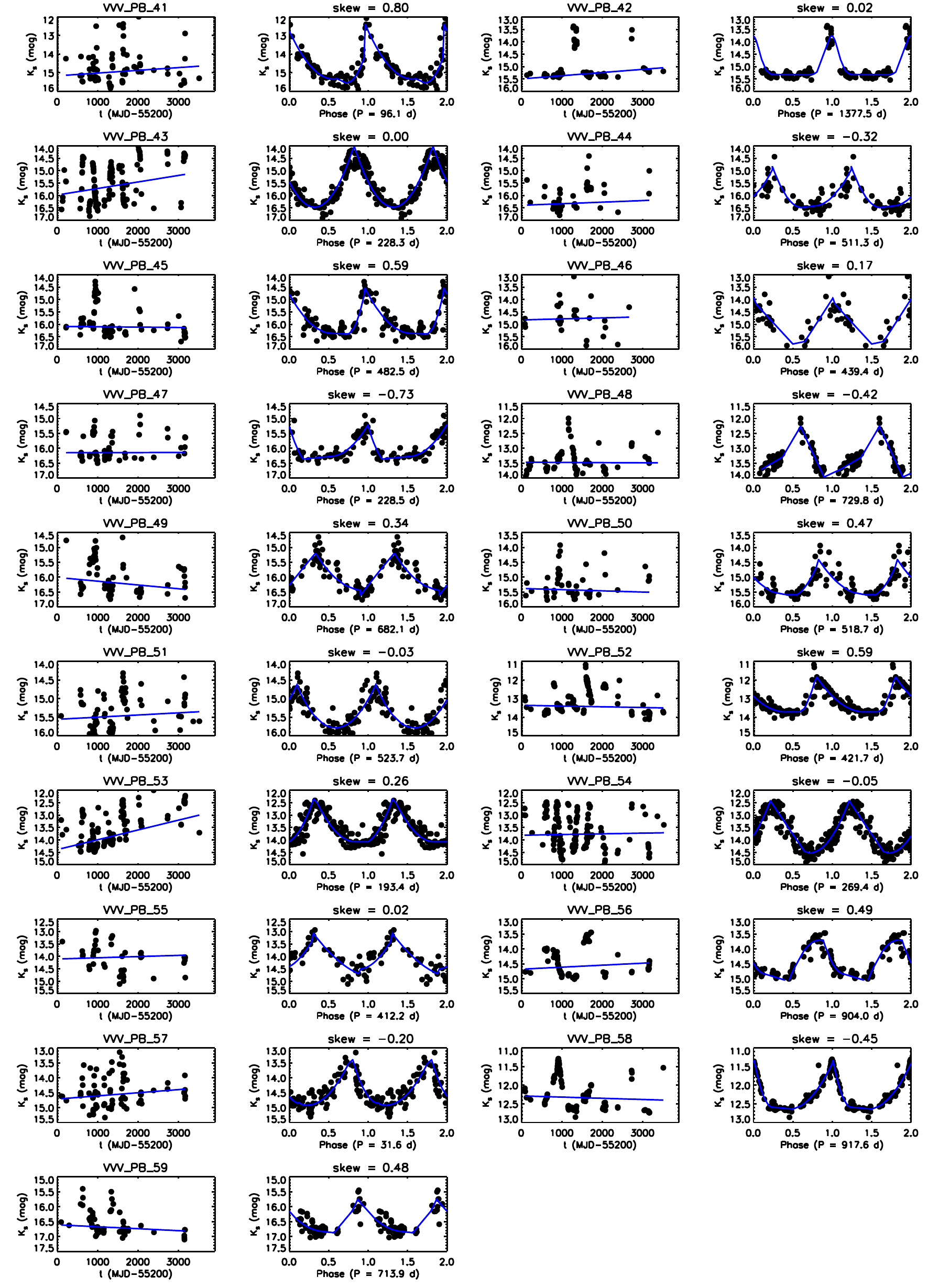}
\caption{Figure~\ref{fig:PB_1} continued.}
\end{figure*}

\begin{figure*}
\centering
\includegraphics[width=6.5in,angle=0]{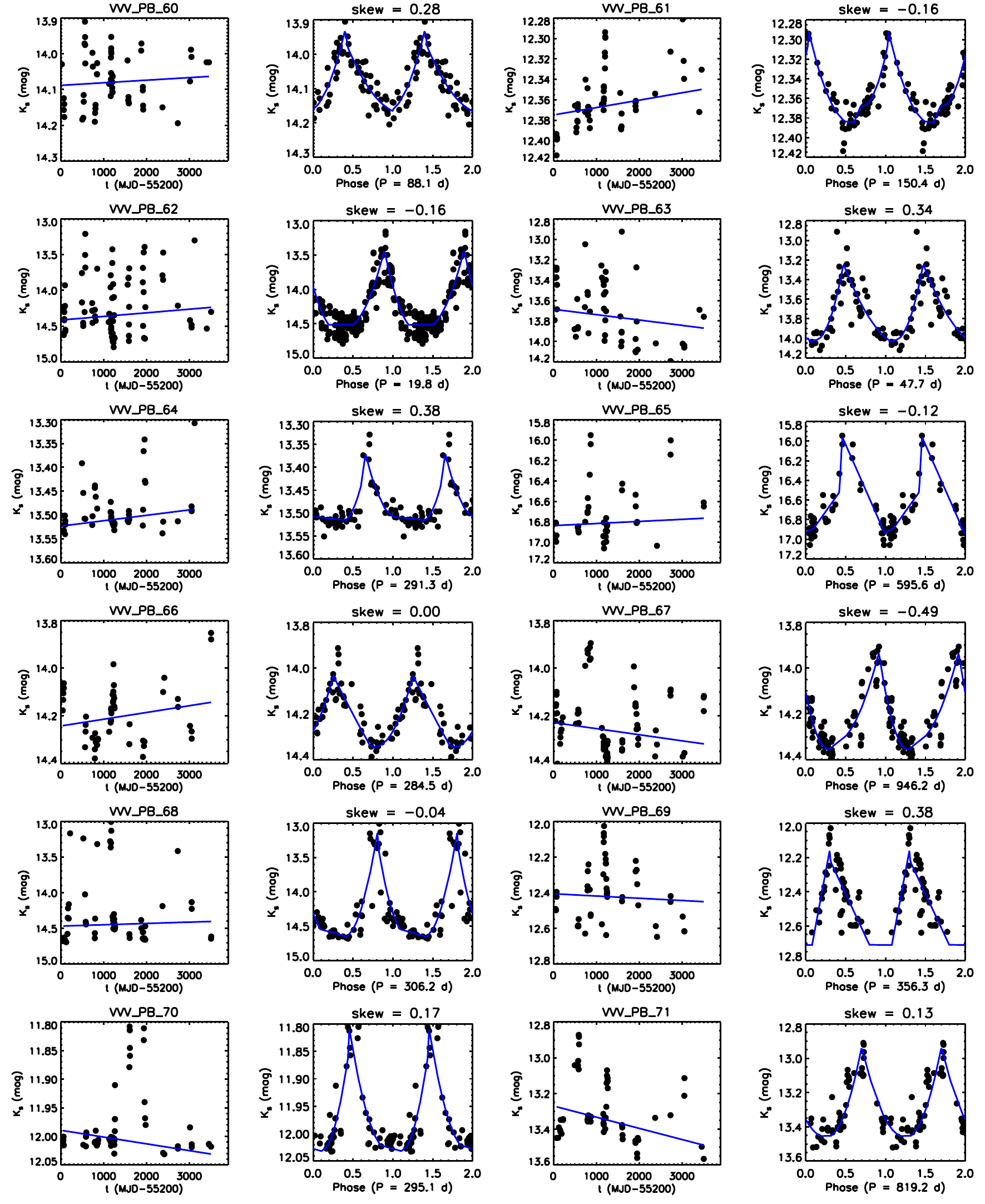}
\caption{VVV/VVVX $K_s$-band light curves and phase-folded light curves of periodic outbursting YSO candidates discovered in the second search. Long-term variation is shown by the blue solid line on light curves, which is removed before the phase-folding. Analytical fittings are shown by the blue curves on the phase-folded light curves.
}
\label{fig:PB_low_1}
\end{figure*}

\begin{figure*}
\centering
\includegraphics[width=6.5in,angle=0]{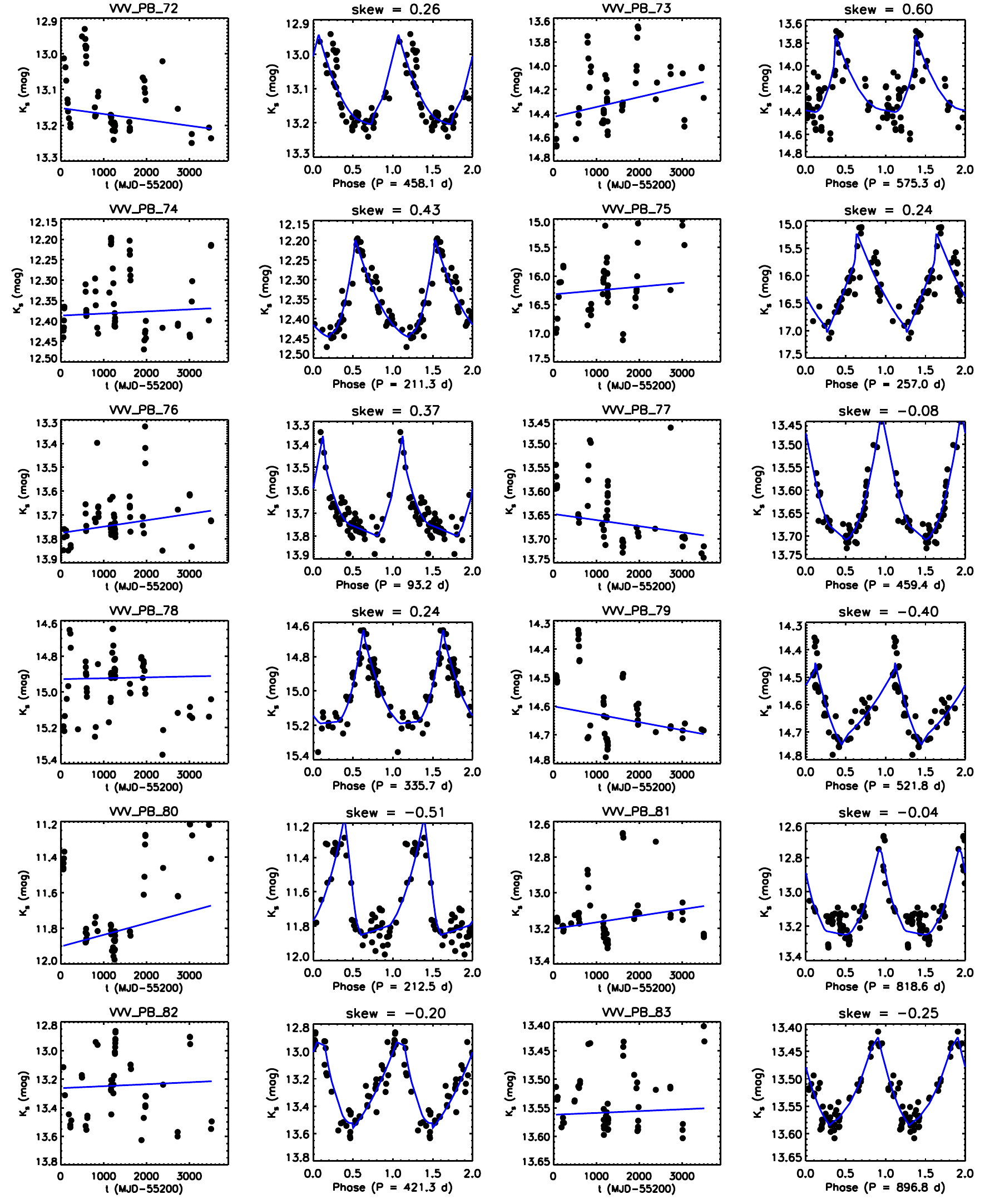}
\caption{Figure~\ref{fig:PB_low_1} continued.}
\label{fig:PB_low_2}
\end{figure*}

\begin{figure*}
\centering
\includegraphics[width=6.5in,angle=0]{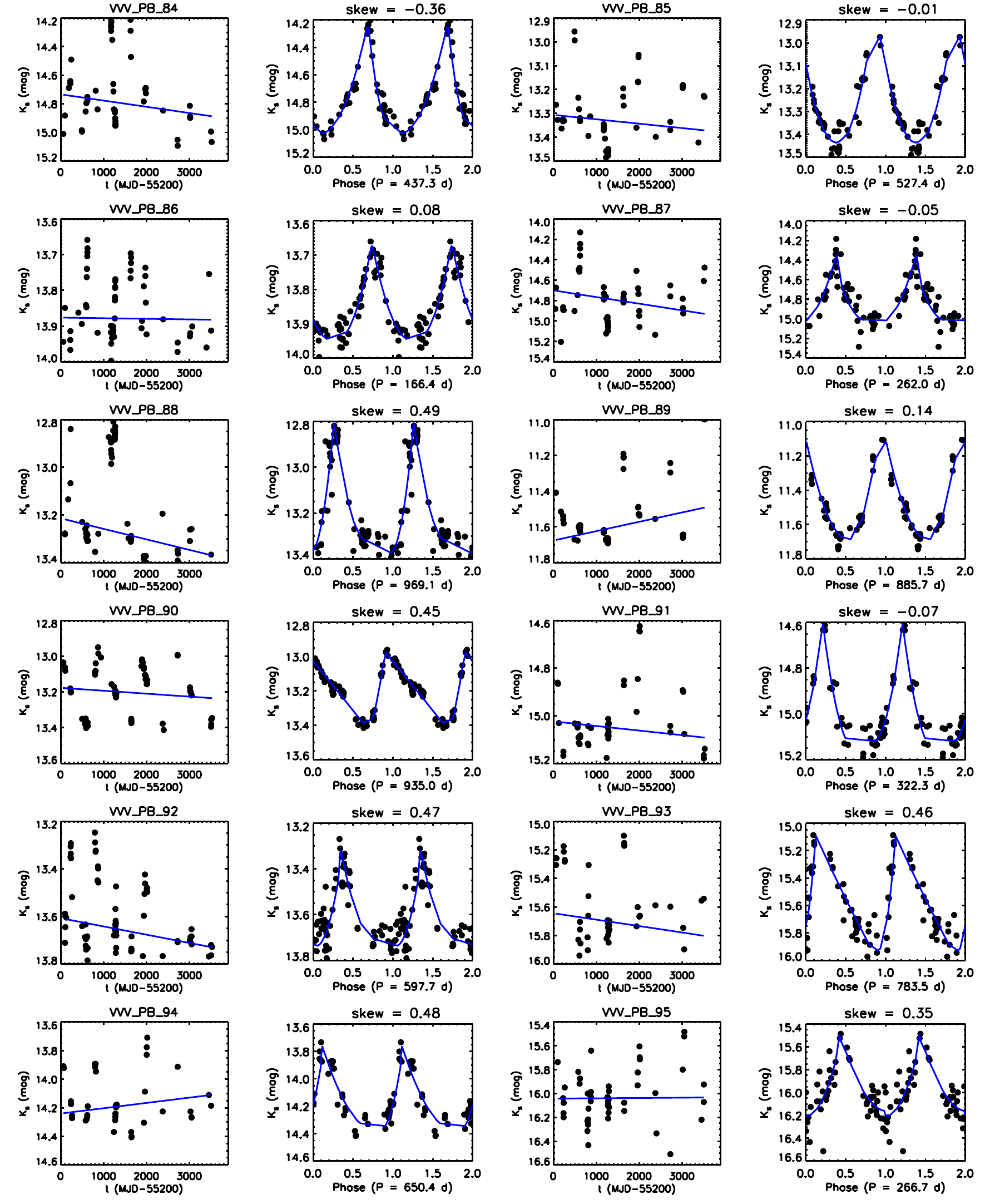}
\caption{Figure~\ref{fig:PB_low_1} continued.}
\label{fig:PB_low_3}
\end{figure*}

\begin{figure*}
\centering
\includegraphics[width=6.5in,angle=0]{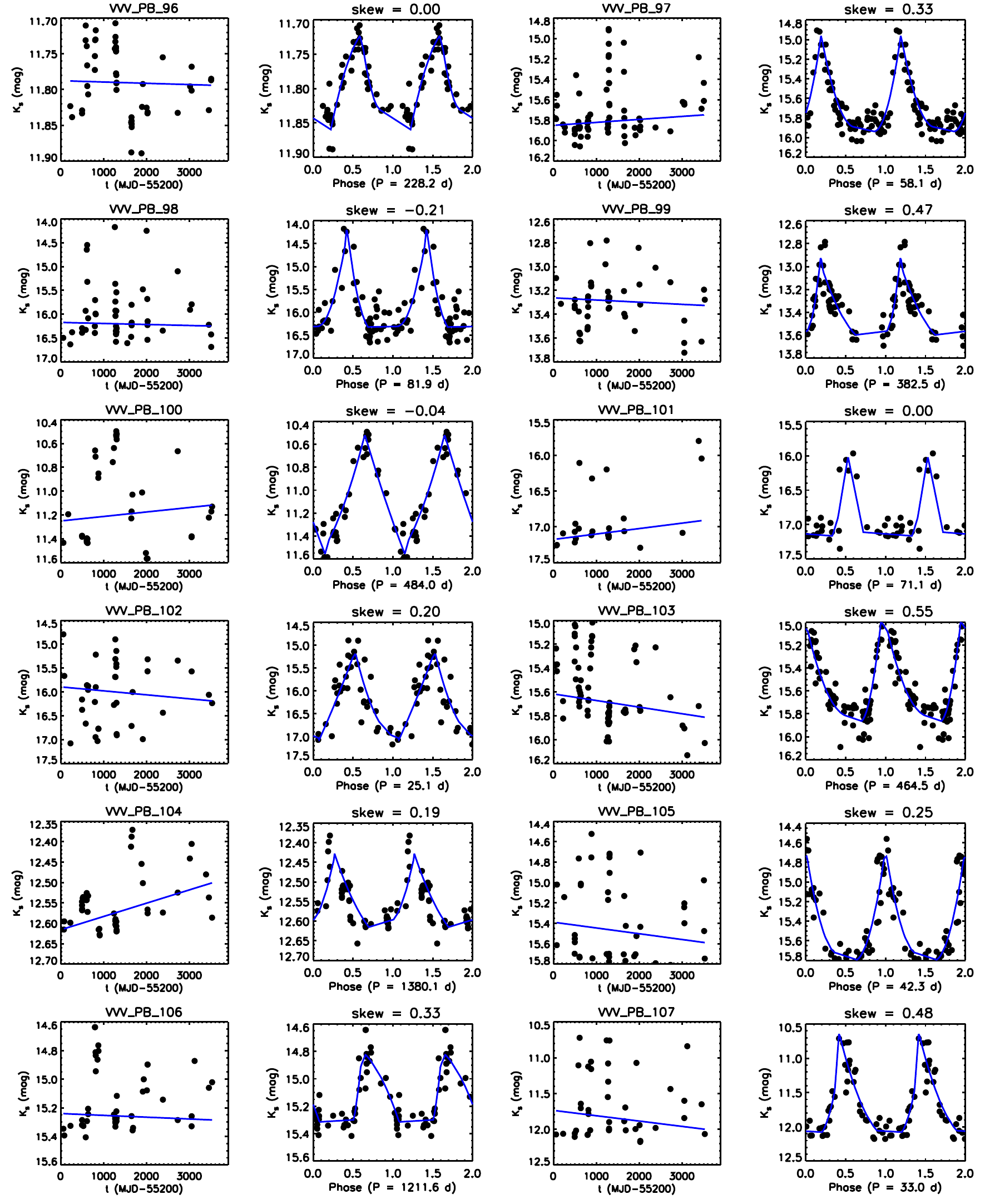}
\caption{Figure~\ref{fig:PB_low_1} continued.}
\label{fig:PB_low_4}
\end{figure*}

\begin{figure*}
\centering
\includegraphics[width=6.5in,angle=0]{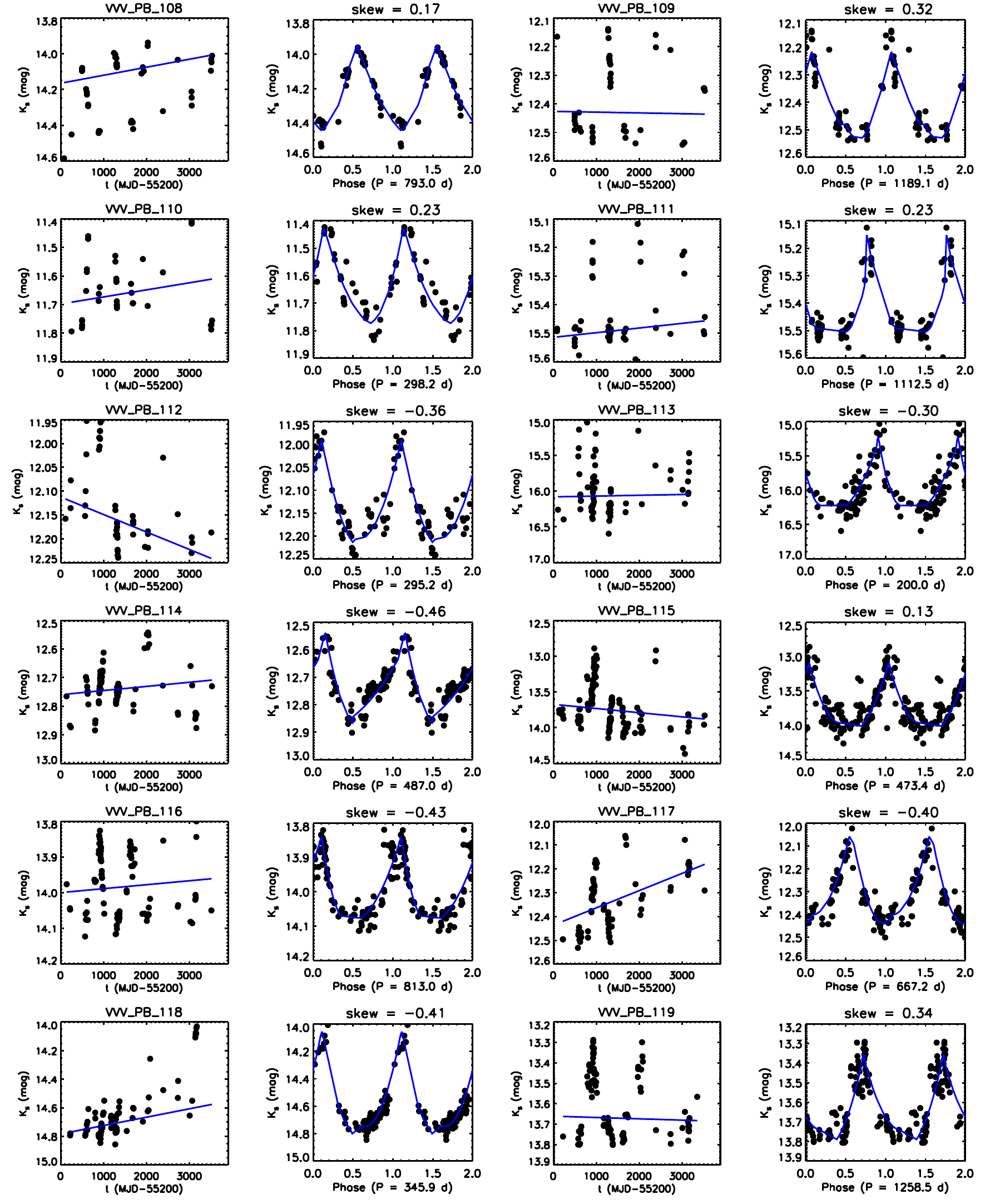}
\caption{Figure~\ref{fig:PB_low_1} continued.}
\end{figure*}

\begin{figure*}
\centering
\includegraphics[width=6.5in,angle=0]{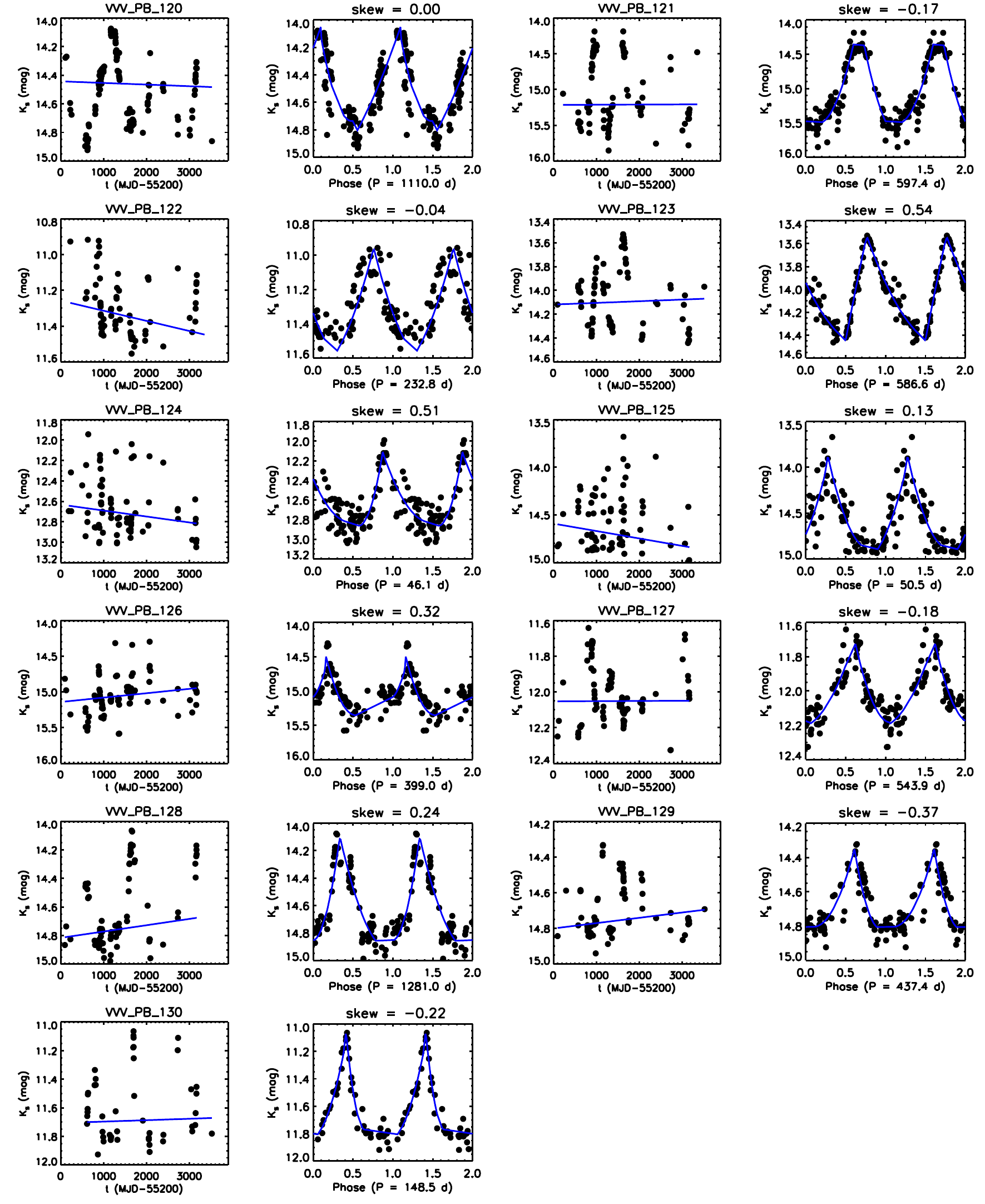}
\caption{Figure~\ref{fig:PB_low_1} continued.}
\end{figure*}